\documentclass[superscriptaddress]{revtex4-2}

\usepackage[utf8]{inputenc}

\date{\today}

\usepackage{url}
\usepackage{amsmath}
\usepackage{natbib}
\usepackage{graphicx}
\usepackage{float}
\usepackage{indentfirst}
\usepackage{geometry}

\usepackage{booktabs}
\usepackage{color}
\usepackage{caption}
\usepackage[position=top, singlelinecheck=false]{subfig}
\usepackage{setspace}
\usepackage{verbatim}
\usepackage{bm}
\usepackage{commath}

\begin{document}
\title{Droplet Bag Formation in Turbulent Airflows}

\author{Kaitao Tang}
\affiliation{Department of Engineering Science, University of Oxford, Oxford OX1 3PJ, UK}

\author{Thomas A. A. Adcock}
\affiliation{Department of Engineering Science, University of Oxford, Oxford OX1 3PJ, UK}

\author{Wouter Mostert}
\affiliation{Department of Engineering Science, University of Oxford, Oxford OX1 3PJ, UK}

\maketitle

We present novel numerical simulations investigating the evolution of liquid droplets into bag-like structures in turbulent airflows. The droplet bag breakup problem is of significance for many multiphase processes in scientific and engineering applications. Turbulent fluctuations are introduced synthetically into a mean flow, and the droplet is inserted when the air-phase turbulence reaches a statistically stationary state. The morphological evolution of the droplet under different turbulence configurations is retrieved and analysed in comparison with laminar aerobreakup results. While the detailed evolution history of individual droplets varies widely between different realisations of the turbulent flow, common dynamic and morphological evolution patterns are observed. The presence of turbulence is found to enhance the drag coefficient of the droplet as it flattens. At late times, the droplet becomes tilted and increasingly corrugated under strong turbulence intensity. We quantify these phenomena and discuss their possible governing mechanisms associated with turbulence intermittency. Lastly, the influences of liquid-gas viscosity ratio are examined and the implications of air-phase turbulence on the later bag film breakup process are discussed.

\section{Introduction}
\label{sec:introduction}

Atomisation involves the morphological evolution of the liquid bulk driven by external forces, followed by the formation of corrugated ligaments subject to capillary pinch-off, and terminates in a number of fragments featuring a broad size distribution \cite{Villermaux2007}. The aerobreakup of spherical liquid droplets in uniform airflows is a canonical problem of atomisation and two-phase fluid dynamics \cite{jackiw2022prediction, tang2022bag}, an indispensable fragmentation mechanism found in numerous natural and industrial processes, including precipitation \cite{Villermaux2009}, sea spume production \cite{Troitskaya2017}, disease transmission \cite{bourouiba2021fluid, kant2022bags}, liquid fuel combustion \cite{boyd2024simulation} and inkjet printing \cite{lohse2022fundamental}. In a uniform flow, droplet aerobreakup can be parameterised by the following four non-dimensional controlling parameters \cite{Jalaal2014}:
\begin{equation}
    We \equiv \frac{\rho_g U_0^2 d_0}{\sigma}, \quad Oh \equiv \frac{\mu_l}{\sqrt{\rho_l d_0 \sigma}}, \quad
    \rho_r \equiv \frac{\rho_l}{\rho_g}, \quad
    \mu_r \equiv \frac{\mu_l}{\mu_g}.
    \label{for:non-dimensional-groups}
\end{equation}
Among these, ${\rm We}$ and ${\rm Oh}$ are respectively the Weber and Ohnesorge numbers quantifying the ratio of inertial to capillary and viscous to capillary forces, and $\rho_r$ and $\mu_r$ are respectively the density and viscosity ratios of the liquid and gas phases. Here, $\rho_l$, $\mu_l$ and $\rho_g$, $\mu_g$ are respectively the density and viscosity for the liquid and gas phase, $d_0$ is the initial diameter of the droplet, $U_0$ is the velocity of the ambient airflow, and $\sigma$ is the surface tension at the liquid-gas interface \cite{Guildenbecher2009}.

The effects of ${\rm We}$ and ${\rm Oh}$ on the morphological evolution of droplets undergoing aerobreakup have been extensively investigated in many experimental and numerical studies \cite{Marcotte2019}. Various aerobreakup regimes, including bag, multi-mode (bag-stamen), sheet-thinning and catastrophic breakup, have been observed and the transition thresholds between them have been delineated using these two non-dimensional parameters \cite{Hsiang1995, Theofanous2011, Yang2017}. Among these regimes, the investigation of the bag breakup regime can potentially shed light on the mechanism of ocean spray production \cite{Veron2015}, as the bag-shaped geometry of the late-time deformed drops bears a strong resemblance to the small-scale sea-surface perturbations, whose breakup dominates sea spume generation under extreme wind conditions \cite{Troitskaya2017, Troitskaya2018}. Most of the early works investigating droplet aerobreakup focus on the analysis of bag formation using hydrodynamic instability models \cite{sharma2022advances}, while discussions on the late-time bag breakup mechanisms and fragment statistics are still scarce \cite{Zhao2011, Jackiw2021}. The recent experimental work by Jackiw and Ashgriz \cite{jackiw2022prediction} discussed in detail the physical mechanisms governing the early-time droplet deformation patterns, which are not captured by the simplified hydrodynamic instability models. They also observed that the major pathways leading to film fragment generation are the destabilisation and collision of receding hole rims under centripetal acceleration, and an analytical model predicting the overall volume probability distribution of aerobreakup fragments is developed and verified at various ${\rm We}$ values \cite{ade2023size}. Numerical investigations into fragment statistics, especially those exploiting the flexible Volume-of-Fluid (VoF) interface reconstruction method, have historically been handicapped by the lack of grid convergence, which arises from uncontrolled numerical perforation of bag films when they reach the minimum grid size. However, recently Chirco \emph{et al.} \cite{chirco2021manifold} developed a Manifold Death (MD) algorithm that prevents such spurious breakup of thin films by artificially perforating them once their thickness reduces to a prescribed critical value. This algorithm improves the grid convergence behaviour of the statistics of large bag film fragments, allowing further investigation into the generation mechanism and behaviour of individual aerobreakup fragments \cite{tang2022bag, kulkarni2024atomizing}.

Nevertheless, despite the progress made against these limitations, a complete accounting of the realistic physical environment around droplet breakup has not yet been achieved. One important effect is that of turbulence in the freestream gas flow. In most of the natural phenomena as well as previous experimental studies, droplet breakup occurs in turbulent gas flows. For example, combustion chamber flows may feature a variety of integral length scales \cite{jiao2019direct}, and spume generation at the air-sea interface occurs within the turbulent atmospheric boundary layer \cite{Troitskaya2018}, which can be further modified over time due to coupled wind-wave interactions \cite{sullivan2010dynamics, wu2022revisiting}. In these scenarios, apart from the aerodynamic force exerted by the mean flow, accelerating, shearing and fluctuating effects of the turbulent flow can also affect the deformation and breakup of the droplet \cite{Xu2020}. In particular, a detailed and quantitative analysis approach is still lacking for the interaction between the droplet surface and the turbulent vortices present in the ambient gas flow. It is also noted that, although droplet-turbulence interactions have seen some investigation in recent years with important findings, e.g., droplets larger and smaller than the Hinze scale demonstrate different behaviour: the former tend to break up through a memoryless process \cite{vela2022memoryless} and absorb energy from the ambient flow, while the latter oscillate rapidly or coalesce to release energy to the environment \cite{crialesi2023interaction}; these studies lack a base flow and are mostly carried out at low density ratios close to 1, which differ from the air-sea interaction scenario featuring finite wind speeds and large density contrasts.

Despite the difficulty in accurately measuring turbulent flow properties \cite{jiao2019direct} and quantifying the response of droplets to external turbulence forcing, there have been a few experimental studies of turbulent aerobreakup in recent years \cite{zhao2019effect, Xu2020, xu2022droplet}. Among these, Zhao \emph{et al.} \cite{zhao2019effect} utilised perforated plates to generate turbulent counterflows, which in turn drives the deformation of droplets falling from above. It is found that increasing turbulence intensity results in larger lengths and widths of the bag, which then breaks up to form smaller fragments compared with uniform flow conditions. Xu \emph{et al.} \cite{Xu2020, xu2022droplet} placed droplets within the shear layer of a turbulent air jet, where two new breakup regimes with complicated morphological patterns are identified: butterfly and swing breakup. 

Numerical studies of the turbulent aerobreakup problem are even more scarce, most likely due to the difficulty in imposing turbulent fluctuations on a uniform mean flow, and the high computational cost of resolving all turbulence length scales apart from the droplet dynamics. The linear-forcing method proposed by Rosales and Meneveau \cite{rosales2005linear} cannot be directly applied to the aerobreakup problem, despite its recent successful application to turbulent bubble breakup studies \cite{perrard2021bubble, riviere2022capillary}. This is because the linear-forcing method is usually implemented with triply-periodic boundary conditions, while in the aerobreakup problem the droplet produces a separated wake region that breaks the fore-aft symmetry of the ambient flow field; in other words, a droplet undergoing aerobreakup can be affected by the wake of its preceding periodic image \cite{loisy2017interaction}. On the other hand, synthetic turbulence generation methods produce pseudo-turbulence fluctuation by summation of random Fourier modes (Synthetic Random Fourier Method), or superposition of spatial and temporal coherence on initially uncorrelated random number series (Synthetic Digital Filtering Method). Jiao \emph{et al.} \cite{jiao2019direct}, to our knowledge the only investigation of the turbulent aerobreakup problem utilising direct numerical simulations, applied the LEMOS inflow generator \cite{kornev2007method} at the inlet with outflow boundary conditions. This generator represents turbulence as a series of random spots whose inner velocity distributions are determined by prescribed autocorrelation functions, which enables generation of turbulence structures with prescribed integral length scales and fluctuating velocities on the mean flow \cite{kornev2007method}. Their results suggest that superimposed air-phase turbulence leads to rapid redistribution of flow pressure around the droplet, which can cause asymmetric deformation patterns coupled with various forms of rotational motion; while the bag film breakup process and associated fragment statistics are not covered due to limited grid resolution level. It is also noted that this work is carried out at a relatively low liquid-gas density ratio $\rho_r = 20$, which means that their results might not be directly applicable to air-water systems where the density ratio is much larger.

Overall, it can be seen that the currently available results of the turbulent aerobreakup studies are very limited and obtained under considerably different flow configurations. A well-defined methodological framework is still lacking, based on which systematic investigations of this problem can be carried out. Droplet bag breakup within a uniform incoming flow with a superimposed, well-characterised turbulent fluctuation features a simple configuration and therefore serves as a first step towards understanding aerobreakup in more complicated turbulent flow environments, e.g. boundary-layer flows where shear effects dominate. We investigate the turbulent aerobreakup problem using two-phase and three-dimensional direct numerical simulations. Our focus is on elucidating the droplet deformation mechanisms up to the point of bag formation, which serves as a basis for future analysis of the breakup behaviour and fragment statistics. Our study is structured as follows. We present in \S\ref{subsec:description} our problem configuration and parameter space, and then introduce the numerical method in \S\ref{subsec:num-method}, with a focus on the synthetic turbulence generation method \cite{xie2008efficient} we adopt for producing turbulent incoming airflows. We proceed to present in \S\ref{sec:overview-drop-deform} a brief overview of the droplet-turbulence interactions. We then discuss in \S\ref{sec:drop-turb-dynamics} the evolution of the droplet centre-of-mass dynamic properties, and the air-phase pressure distributions giving rise to such evolution trends. Afterwards, we analyse in \S\ref{sec:drop-deform} various droplet deformation patterns, including the aspect ratio (\S\ref{subsec:global}), the tilting behaviour (\S\ref{subsec:tilting}) and the surface curvature and velocity increment distributions (\S\ref{subsec:corrug-form}). Finally, we discuss the possible influences of turbulent airflows on the bag film fragmentation process, and conclude the study in \S\ref{sec:discussions} with some remarks on future work.

\section{Formulation and methodology}
\label{sec:formulation}

\subsection{Problem description}
\label{subsec:description}
\begin{figure}[htbp]
	\centering
	\includegraphics[width=.6\textwidth]{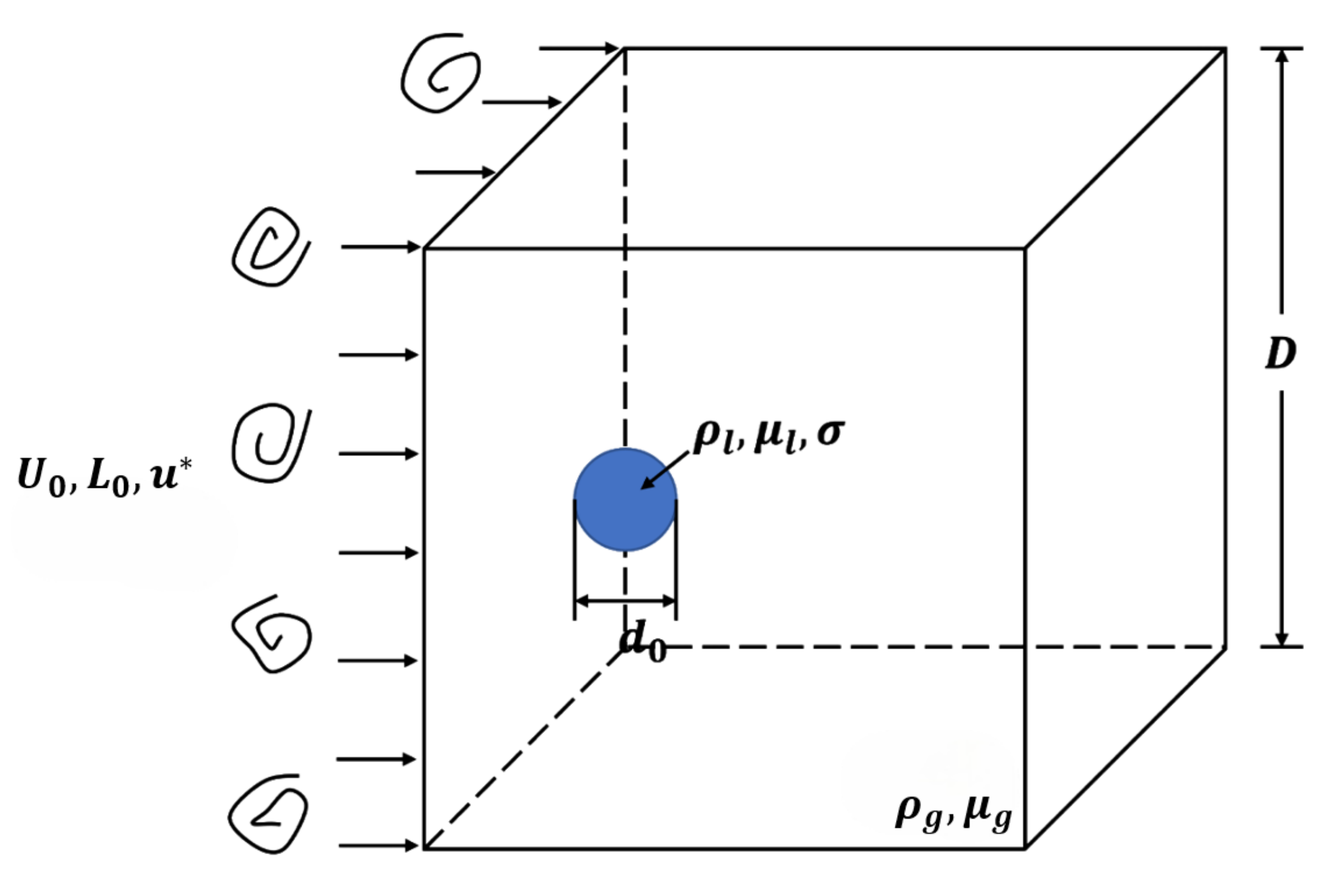}
	\caption{Sketch showing the configuration of the turbulent droplet aerobreakup problem.}
	\label{fig:flow-config-turb}
\end{figure}

The configuration for our fully three-dimensional direct numerical simulations is shown in fig.~\ref{fig:flow-config-turb}. A liquid droplet with initial diameter $d_0$, density $\rho_l$ and viscosity $\mu_l$ is inserted close to the left boundary, within a turbulent gas phase featuring density $\rho_g$ and viscosity $\mu_g$. The domain width $D$ is set as $10d_0$, at which the influence of finite domain size on the aerobreakup process can be considered negligible. Outlet flow conditions are applied at the right boundary and a turbulent inflow with mean velocity $U_0$, fluctuating velocity $u^*$ and injection length scale $L_0$ is introduced at the left boundary, while zero-gradient conditions are applied at the other domain boundaries.

Among the dimensional parameters introduced above, the fluctuation velocity $u^*$ and the injection length scale $L_0$ characterises the swirling motion of the largest eddies within the synthesised air-phase turbulence. These in turn lead to two additional non-dimensional groups \cite{riviere2021sub} apart from those already introduced in Eq.~\eqref{for:non-dimensional-groups},
\begin{gather}
    \widetilde{\rm We} \equiv \frac{2\rho_g \varepsilon^{2/3} d_0^{5/3}}{\sigma} = \frac{2\rho_g {u^*}^2 d_0^{5/3}}{\sigma L_0}^{2/3}, \\ 
    \frac{d_h}{d_0} \propto {\left( \frac{\sigma}{\rho_g} \right)}^{3/5} \varepsilon^{-2/5} = {We}^{-3/5} {\left( \frac{u^*}{U_0} \right)}^{-6/5} {\left( \frac{L_0}{d_0} \right)}^{2/5},
\end{gather}
where $\widetilde{\rm We}$ and $d_h/d_0$ are respectively the turbulent Weber number and the Hinze scale non-dimensionalised by the initial droplet diameter; and $\varepsilon$ is the turbulence dissipation rate per unit volume, which we scale using the nominal value of ${u^*}^3/L_0$ for the turbulent fluctuations we produce. The resulting turbulence reaches statistical stationarity after an initial startup transient, is unforced in the interior of the numerical domain, and shows spatially a gradual decay in the streamwise direction (parallel to the mean flow), but is homogeneous in the spanwise plane. The turbulence is further characterised in \S\ref{subsec:num-method} below.

As for the four non-dimensional parameters defined in \eqref{for:non-dimensional-groups}, namely the mean flow Weber number ${\rm We}$, the Ohnesorge number ${\rm Oh}$, the density ratio $\rho_r$ and the viscosity ratio $\mu_r$, we fix ${\rm We}, \, {\rm Oh}$ and $\rho_r$ as 15, 0.005 and 833, as is typical for air-water droplet bag breakup. We test three values of $\mu_r$, namely 55, 20 and 10, which allows us to analyse the influence of the liquid-gas viscosity ratio on turbulent aerobreakup in \S\ref{sec:visc-ratio}. It should be noted that the single configuration with $\mu_r = 10$ only serves to examine the effects of $\mu_r$ and can be difficult to obtain for liquid-gas systems under experimental conditions. For reference to the readers, we note that these three viscosity ratios correspond to three different gas-phase Reynolds numbers $Re \equiv \rho_g U_0 d_0/\mu_g = 1476$, 537 and 268. At this magnitude of $Re$, flows past a solid sphere would begin to transit to turbulence \cite{rodriguez2011direct, tang2022bag}. A full list showing the parameter space of this study is provided as Table~\ref{tab:configs}, featuring a wide range of $\widetilde{\rm We}$ and $d_h/d_0$ values including both super- and sub-Hinze-Scale droplets. The presence of significant variations within the turbulent ambient flow necessitates ensemble averaging across multiple realisations, thus enabling us to determine the statistically converged evolution patterns or distributions for physical properties under a certain turbulence configuration. We have therefore provided the number of ensemble realisations in the last column of Table~\ref{tab:configs}.

\begin{table}
\setlength{\tabcolsep}{1em}
\centering
\begin{tabular}{c|c|c|c|c|c|c|c|c|c}
    \toprule
    $u^*/U_0$ & $L_0/R_0$ & $\mu_r$ & ${\rm Re}$ & $\widetilde{\rm We}$ & $d_h/d_0$ & $\lambda/d_0$ & $Re_{\lambda}$ & $\eta/d_0$ & Real. No. \\ [0.5ex] 
    \midrule
    0 & 0 & 55 & 1476 & - & - & - & - & - & 1 \\
    0.25 & 1 & 55 & 1476 & 2.976 & 1.008 & 0.142 & 52.8 & 0.0100 & 5 \\
    0.25 & 1.5 & 55 & 1476 & 2.271 & 1.186 & 0.174 & 64.6 & 0.0110 & 5 \\
    0.25 & 2 & 55 & 1476 & 1.875 & 1.331 & 0.202 & 74.6 & 0.0119 & 5 \\
    0.25 & 4 & 55 & 1476 & 1.181 & 1.756 & 0.284 & 106 & 0.0141 & 5 \\
    0.25 & 8 & 55 & 1476 & 0.744 & 2.317 & 0.402 & 149 & 0.0167 & 5 \\
    0.5 & 1 & 55 & 1476 & 11.90 & 0.439 & 0.101 & 74.6 & 0.0059 & 5 \\
    0.65 & 1 & 55 & 1476 & 20.12 & 0.320 & 0.088 & 85.0 & 0.0049 & 5 \\
    0.8 & 1 & 55 & 1476 & 30.48 & 0.249 & 0.080 & 94.4 & 0.0042 & 5 \\
    0.25 & 1 & 20 & 537 & 2.976 & 1.008 & 0.236 & 31.7 & 0.0213 & 3 \\
    0.5 & 1 & 20 & 537 & 11.90 & 0.439 & 0.167 & 44.9 & 0.0127 & 3 \\
    0.65 & 1 & 20 & 537 & 20.12 & 0.320 & 0.146 & 51.2 & 0.0104 & 3 \\
    0.8 & 1 & 20 & 537 & 30.48 & 0.249 & 0.132 & 56.8 & 0.0089 & 3 \\
    0.25 & 1 & 10 & 268 & 2.976 & 1.008 & 0.334 & 11.2 & 0.0359 & 3 \\
    \bottomrule
\end{tabular}
\caption{List showing the parameter space for three-dimensional numerical simulations of turbulent bag formation carried out in this work.}
\label{tab:configs}
\end{table}
Among the physical properties introduced in Table~\ref{tab:configs}, the Taylor microscale $\lambda$, the Taylor Reynolds number $Re_{\lambda}$, and the Kolmogorov microscale $\eta$ are defined as follows \cite{riviere2021sub},
\begin{gather}
    \lambda \equiv \sqrt{\frac{15 \nu_g}{\varepsilon}} u^*, \quad
    Re_\lambda \equiv \frac{u^* \lambda}{\nu_g}, \quad 
    \eta \equiv {\left( \frac{\nu_g^3}{\varepsilon} \right)}^{1/4},
    \label{for:lambda-re-eta-defs}
\end{gather}
where $\nu_g \equiv \mu_g / \rho_g$ is the kinematic viscosity of the gas phase. According to Table~\ref{tab:configs}, our turbulence configurations feature Taylor Reynolds number $Re_\lambda$ ranging from 31.7 to 149, which becomes stronger than the turbulence-bubble interaction studies carried out by Rivi{\`e}re \emph{et al.} \cite{riviere2021sub} for simulations with large $u^*$ or $L_0$, but remains typical of two-phase turbulence simulations \cite{dodd2016interaction, Elghobashi2019}.

\subsection{Numerical method}
\label{subsec:num-method}
We use the open-source Basilisk numerical library \cite{Popinet2019basilisk} to solve the two-phase incompressible Navier-Stokes equations, which utilises a finite-volume numerical scheme coupled with the geometric volume-of-fluid (VOF) method for fluid interface reconstruction \cite{popinet2018numerical}. Usage of an octree-based adaptive mesh refinement (AMR) technique significantly reduces the computational cost for such two-phase simulations. Readers are referred to our previous works \cite{tang2022bag, tang2024fragmentation} for a detailed description of the aforementioned numerical methods. Here we focus on the synthetic digital filtering method proposed by Xie and Castro \cite{xie2008efficient}, which we adopt for generating turbulent fluctuations at the inlet.

As is mentioned in \S\ref{sec:introduction}, the linear forcing method of Rosales and Meneveau \cite{rosales2005linear} has seen successful applications in turbulence-bubble interaction studies \cite{riviere2021sub, perrard2021bubble}. It adds a volumetric forcing term to the Navier-Stokes equation in the physical space, which leads to the transition from an ABC-flow at initialisation to well-characterised HIT fluctuations at steady-state. However, this method cannot be applied to the present turbulent aerobreakup problem, otherwise the deformation and breakup of the drop will be affected by the wake of its preceding mirror image. To show this, a scaling analysis can be made based on existing works on wake velocity defects (the difference between velocities at the same location with and without the obstacle) of spherical obstacles immersed in a turbulent carrier phase \cite{legendre2006wake, loisy2017interaction}. Namely, as the distance to a solid spherical obstacle $z$ increases, the magnitude of the velocity defect $\Delta u$ decays following a power law of $z^{-1}$ within the range of $z \leq 13d_0$, which asymptotes to $z^{-2}$ at larger $z$ values. In our case where the liquid-gas density ratio $\rho_r$ is large, the droplet can be approximated as a solid obstacle, and the velocity defect caused by the droplet's preceding mirror image reads
\begin{equation}
    \frac{\Delta u_i}{u^*} \propto \frac{U_0}{u^*} {\left( \frac{D}{d_0} \right)}^{-1}.
    \label{for:vel-defect-estimation}
\end{equation}
For a typical simulation configuration considered in the current work, where $U_0/u^* \approx 10$ and $D/d_0 = 10$, Eq.~\eqref{for:vel-defect-estimation} suggests that the magnitude of velocity defect caused by the preceding periodic image of the droplet is the same as the turbulent fluctuating velocity of the ambient airflow. In other words, significant mutual interactions between neighbouring drops would arise across periodic boundaries. While such interactions may be eliminated by further increasing the domain size $D$, this would incur higher computational costs for resolving turbulence in an enlarged simulation domain, which suggests that the linear forcing method cannot be applied directly with our current computational capabilities. In addition, the turbulent flows generated by this method feature a fixed injection length scale $L_0 = 0.19D$, which does not meet our need to change it as a variable.

Based on the discussions above, it would be ideal to generate and maintain homogeneous and isotropic airphase turbulence in a simulation domain with inflow and outflow conditions. Wu \cite{wu2017inflow} provides an overview of currently available inlet turbulence generation methods and their applications, where at each timestep the inlet velocity condition can be set using flow-field slices imported from an external auxiliary simulation (Strong Recycling Method), slices extracted from a downstream location within the current simulation domain (Weak Recycling Method), or synthesised using random number series or Fourier modes filtered by an approximate correlation function (Synthetic Turbulence Generation). Here we select the synthetic turbulence generation method proposed by Xie and Castro \cite{xie2008efficient} due to its simplicity, high efficiency and well-validated turbulence statistics including inertial subrange and two-point velocity correlations \cite{chen2022freestream}.

The framework of this turbulence generation method is briefly reviewed as follows. Firstly, at each timestep, a two-dimensional random matrix $R_{ij}$ with zero mean and unit variance is generated, and then filtered and normalised to produce two-dimensional space-correlated velocity fluctuation matrices $\psi_{ij}$ for each velocity component,
\begin{equation}
    \psi_{m,l} = \sum_{j=-N}^N \sum_{k=-N}^N b_j b_k R_{m+j, \, l+k}, \quad 0 \leq m, \, l \leq N_{grid},
\end{equation}
where $N_{grid} \equiv 2^L$ is determined by the maximum grid level $L$ for resolving turbulence, and $N = 3n = 3L_0 / \Delta$ is proportional to the number of grid cells spanned by the turbulence integral length $L_0$. The filter coefficients $b_j$ are prescribed as follows to approximate exponential velocity correlation functions,
\begin{equation}
    b_j = \frac{e^{-\frac{\pi|j|}{2n}}}{\sqrt{\sum_{i=-N}^N e^{-\frac{\pi|i|}{n}}}}.
\end{equation}

Time correlation is imposed between turbulent inlet conditions at successive timesteps by computing the new normalised fluctuating velocity components $\Psi(t)|_{m,l}$ based on that of the previous timestep $\Psi(t-\Delta t)|_{m,l}$ and newly synthesised fluctuation matrices $\psi_{m,l}$,
\begin{equation}
    \Psi(t)|_{m,l} = \Psi(t-\Delta t)|_{m,l} e^{-\frac{\pi \Delta t}{4T_{La}}} + \psi_{m,l} \sqrt{1 - e^{-\frac{\pi \Delta t}{2T_{La}}}}, \quad 0 \leq m, \, l \leq N_{grid},
\end{equation}
where $T_{La} \equiv L_0/U_0$ is the injection time scale. Afterwards, the full inlet velocity profile is generated as follows:
\begin{equation}
    U_{in}|_{m,l} = U_0 + u^2_{rms} \Psi(t)|_{m,l}, \quad 0 \leq m, \, l \leq N_{grid},
\end{equation}
where the constant coefficient $u^2_{rms}$ is a simplification of the Cholesky decomposition of the Reynolds stress tensor \cite{kim2013divergence, chen2022freestream} for homogeneous and isotropic turbulence (HIT) fluctuations. $U_{in}$ is then corrected following Kim \emph{et al.} \cite{kim2013divergence} to ensure constant mass influx, and superimposed at the left boundary of the simulation domain as a time-variant Dirichlet boundary condition.

Following Rivi{\`e}re \emph{et al.} \cite{riviere2021sub}, we first run precursor simulations without droplets to create statistically stationary turbulent air-phase flows within the entire simulation domain, where the maximum grid resolution level for the AMR scheme is set as $L_{turb} = 8$ for simulations with $\mu_r = 55$, and 9 for those with $\mu_r = 20$. In the latter case, we reach a minimum grid size of $\Delta_{turb} = D/2^{L_{turb}} = 0.0195d_0$, allowing us to resolve the air-phase turbulence down to the Kolmogorov microscale $\eta$ as shown in Table~\ref{tab:configs}. Afterwards, we insert the droplet which is allowed to interact freely with the ambient air-phase turbulence, and in the meantime increase the maximum grid resolution level to $L_{drop} = 12$ for resolving the droplet surface, thus obtaining high-fidelity droplet deformation patterns up to the initiation of bag formation.

Finally, the droplet diameter $d_0$, incoming flow velocity $U_0$, dynamic flow pressure $p_0 \equiv \rho_g U_0^2 /2$, and the characteristic droplet deformation time $\tau \equiv \sqrt{\rho_l/\rho_g} d_0/U_0$ provide the natural reference scales for the length, mass and time quantities associated with the turbulent aerobreakup process. We will henceforth use these to non-dimensionalise the numerical results in the remainder of this study unless otherwise specified.

\subsubsection{Verification of air-phase turbulence statistics}
\label{subsubsec:turb-gen}
Upon the start of the simulation, a turbulent airflow is produced at the left boundary and transported downstream through the initially quiescent simulation domain, eventually reaching a statistical stationary state before the droplet is inserted. Here we inspect a few characteristics of the stationary-state turbulent airflow to evaluate its homogeneity and isotropy. In particular, Table~\ref{tab:configs} shows that the Kolmogorov microscale $\eta/d_0$ can be as small as 0.004 for simulations with viscosity ratios $\mu_r=55$, while the minimum bulk grid size is only $\Delta_{turb}/d_0 = D/d_0/2^{L_{turb}} = 0.078$. This suggests that for simulations with the most severe turbulence fluctuations, we are not yet able to resolve the smallest turbulence length scales where viscous dissipation dominates. Previous studies conclude that only strong airphase eddies at the large turbulence integral scale will impact droplet deformation \cite{jiao2019direct, zhao2019effect, xu2022droplet}, which indicates that under-resolved small eddies at the Kolmogorov length scale will not be likely to affect the fidelity of the droplet deformation data. Nonetheless, we evaluate the influence of reducing $\mu_r$ on the turbulence statistics here, and further investigate its effects on droplet surface statistics in \S\ref{sec:visc-ratio}.

\begin{figure}[htbp]\textit{}
	\centering
	\subfloat[]{
		\label{fig:tke-grid-converge}
		\includegraphics[width=.48\textwidth]{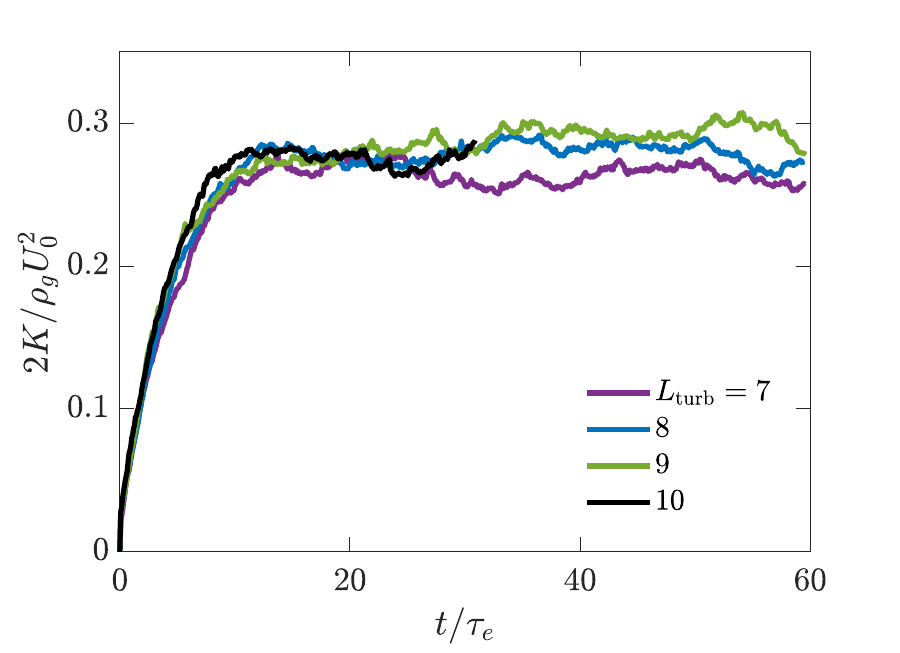}}
	\centering
	\subfloat[]{
		\label{fig:turb-spect}
		\includegraphics[width=.48\textwidth]{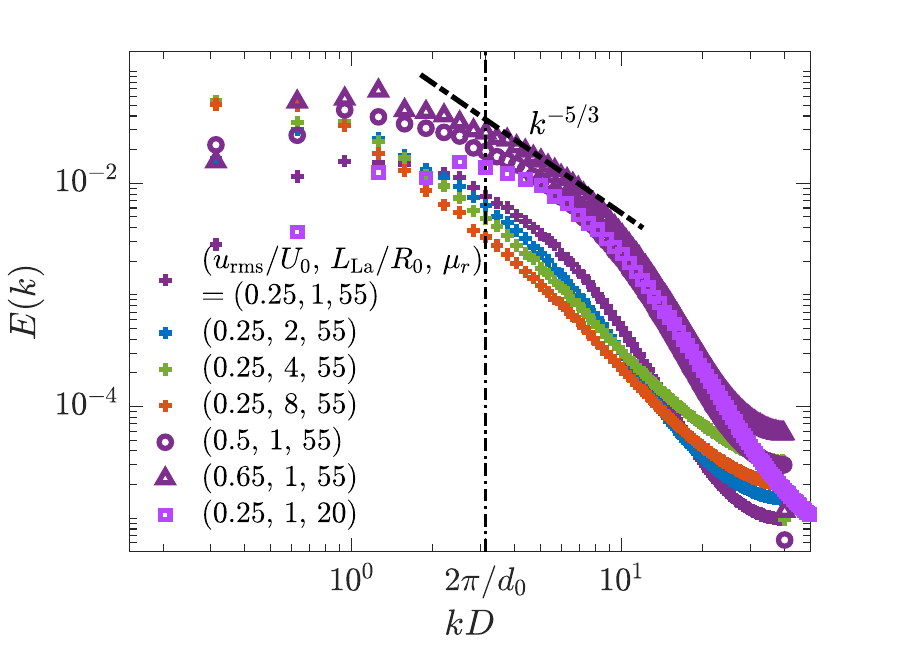}}

        \centering
	\subfloat[]{
		\label{fig:struct-dll}
		\includegraphics[width=.48\textwidth]{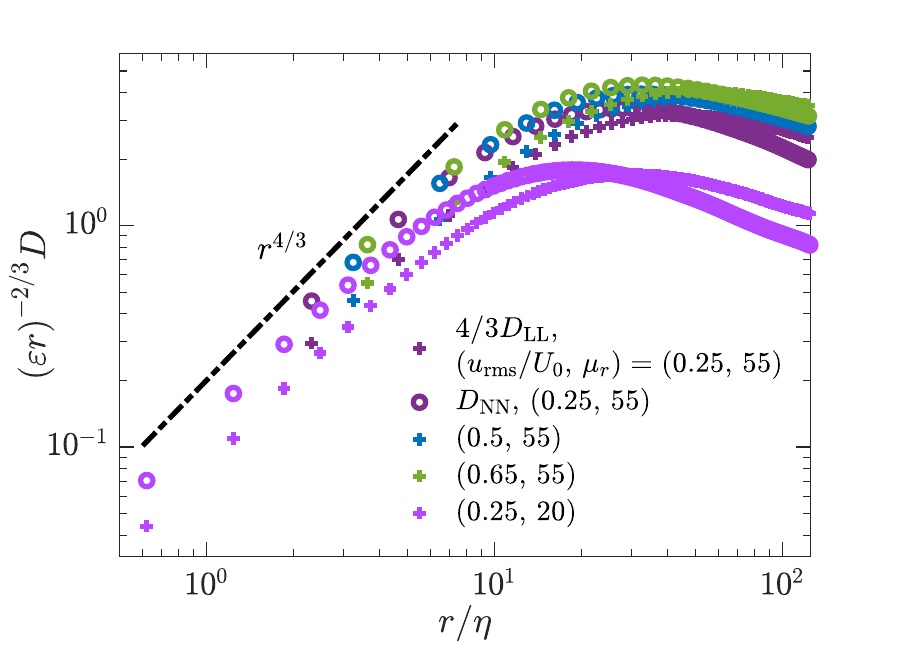}}
        \centering
	\subfloat[]{
		\label{fig:struct-dnn}
		\includegraphics[width=.48\textwidth]{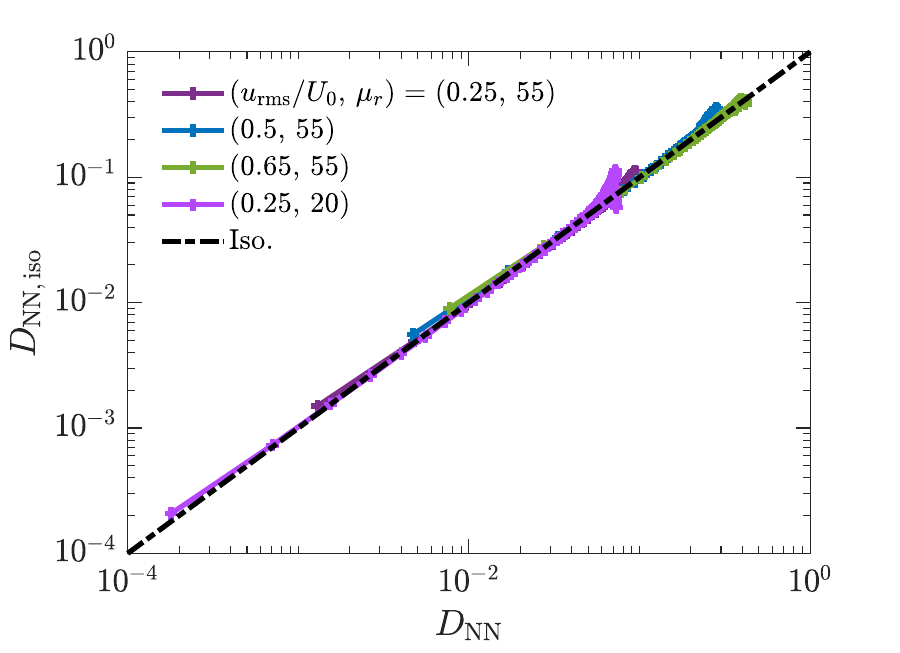}}
	\caption{(a): Grid convergence test for the air-phase turbulent kinetic energy $K$ for the turbulence configuration of $u^*/U_0 = 0.8$ and $L_0/R_0 = 1$. (b): Power spectrum density of fluctuating velocities $u$, $v$ and $w$ calculated from time series recorded near the inlet. (c): Compensated second-order turbulence structure functions $D_{LL}$ and $D_{NN}$ as a function of the dimensionless distance $r/\eta$, where $\eta$ is the Kolmogorov length scale. (d): Comparison between $D_{NN}|_{iso}$ and $D_{NN}$ at different values of $u^*$ as a test for flow isotropy.}
	\label{fig:turb-charact}
\end{figure}

Fig.~\ref{fig:tke-grid-converge} shows the evolution of the averaged air-phase turbulent kinetic energy $K$ within the simulation domain at various values of $L_{turb}$ for $u^*/U_0 = 0.8$ and $L_0/R_0 = 1$, where time is non-dimensionalised by the turnover time $\tau_e \equiv d_0^{2/3} \varepsilon^{-1/3}$ of eddies comparable to the size of the droplet. The turbulent kinetic energy first increases and then saturates as the turbulent airflow fully develops, with the steady-state value reaching grid convergence at a bulk resolution level of $L_{turb} = 8$. This suggests that at $L_{turb} = 8$, the mid-to-large-scale turbulent velocity fluctuations are already well-resolved, although the dissipation behaviour associated with smaller length scales requires further examination. Fig.~\ref{fig:turb-spect} shows the turbulence power spectrum $E(k)$ sampled from the entire simulation domain; where we observe the $-5/3$ power law characteristic of the inertial subrange, and the wavelength corresponding to the droplet diameter $2\pi/d_0$ is found to either fall within or remain close to the inertial regime. The slope of $E(k)$ steepens at higher frequencies, and diminishes again towards the upper bound of the wavenumber $k$ for spectra with $\mu_r = 55$; the latter behaviour is also present in fig.~13 of Xie and Castro's original paper \cite{xie2008efficient}. Indeed, this is most likely because the Kolmogorov microscale and the associated dissipative action at that scale are not fully resolved for $\mu_r = 55$. This behaviour in the spectrum does not appear for $\mu_r = 20$, where the Kolmogorov microscale and dissipation behaviour are well resolved.

We now proceed to assess the isotropy of the turbulence we generate. For this purpose, we introduce the second-order structure functions measured in both the longitudinal ($D_{LL}$) and transverse ($D_{RR}$) directions \cite{pope2000turbulent, riviere2021sub, perrard2021bubble} to characterise turbulence fluctuations,
\begin{gather}
    D_{LL}(r) = \frac{1}{3} \sum_i \left< {\left[ u_i(\boldsymbol{x},t) - u_i(\boldsymbol{x} + r \boldsymbol{\hat{x}_i},t) \right]}^2 \right>, \\
    D_{NN}(r) = \frac{1}{6} \sum_{i \neq j} \left< {\left[ u_i(\boldsymbol{x},t) - u_i(\boldsymbol{x} + r \boldsymbol{\hat{x}_j},t) \right]}^2 \right>,
    \label{for:struct-func-def}
\end{gather}
Fig.~\ref{fig:struct-dll} shows the structure functions $D_{LL}$ and $D_{NN}$ compensated by the theoretical scaling ${(\varepsilon r)}^{2/3}$ for HIT fluctuations. Similar to Rivi{\`e}re \emph{et al.} \cite{riviere2021sub}, we recover the $D_{LL} = 3/4D_{NN}$ relationship and the $D_{LL} \propto r^2$ scaling (corresponding to the $r^{4/3}$ scaling observed for the compensated structure functions in fig.~\ref{fig:struct-dll}) at small length scales regardless of the value of $\mu_r$. The compensated structure functions reach their maxima within the range of $20 \leq r/\eta \leq 40$, and for $\mu_r =20$ their peak values are close to 2, agreeing with theoretical predictions \cite{pope2000turbulent, riviere2021sub}. For $\mu_r = 55$, the peak values become greater than 2, which suggests that the turbulence dissipation $\varepsilon$ is underpredicted at the set grid resolution. This is because we are not yet resolving the air-phase turbulence down to the Kolmogorov microscale $\eta$ in these simulations. At larger values of $r$ both $D_{LL}$ and $D_{NN}$ start to decrease, where flow isotropy is no longer expected, especially along the streamwise direction due to the decay of turbulence intensity.

As an additional test for flow isotropy, we calculate the transverse structure function $D_{NN}|_{iso}$ corresponding to HIT fluctuations from its longitudinal counterpart,
\begin{equation}
    D_{NN}|_{iso} = D_{LL} + \frac{r}{2} \frac{\partial}{\partial r} D_{LL} (r).
    \label{for:dnn-iso-def}
\end{equation}
Fig.~\ref{fig:struct-dnn} compares the computed values of $D_{NN}|_{iso}$ according to \eqref{for:dnn-iso-def} and those of $D_{NN}$ directly determined from our turbulence data according to \eqref{for:struct-func-def}, showing an excellent agreement between the two regardless of the specific value of $\mu_r$. 

Overall, the results presented in fig.~\ref{fig:turb-charact} suggest that the air-phase turbulence generated using the Synthetic Turbulence Generation method introduced in \S\ref{subsec:num-method} features satisfactory homogeneity and isotropy. The turbulent energy budget might not be fully recovered due to under-resolved viscous dissipation at small length scales, especially with large values of $u^*$ and $\mu_r$. Similar to grid-generated turbulence \cite{zhou2014relevance}, the intensity of synthetically generated turbulence also decays further away from the inlet due to the lack of turbulence forcing in the simulation domain, causing large-scale inhomogeneity in the spanwise direction. However, we anticipate that these limitations should not affect the integrity of our results at $\mu_r=55$, since eddies at the Kolmogorov microscales do not have any discernible influence on droplet deformation, and we do not consider energy exchange between the droplet and ambient turbulent flow in this work.

\section{Phenomenology of droplet-turbulence interactions}
\label{sec:overview-drop-deform}
After the air-phase turbulence reaches the statistically stationary state, we insert the droplet and immediately release it to freely interact with the surrounding airflow, while recording its motion and deformation history. In this section, we discuss the influence of air-phase turbulence on droplet morphology qualitatively by inspecting snapshots produced from various individual simulations, with the aim of summarising general patterns of such influences and providing the basis for further quantitative analyses in the following sections.

\begin{figure}[htbp]
	\centering
	\subfloat[]{
		\label{fig:side-lam-16}
		\includegraphics[width=.24\textwidth]{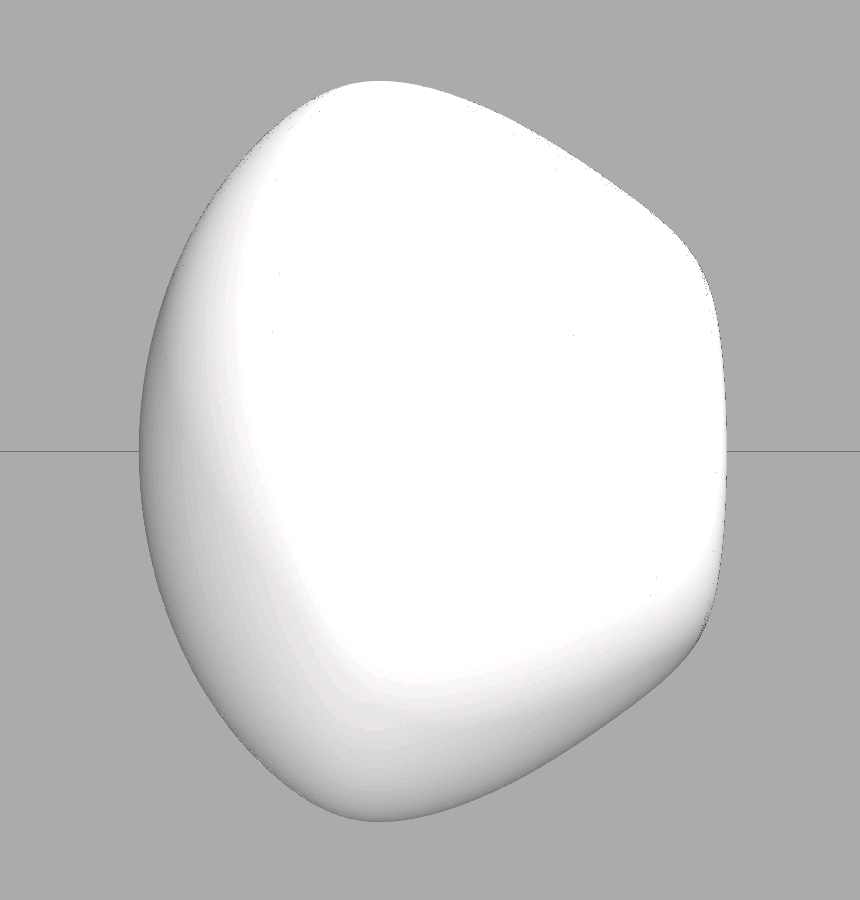}}
	\centering
	\subfloat[]{
		\label{fig:side-lam-40}
		\includegraphics[width=.24\textwidth]{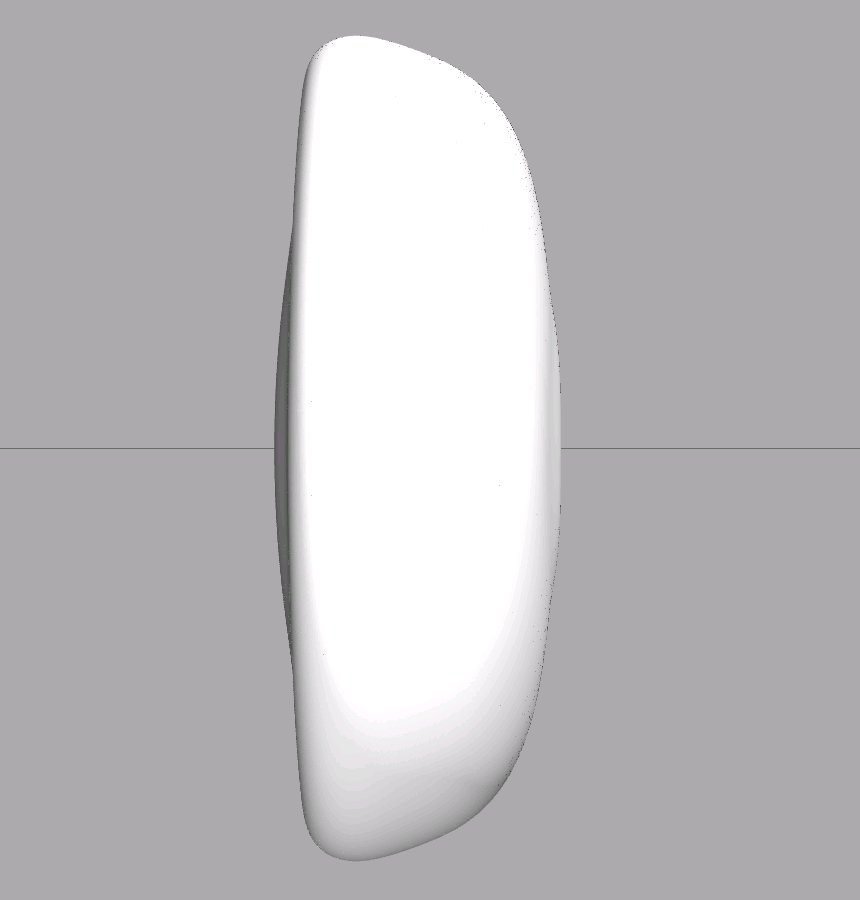}}
    \centering
	\subfloat[]{
		\label{fig:side-lam-64}
		\includegraphics[width=.24\textwidth]{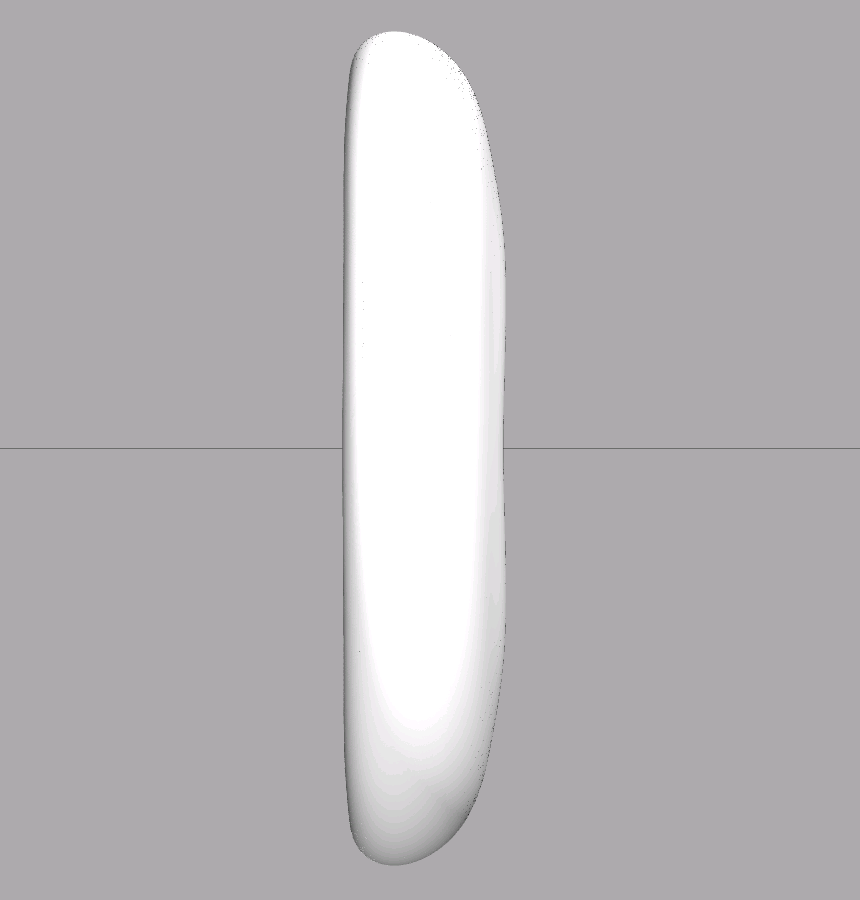}}
    \centering
	\subfloat[]{
		\label{fig:side-lam-88}
		\includegraphics[width=.24\textwidth]{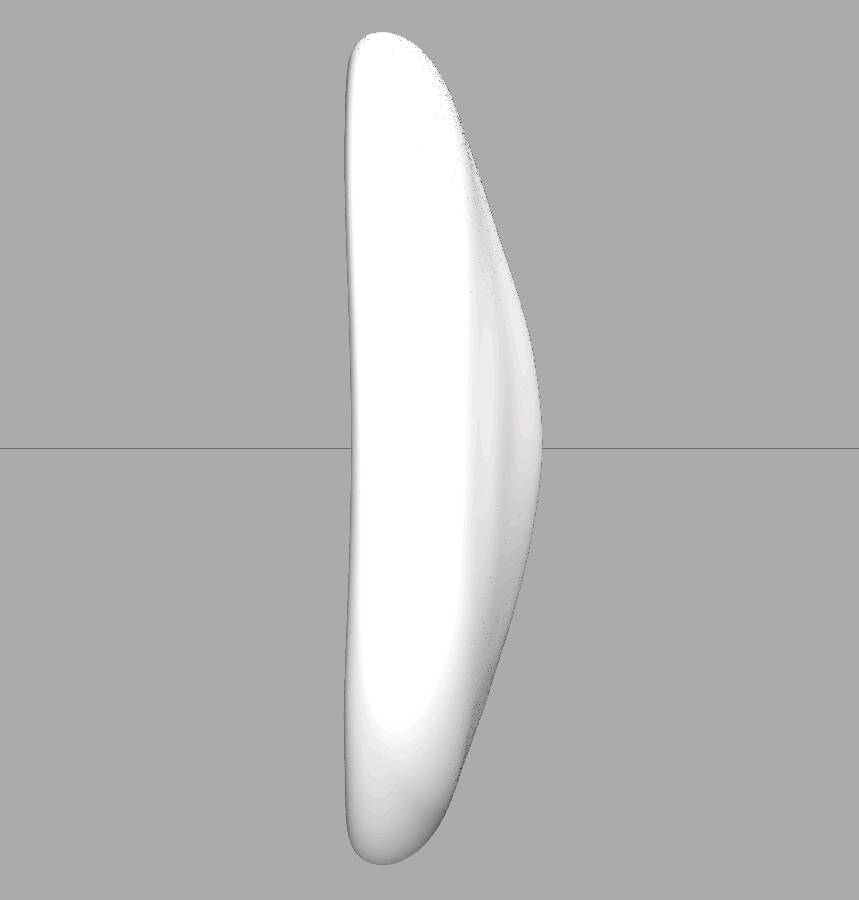}}

    \centering
	\subfloat[]{
		\label{fig:side-0.5-1-16}
		\includegraphics[width=.24\textwidth]{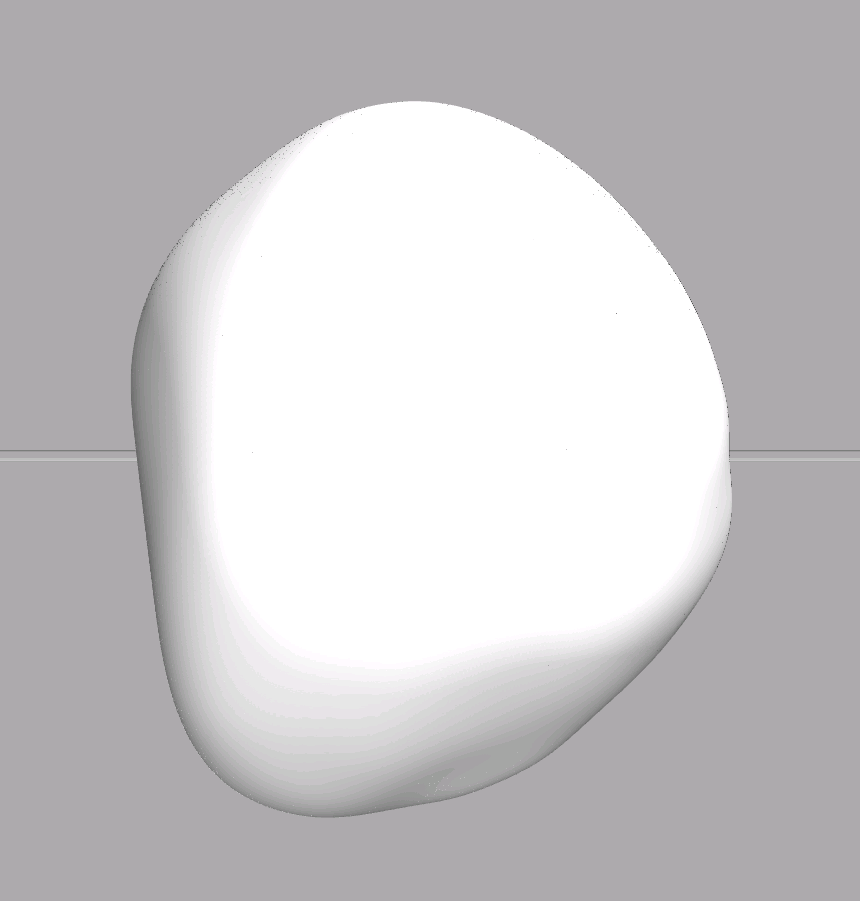}}
	\centering
	\subfloat[]{
		\label{fig:side-0.5-1-40}
		\includegraphics[width=.24\textwidth]{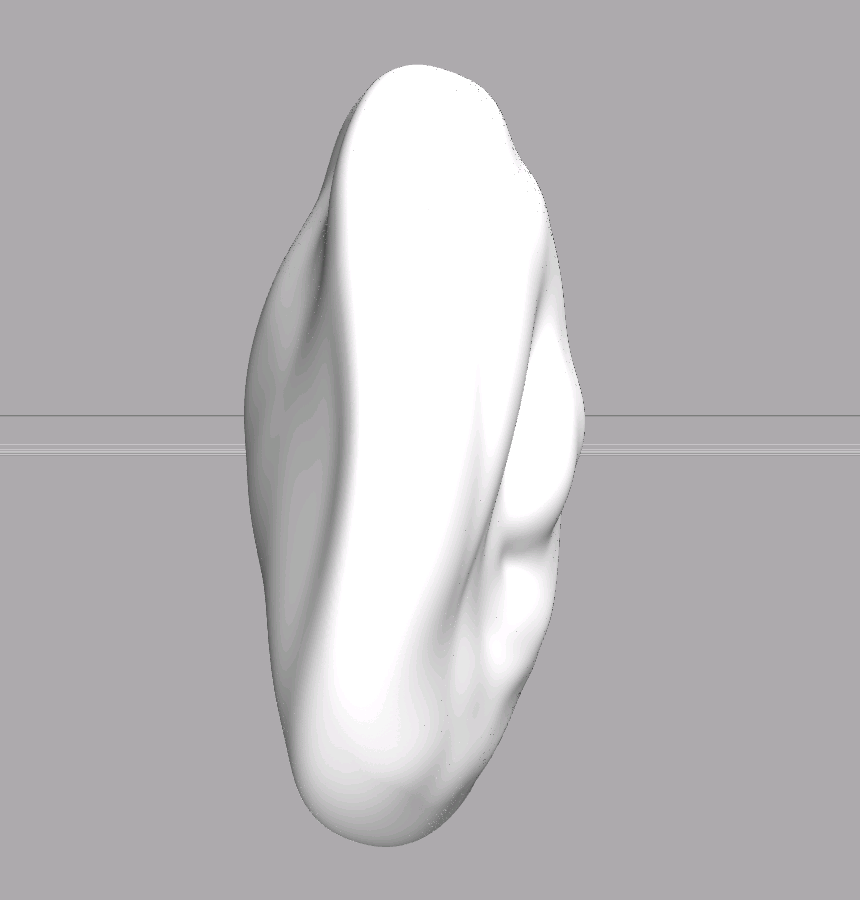}}
    \centering
	\subfloat[]{
		\label{fig:side-0.5-1-64}
		\includegraphics[width=.24\textwidth]{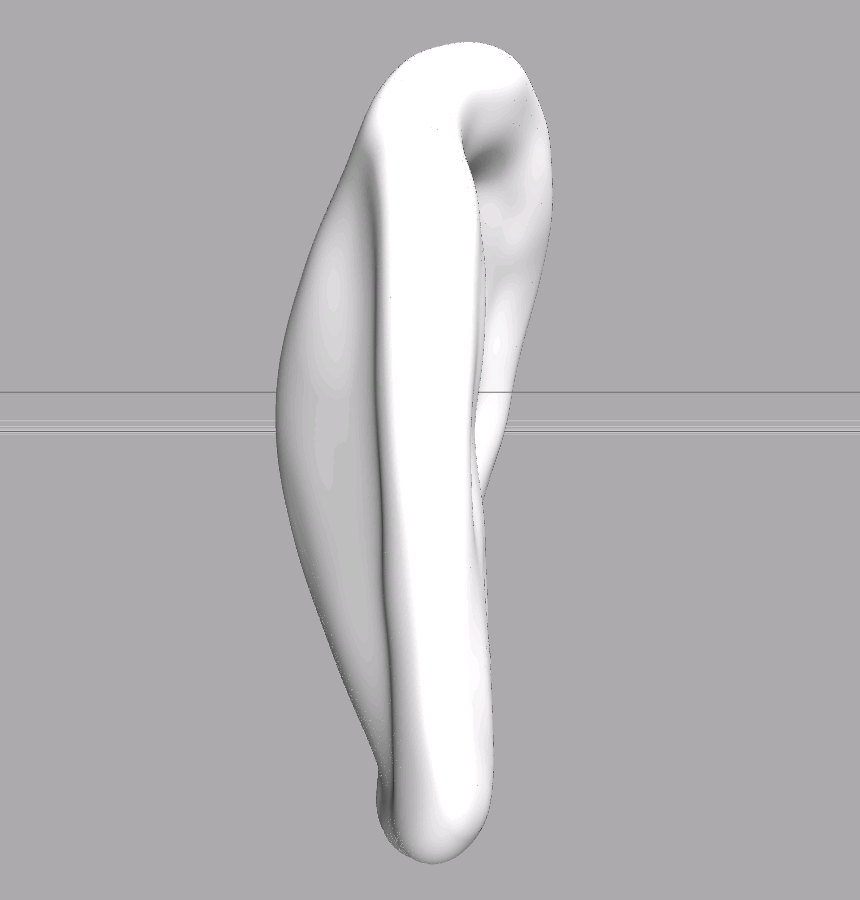}}
    \centering
	\subfloat[]{
		\label{fig:side-0.5-1-88}
		\includegraphics[width=.24\textwidth]{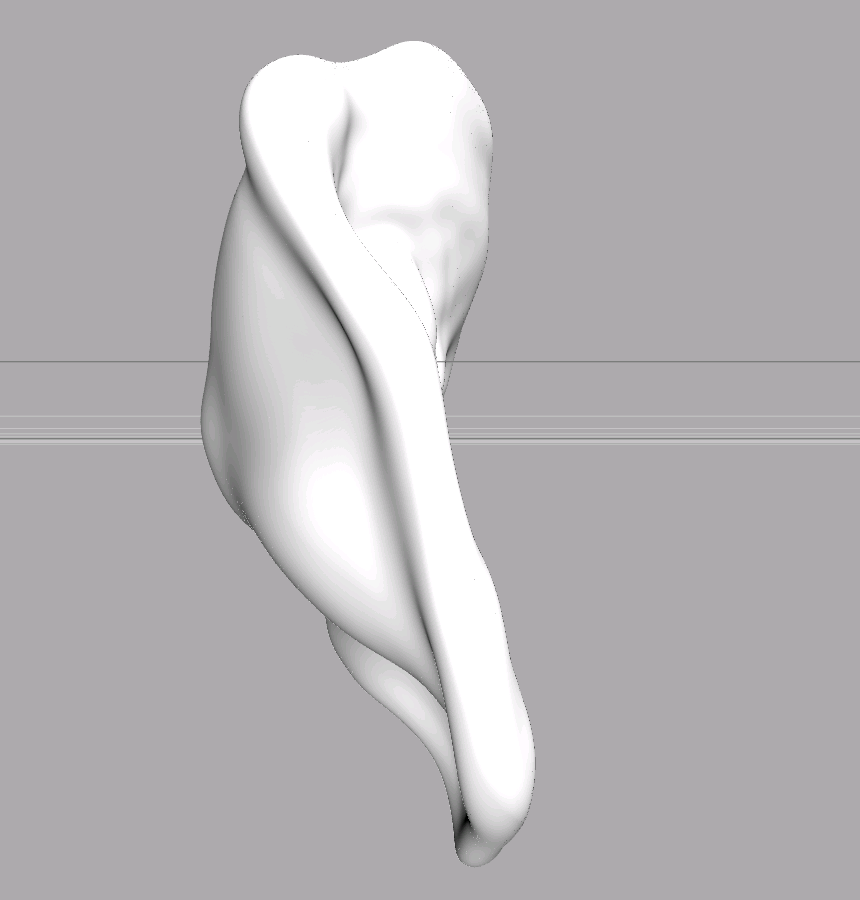}}
	\caption{Snapshots showing the bag formation process in the laminar airflow case (a-d) in comparison with a typical turbulent airflow case with $u^*/U_0 = 0.5$, $L_0/R_0 = 1$ (e-h). The snapshots are taken at $t/\tau = 0.28$ (a,e), 0.69 (b,f), 1.11 (c,g) and 1.52 (d,h).}
	\label{fig:lam-turb-comp-snapshots}
\end{figure}

In fig.~\ref{fig:lam-turb-comp-snapshots} we first compare the morphological evolution of droplets within a laminar airflow (figs.~\ref{fig:side-lam-16}-\ref{fig:side-lam-88}) and a typical turbulent airflow (figs.~\ref{fig:side-0.5-1-16}-\ref{fig:side-0.5-1-88}) with $u^*/U_0 = 0.25$ and $L_0/R_0 = 1$. It can be observed that while the droplet shape remains axisymmetric until the blow-out of the bag in the laminar aerobreakup case, this symmetry about the streamwise minor axis of the drop can be lost at very early times when turbulence is introduced. As the droplet flattens in the turbulent airflow, its minor axis no longer aligns with the streamwise direction in fig.~\ref{fig:side-0.5-1-88}, which will be covered in \S\ref{subsec:tilting}. Furthermore, significant corrugations arise across the entire droplet surface and disrupt the well-defined bag shape observed in laminar aerobreakup scenarios, which we will discuss in detail in \S\ref{subsec:corrug-form}.

\begin{figure}[htbp]
	\centering
	\subfloat[]{
		\label{fig:back-0.25-1-ens-1}
		\includegraphics[width=.24\textwidth]{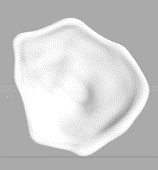}}
	\centering
	\subfloat[]{
		\label{fig:back-0.25-1-ens-2}
		\includegraphics[width=.24\textwidth]{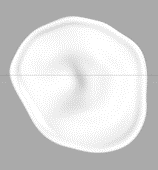}}
    \centering
	\subfloat[]{
		\label{fig:back-0.25-1-ens-3}
		\includegraphics[width=.24\textwidth]{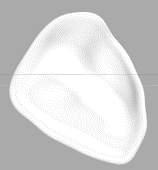}}
    \centering
	\subfloat[]{
		\label{fig:back-0.25-1-ens-4}
		\includegraphics[width=.24\textwidth]{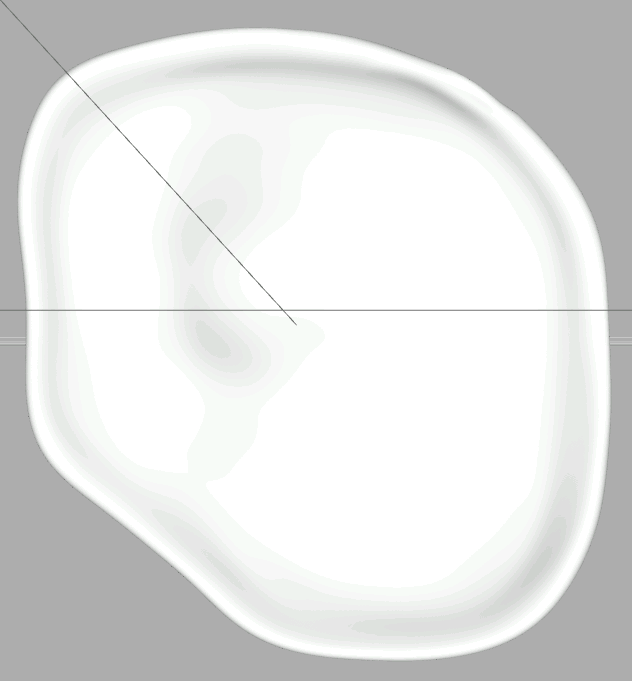}}

    \centering
	\subfloat[]{
		\label{fig:back-0.65-1-ens-1}
		\includegraphics[width=.24\textwidth]{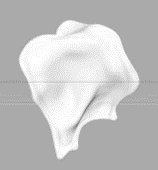}}
	\centering
	\subfloat[]{
		\label{fig:back-0.65-1-ens-2}
		\includegraphics[width=.24\textwidth]{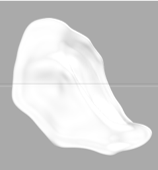}}
    \centering
	\subfloat[]{
		\label{fig:back-0.65-1-ens-3}
		\includegraphics[width=.24\textwidth]{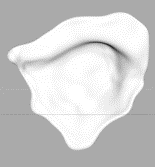}}
    \centering
	\subfloat[]{
		\label{fig:back-0.65-1-ens-4}
		\includegraphics[width=.24\textwidth]{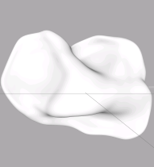}}
    
	\caption{Snapshots showing the bags formed at late times, retrieved from different ensemble realisations with the same turbulence configurations. For (a) to (d): $(u^*/U_0, \, L_0/R_0) = (0.25, \, 1)$. For (e) to (h): $(u^*/U_0, \, L_0/R_0) = (0.65, \, 1)$.}
	\label{fig:back-snapshots-ens-realise}
\end{figure}

Due to the chaotic nature of the ambient airflow, droplets forced by turbulence with identical configurations may exhibit different deformation histories. Fig.~\ref{fig:back-snapshots-ens-realise} shows the bag shapes from two different turbulence configurations: $(u^*/U_0, \, L_0/R_0) = (0.25, \, 1)$ (a-c), and $(0.65, \, 1)$ (d-f), where the snapshots in the same row correspond to bag shapes taken from different individual ensemble realisations at the same simulation time. While the bag shapes and orientations vary significantly among different realisations, some common features are still observed for those with the same turbulence configurations. Namely, at $u^*/U_0 = 0.25$ we find the bags to feature well-defined peripheral rims and inflating central films similar to laminar bag breakup phenomena observed in previous studies \cite{tang2022bag, ling2023detailed}, despite mild stretching across the entire bag. On the other hand, we find the flattened drops severely distorted at $u^*/U_0 = 0.65$, and the distinction between the peripheral rim and the central bag becomes much less prominent. Ensemble-averaging across different turbulence realisations is therefore necessary for obtaining statistically meaningful data \cite{Mostert2021, tang2022bag}.

\begin{figure}[htbp]
	\centering
	\subfloat[]{
		\label{fig:back-0.25-1}
		\includegraphics[width=.24\textwidth]{Figures/back-0.25-1.png}}
	\centering
	\subfloat[]{
		\label{fig:back-0.5-1}
		\includegraphics[width=.24\textwidth]{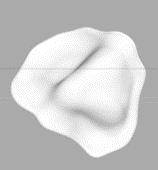}}
    \centering
	\subfloat[]{
		\label{fig:back-0.65-1}
		\includegraphics[width=.24\textwidth]{Figures/back-0.65-1.png}}
    \centering
	\subfloat[]{
		\label{fig:back-0.8-1}
		\includegraphics[width=.24\textwidth]{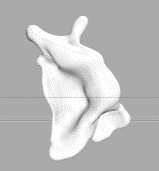}}

    \centering
	\subfloat[]{
		\label{fig:back-0.25-1.5}
		\includegraphics[width=.24\textwidth]{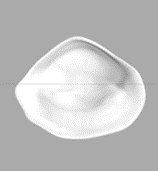}}
	\centering
	\subfloat[]{
		\label{fig:back-0.25-2}
		\includegraphics[width=.24\textwidth]{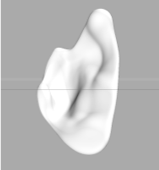}}
    \centering
	\subfloat[]{
		\label{fig:back-0.25-4}
		\includegraphics[width=.24\textwidth]{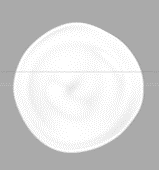}}
    \centering
	\subfloat[]{
		\label{fig:back-0.25-8}
		\includegraphics[width=.24\textwidth]{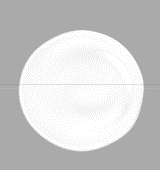}}
	\caption{Snapshots showing the bags formed at late times under different turbulence configurations. For (a) to (d): $L_0/R_0 = 1$, while $u^*/U_0 = 0.25$ (a), 0.5 (b), 0.65 (c), and 0.8 (d). For (e) to (h): $u^*/U_0 = 0.25$, while $L_0/R_0 = 1.5$ (e), 2 (f), 4 (g), and 8 (h). The Hinze scales $d_h/d_0$ for (a)-(h) are respectively: 1.008 (a), 0.439 (b), 0.320 (c), 0.249 (d), 1.186 (e), 1.331 (f), 1.756 (g), 2.317 (h).}
	\label{fig:back-snapshots}
\end{figure}

Apart from the fluctuating velocity $u^*$, variations in the injection length scale $L_0$ also lead to qualitative changes in droplet deformation. Fig.~\ref{fig:back-snapshots} shows the typical late-time bag shapes across a wide range of turbulence configurations $(u^*, \, L_0)$. It is observed that as the turbulence intensity $u^*$ increases (figs.~\ref{fig:back-0.25-1}-\ref{fig:back-0.8-1}) or the injection length scale $L_0$ decreases (figs.~\ref{fig:back-0.25-1.5}-\ref{fig:back-0.25-8}) such that the non-dimensionalised Hinze scale $d_h/d_0$ becomes smaller than unity, the shape of the deformed droplet deviates further away from the well-defined laminar aerobreakup morphology, as is already observed in fig.~\ref{fig:back-snapshots-ens-realise}. Instead, large-scale distortion arises across the bag, and the peripheral rim becomes increasingly corrugated, forming sharp protrusions (fig.~\ref{fig:back-0.65-1}) which evolve into liquid nodes (fig.~\ref{fig:back-0.8-1}) \cite{jackiw2022prediction} bearing some resemblance to the `handle' structures observed by Cannon \emph{et al.} \cite{cannon2024morphology} for $\rho_r = \mu_r = 1$. This also matches qualitatively with previous multiphase turbulence studies \cite{riviere2021sub, riviere2022capillary}, where the morphological evolution and breakup behaviour of the bubble or droplet relies strongly on $d_h/d_0$. Similar rim protrusions are not observed in previous numerical studies of laminar aerobreakup \cite{tang2022bag, ling2023detailed} but reported in experimental bag breakup studies, as they eventually pinch off to produce node drops. It is found that either the Rayleigh-Plateau (RP) or the Rayleigh-Taylor (RT) instability can explain their formation \cite{zhao2010morphological, jackiw2022prediction}.

It is also noted that as the injection length scale $L_0$ increases, the late-time bags recover the laminar axisymmetric shape with only an overall tilting motion, resembling the bag morphologies reported by Jiao \emph{et al.} \cite{jiao2019direct}. This is expected as the droplet now interacts with turbulent eddies of larger size in individual realisations, and therefore experiences less severe ambient flow velocity variation across its diameter with a fixed turbulent fluctuation intensity. Indeed, this agrees with Wang \emph{et al.} \cite{wang2019rotational} and Xu \emph{et al.} \cite{xu2024intermittency} that eddies interact preferentially with particles featuring diameters closest to their sizes. Overall, the bag morphologies observed in fig.~\ref{fig:back-snapshots} suggest that `mild' or `severe' modifications of the bag shape by ambient turbulence can be predicted based on the non-dimensionalised Hinze scale value $d_h/d_0$: the former occurs when $d_h/d_0 > 1$, whereas the latter occurs when $d_h/d_0 \leq 1$. The validation of this heuristic criterion requires ensemble realisations at more ambient turbulence configurations, and is therefore outside the scope of the current work.

\section{Droplet dynamics}
\label{sec:drop-turb-dynamics}

We now seek to inspect the evolution of the centre-of-mass dynamic properties of the droplet, including both velocity and acceleration, which is closely tied to the droplet deformation history and will shed light on the overall effect of the ambient turbulence on the droplet.

\begin{figure}[htbp]
	\centering
	\subfloat[]{
		\label{fig:ux-urms-sweep}
		\includegraphics[width=.48\textwidth]{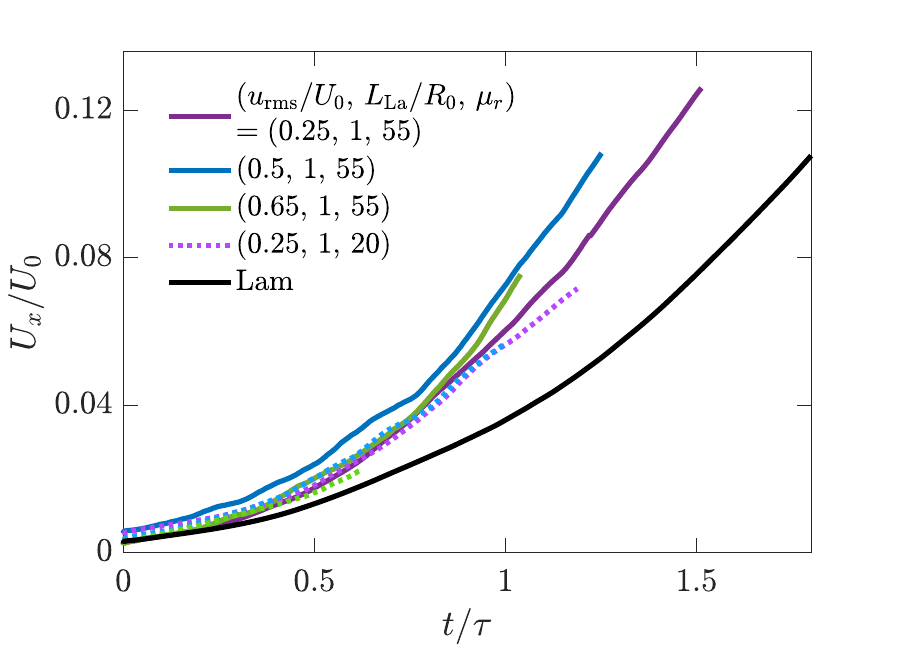}
  }
	\centering
	\subfloat[]{
		\label{fig:rdot-urms-sweep}
		\includegraphics[width=.48\textwidth]{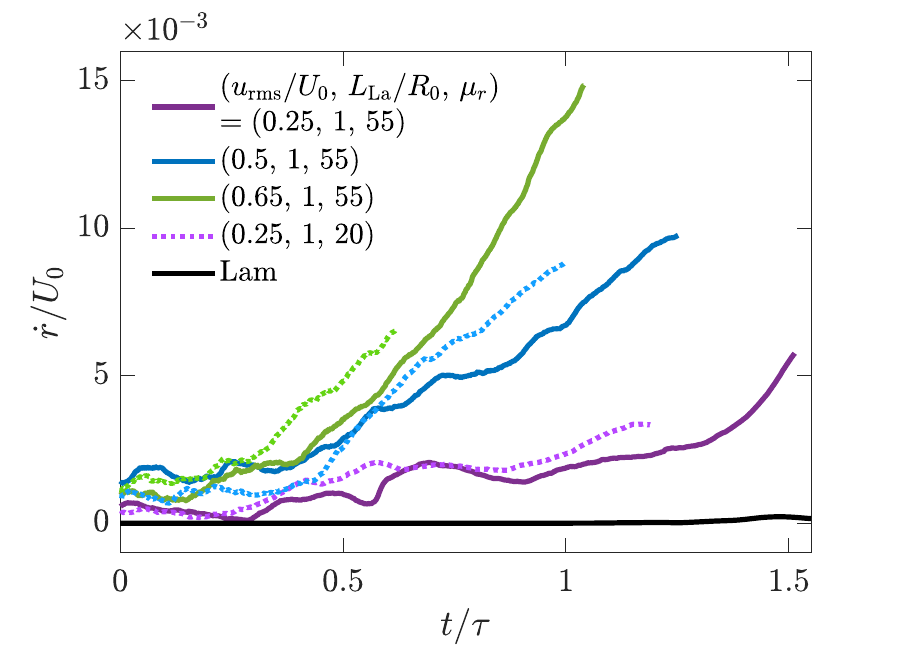}}

    \centering
	\subfloat[]{
		\label{fig:ux-lla-sweep}
		\includegraphics[width=.48\textwidth]{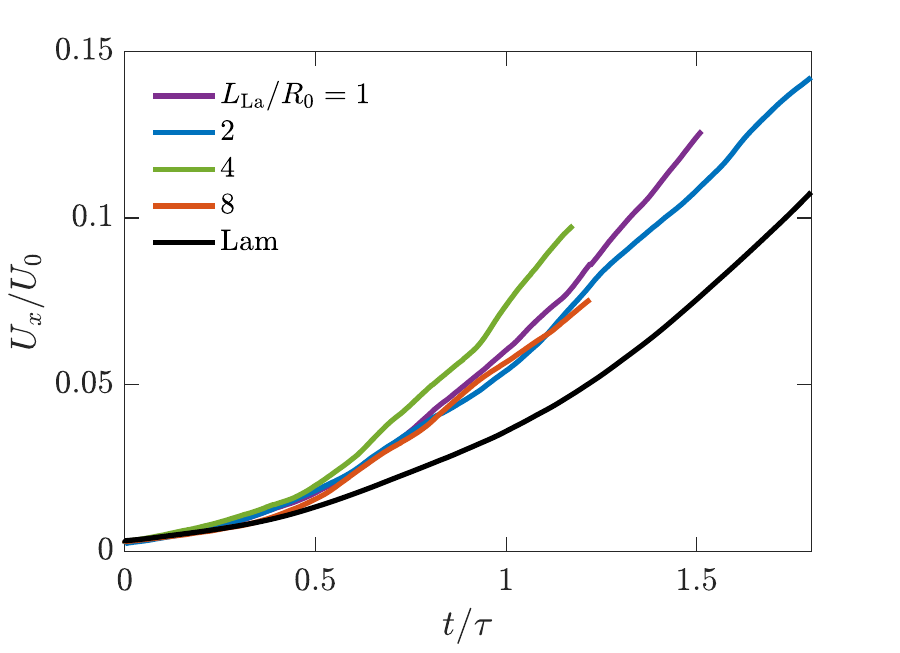}
  }
	\centering
	\subfloat[]{
		\label{fig:rdot-lla-sweep}
		\includegraphics[width=.48\textwidth]{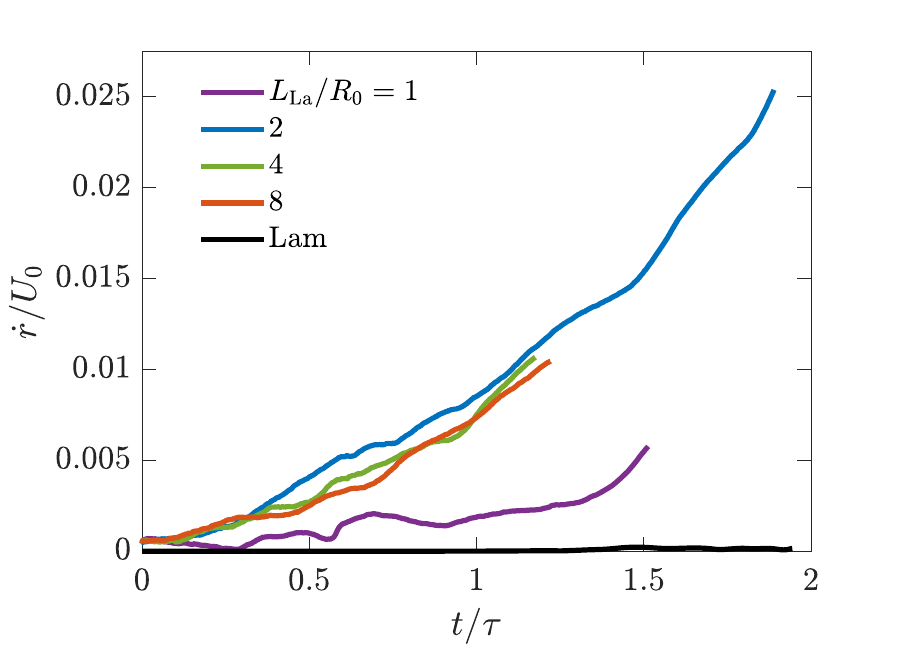}}
	\caption{Evolution of the ensemble-averaged droplet centre-of-mass velocity $U_x$ (a,c) and radial velocity $\Dot{r}$ (b,d), measured at different fluctuating velocities $u^*$ (a,b) and injection length scales $L_0$ (c,d).}
	\label{fig:ux-rdot-urms-lla-sweep}
\end{figure}

In fig.~\ref{fig:ux-rdot-urms-lla-sweep}, we present the streamwise and radial components of the droplet centre-of-mass velocity $U_x$ and $\Dot{r} \equiv \sqrt{U_y^2 + U_z^2}$ at both viscosity ratios $\mu_r = 55$ and 20. Here, $\Dot{r}$ is the component of the droplet velocity orthogonal to the mean flow. For example, in a laminar freestream flow which remains strictly axisymmetric, e.g. at early times, this component would be zero. The results have been ensemble-averaged to reduce variations across different individual realisations. It is observed that the introduction of air-phase turbulence leads to an increase of both $U_x$ and $\Dot{r}$; in other words, ambient turbulence intermittency contributes to droplet acceleration in both the axial and radial directions. It is particularly interesting to note that according to figs.~\ref{fig:ux-urms-sweep} and \ref{fig:ux-lla-sweep}, the evolution patterns of the streamwise velocity $U_x$ do not depend strongly on the air-phase turbulence configurations within our parameter space, including $u^*$, $L_0$ and $\mu_r$; and compared with the laminar aerobreakup case, a consistent enhancement in the increasing rate of $U_x$ is reached at around $t = 0.6\tau$ for all turbulent aerobreakup cases, when the droplet is about to become fully flattened. On the other hand, fig.~\ref{fig:ux-lla-sweep} suggests that increasing $u^*$ causes the radial component of the droplet velocity $\Dot{r}$ to increase; although the magnitude of $\Dot{r}$ still remains much smaller than its streamwise counterpart $U_x$, meaning that the droplet is transported mostly in the mean flow direction. This is expected since the large liquid-gas density ratio $\rho_r$ ensures flow separation and hence a large pressure difference between the frontal and leeward sides of the droplet, which dominates droplet acceleration even when large turbulence fluctuations in the cross-stream directions are present. It is also likely that the increase in $\Dot{r}$ incorporates the contribution of irregularity in the droplet internal flows, which causes the droplet morphology to deviate further away from axisymmetry at large turbulence fluctuating speeds $u^*/U_0$.

To allow for detailed analysis of the droplet streamwise acceleration patterns, especially around $t/\tau = 0.5$ where the significant acceleration by ambient turbulence is observed, we differentiate $U_x$ numerically in time to obtain streamwise acceleration $a_x$, and non-dimensionalise the results as a drag coefficient $C_{D, \, x}$ \cite{ling2023detailed},
\begin{equation}
    C_{D, \, x} \equiv \frac{2m_d}{\rho_g U_0^2 \pi R_0^2} a_x = \frac{8}{3} \frac{\rho_l}{\rho_g} \frac{R_0}{U_0^2} a_x.
    \label{for:drag-coeff-def}
\end{equation}

\begin{figure}[htbp]
	\centering
	\subfloat[]{
		\label{fig:ax-urms-sweep-early}
		\includegraphics[width=.48\textwidth]{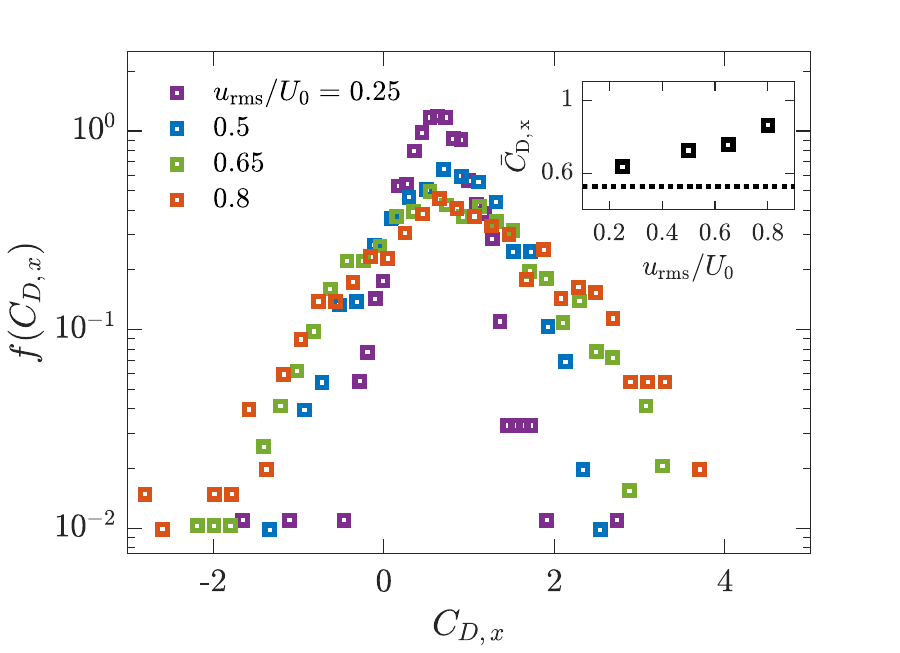}}
	\centering
	\subfloat[]{
		\label{fig:ax-urms-sweep-mid}
		\includegraphics[width=.48\textwidth]{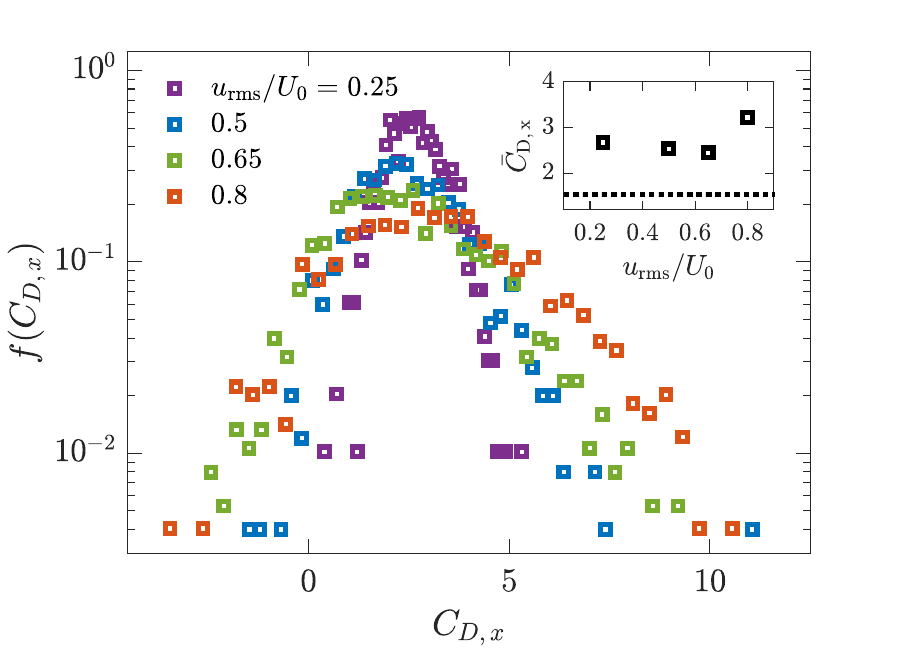}}
  
    \centering
	\subfloat[]{
		\label{fig:ax-lla-sweep-early}
		\includegraphics[width=.48\textwidth]{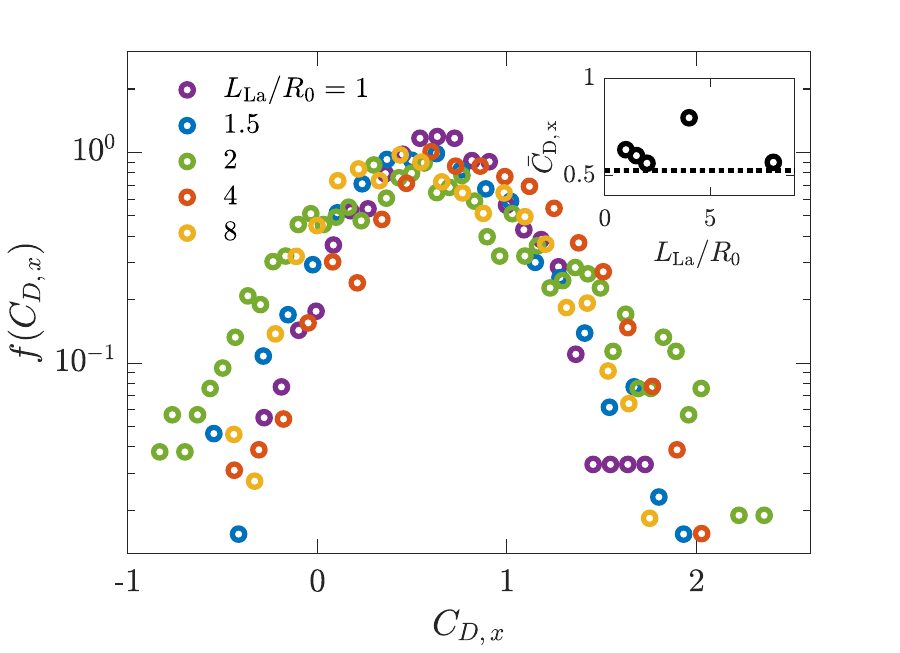}}
    \centering
	\subfloat[]{
		\label{fig:ax-lla-sweep-mid}
		\includegraphics[width=.48\textwidth]{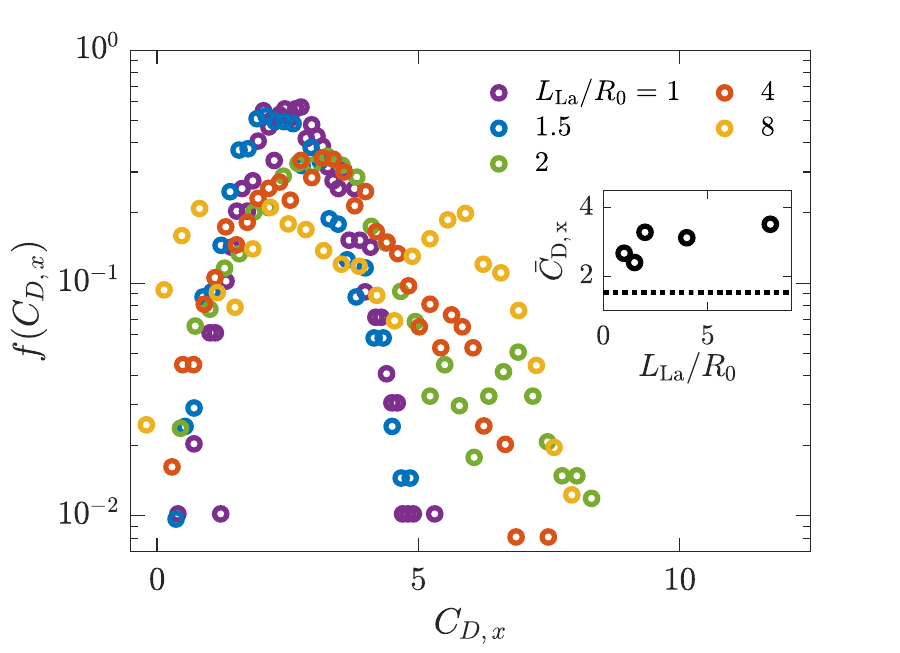}}

	\caption{Distributions of droplet streamwise drag coefficients $C_{D, \, x}$ measured at different turbulence configurations. (a,c) and (b,d) are sampled within time windows $0 \leq t/\tau \leq 0.173$ and $0.485 \leq t/\tau \leq 0.693$, respectively. For (a,b), $L_0/R_0 = 2$ and for (c,d) $u^*/U_0 = 0.25$. The insets show averaged drag coefficients of turbulent aerobreakup cases (scattered data) compared with their laminar counterparts (dashed line).}
	\label{fig:acc-dist-evol}
\end{figure}

Due to the presence of strong fluctuations in the instantaneous streamwise droplet acceleration $a_x$, we select two time-windows for comparison: $0 \leq t/\tau \leq 0.173$ and $0.485 \leq t/\tau \leq 0.693$, which correspond to the early and intermediate deformation stage respectively, and present in fig.~\ref{fig:acc-dist-evol} the distributions of instantaneous streamwise drag coefficients $C_{D, \, x}$ sampled across all ensemble realisations, rather than directly showing the original time sequence of streamwise acceleration. Fig.~\ref{fig:ax-urms-sweep-early} show that at early times, the distributions of $C_{D, \, x}$ are largely symmetric. The mean values of $C_{D, \, x}$ increase over time, as indicated by the shift of the maxima of $C_{D, \, x}$ towards larger values. Fig.~\ref{fig:ax-urms-sweep-mid} suggests that while the distribution of $C_{D, \, x}$ is still largely symmetric for $u_{rms/U_0} \leq 0.5$, a skew towards larger $C_{D, \, x}$ values is observed at larger turbulence fluctuating velocities. Interestingly, while increasing turbulence intensity $u^*$ causes the $C_{D, \, x}$ distribution to broaden, it does not significantly influence the mean value of $C_{D, \, x}$ as shown in the inset of fig.~\ref{fig:ax-urms-sweep-mid}, agreeing with our earlier observations of fig.~\ref{fig:ux-urms-sweep}. On the other hand, fig.~\ref{fig:ax-lla-sweep-early} suggests that increasing $L_0$ does not cause any consistent change in the shape of the $C_{D, \, x}$ distribution at early times. Interestingly, for turbulent aerobreakup cases with $L_0/R_0 \geq 2$, fig.~\ref{fig:ax-lla-sweep-mid} suggests that the $C_{D, \, x}$ distributions become double-peaked, which is especially prominent for $L_0/R_0 = 2$ and 8. The distribution for $L_0/R_0 = 4$ also features a skew towards large $C_{D, \, x}$, while the mean values for $C_{D, \, x}$ remain largely unaffected by $L_0$. We consider this change in the distribution shape of $C_{D, \, x}$ to be most likely associated with the increased occurrence of large-scale velocity fluctuations within the ambient turbulence. Indeed, as the injection length scale of the turbulence becomes larger than the droplet, it is expected that the interior structure of the turbulent eddies becomes more important in controlling the instantaneous droplet acceleration. A detailed investigation of this phenomenon is left for future work.

Agreeing with our observations in fig.~\ref{fig:ux-rdot-urms-lla-sweep}, the insets of fig.~\ref{fig:acc-dist-evol} show that turbulent aerobreakup cases (shown in scattered data points) indeed feature larger $\Bar{C}_{D, \, x}$ values compared with their laminar aerobreakup counterparts, while our current results do not indicate any clear dependence of $\Bar{C}_{D, \, x}$ on the turbulence parameters $u^*$ and $L_0$. We also find that while the magnitudes of instantaneous $C_{ D, \, y}$ and $C_{ D, \, z}$ values are comparable with $C_{ D, \, x}$ \cite{botto2012fully}, their mean values remain close to zero without any obvious evolution over time (not shown here). Overall, these results confirm our observation in fig.~\ref{fig:ux-rdot-urms-lla-sweep} that introducing air-phase turbulence causes a nontrivial streamwise mean acceleration effect, which most likely arises from a qualitative change in the ambient flow field that does not depend significantly on specific turbulence configurations. The recorded extreme values in $C_{D, \, x}$, on the other hand, are found to appear more frequently at large values of $u^*$ and $L_0$, which are probably associated with more extreme fluctuations in the ambient turbulence.

\begin{figure}[htbp]
	\centering
	\subfloat[]{
		\label{fig:p-lam}
		\includegraphics[width=.48\textwidth]{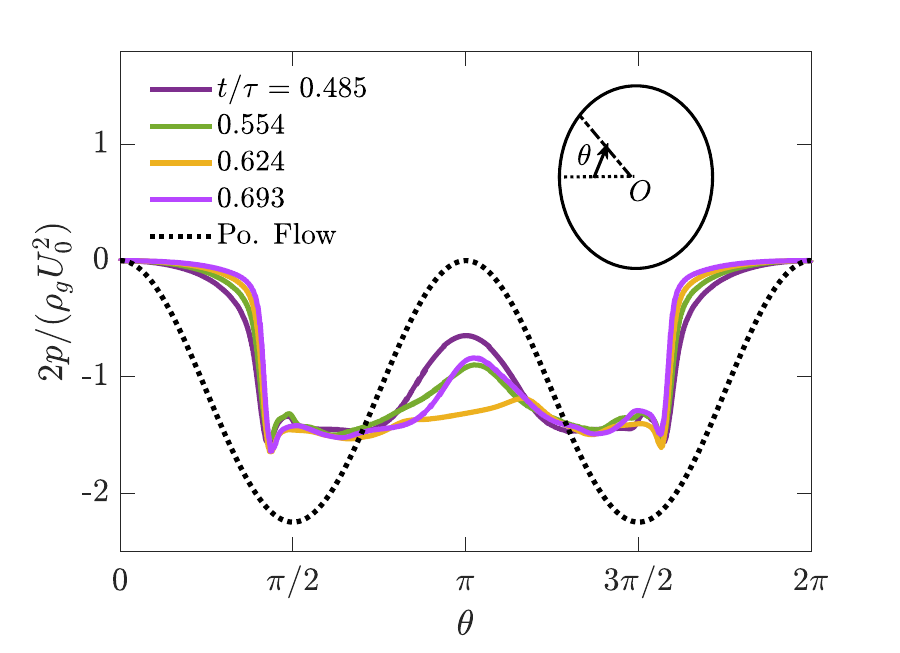}}
	\centering
	\subfloat[]{
		\label{fig:p-0.25-1}
		\includegraphics[width=.48\textwidth]{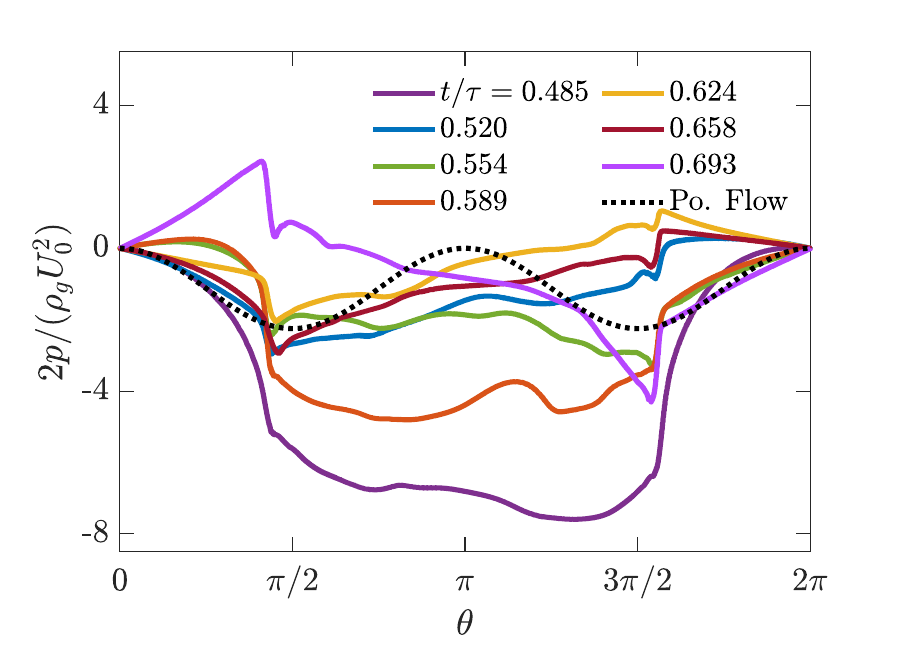}}

    \centering
	\subfloat[]{
		\label{fig:p-0.5-1}
		\includegraphics[width=.48\textwidth]{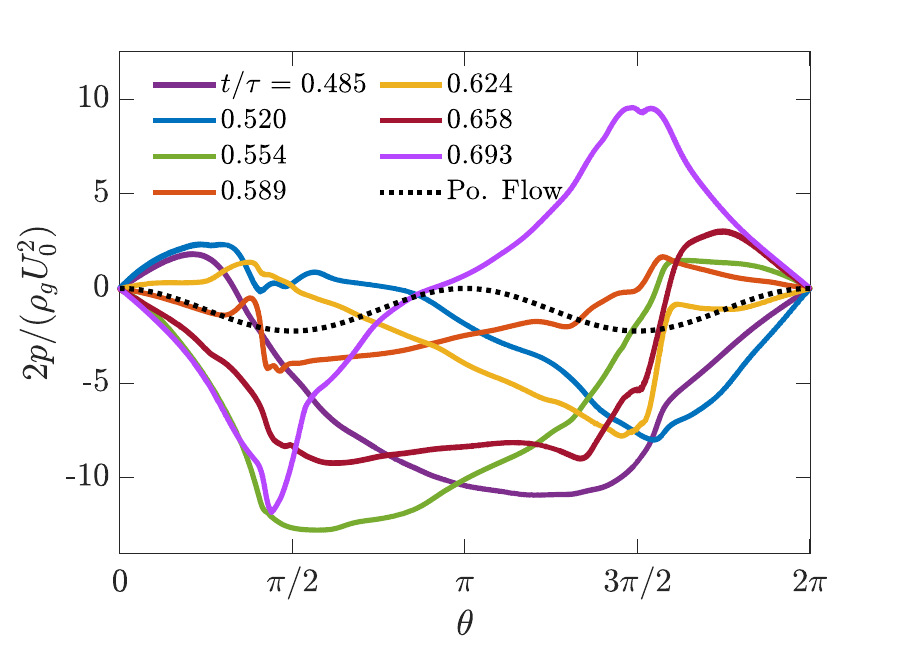}}
	\centering
	\subfloat[]{
		\label{fig:p-0.65-1}
		\includegraphics[width=.48\textwidth]{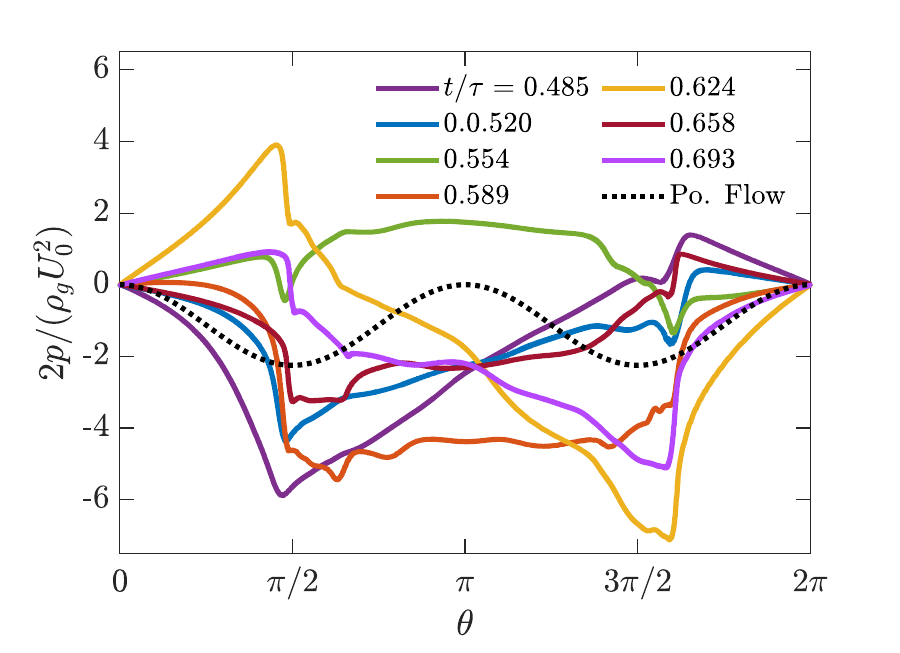}}
	\caption{The air-phase pressure profiles close to the droplet surface, sampled within the plane $z=0$ at different turbulent fluctuating velocities $u^*$. From (a) to (d): laminar airflow (a), $(u^*/U_0, \, L_0/R_0) = (0.25, \, 1)$ (b), (0.5, 1) (c) and (0.25, 4) (d). The inset of fig.~\ref{fig:p-lam} shows how the angle $\theta$ is calculated along the drop surface, and the potential flow solution for laminar airflow around a sphere is included in all subfigures for comparison.}
	\label{fig:p-traverse}
\end{figure}

To further elucidate the physical mechanism leading to the droplet acceleration patterns we observed, we now inspect the distribution of air-phase pressure around the droplet. In fig.~\ref{fig:p-traverse} we show the evolution of pressure profiles within the period of $0.485 \leq t/\tau \leq 0.693$ for individual realisations, when the flattened droplet is further accelerated by airphase turbulence. These pressure profiles are sampled close to the droplet surface within the reference plane $z=0$, and plotted as a function of polar angle $\theta$ traversing the droplet surface in the clockwise direction as shown in the inset of fig.~\ref{fig:p-lam}. Here, $\theta = 0$ corresponds to the leftmost point on the windward surface, where the pressure values have been referenced to $p=0$ to facilitate comparison across different turbulence configurations. Fig.~\ref{fig:p-lam} shows the pressure profiles obtained from laminar aerobreakup, where the minima at $\theta = \pi/2$ and $3\pi/2$ increase compared with the potential flow solution, a known effect of wake flow separation \cite{tang2022bag}. These laminar profiles are relatively stable without significant changes over time. Unique to the laminar aerobreakup case is the pressure maximum at $\theta = \pi$, which is present at all times therein except $t/\tau = 0.624$, when it briefly disappears. This most likely corresponds to the attached wake vortices, which oppose the movement of the leeward side of the droplet at early times \cite{tang2022bag, Jain2015}. 

As airphase turbulence is introduced, the pressure maxima at $\theta = \pi$ are no longer consistently observed, indicating that the attached vortices are now disrupted by external turbulent forcing. Moreover, figs.~\ref{fig:p-0.25-1} and \ref{fig:p-0.5-1} show that the minima in the wake region decrease and become significantly more negative than the potential-flow prediction as the turbulent fluctuating velocity $u^*$ increases, corresponding to an increase in the instantaneous pressure-induced force felt by the droplet. While our problem is transient in nature and differs from the configuration of Peng and Wang \cite{peng2023mechanisms} where the spherical particles are fixed in turbulent ambient flows, our observations match their report of pressure-drag amplification by external turbulence. An earlier study by Rind and Castro \cite{rind2012effects} discussed this pressure decrease in the wake of a disc-shaped obstacle in detail, where it is ascribed to increased entrainment of external turbulent fluid into the separated region and the wake vortices forming closer to the drop surface. The pressure profiles are also found to change significantly within and between the sampling periods when ambient turbulence is present. While our limited pressure-field data do not permit high-frequency time averaging to fully confirm this, the time-averaged turbulent flow field in the droplet wake might feature similar pressure profiles that do not depend strongly on either $u^*$ or $L_0$, thus preventing accumulation of streamwise acceleration at larger $u^*$ or $L_0$ values and leading to the universal acceleration rate we observed in fig.~\ref{fig:acc-dist-evol}. Indeed, while increased turbulent fluctuations lead to temporary peaks in the pressure profiles (e.g. as seen in the $t/\tau=0.693$ curve of fig.~\ref{fig:p-0.5-1}), the rate of pressure redistribution around the droplet increases as the ambient turbulence becomes more severe \cite{jiao2019direct}, so that these momentary peaks are approximately smoothed out over time, thus ensuring a net acceleration effect largely insensitive to the external turbulent forcing conditions as we observed here.

\begin{figure}[htbp]
	\centering
	\subfloat[]{
		\label{fig:mz-urms-sweep}
		\includegraphics[width=.48\textwidth]{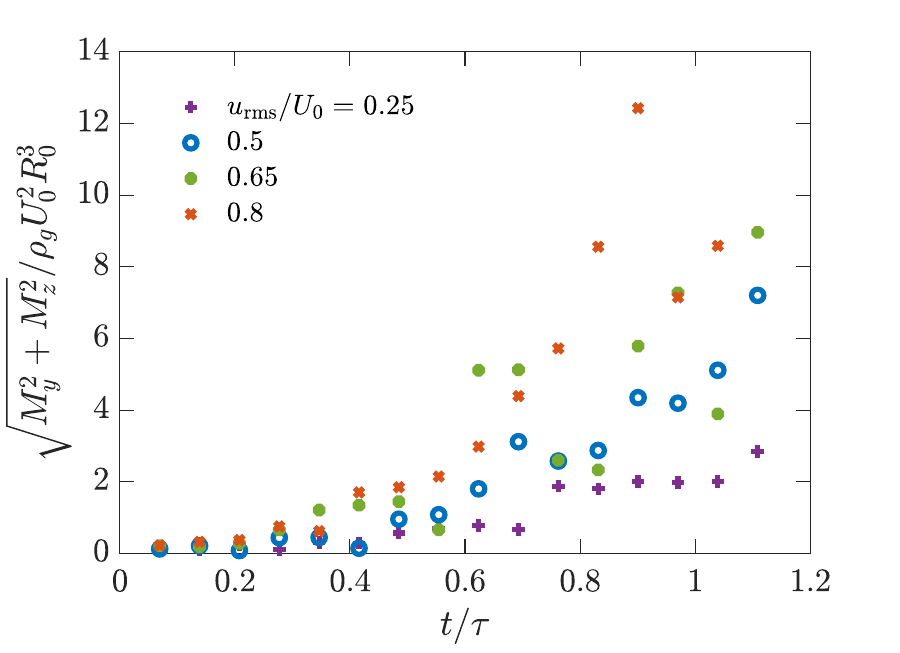}}
	\centering
	\subfloat[]{
		\label{fig:mz-lla-sweep}
		\includegraphics[width=.48\textwidth]{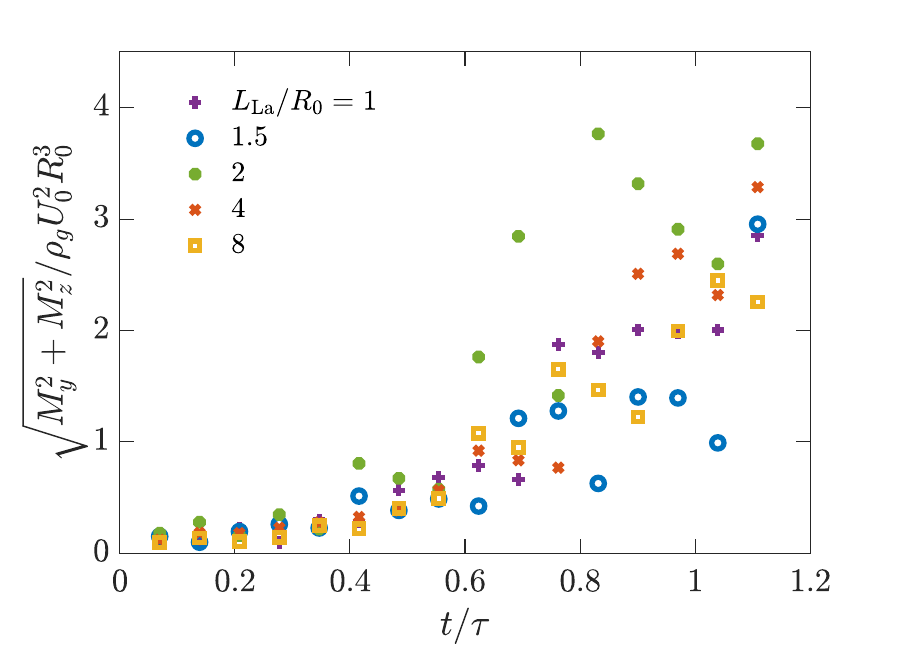}}

	\caption{The evolution of the hydrodynamic torque $\sqrt{M_y^2 + M_z^2}$ acting on the droplet at different turbulence fluctuating velocities $u^*$ (a) and $L_0$ (b), calculated using the pressure field in the vicinity of the droplet surface.}
	\label{fig:mz-info}
\end{figure}

Another interesting observation from fig.~\ref{fig:p-traverse} is that the pressure maxima do not always occur at $\theta = 0$. In other words, the stagnation point may drift away from the centre of the droplet frontal face under the disruption of ambient turbulence. Moreover, many instantaneous pressure profiles show extreme values appearing at $\theta = \pi/4$ and $3\pi/4$ in an anti-symmetric pattern (e.g., $t/\tau = 0.693$ in fig.~\ref{fig:p-0.25-1}), suggesting a net torque acting on the flattened droplet. To further explore this, we compute the evolution of the hydrodynamic torque $\bm{M}$ acting upon the droplet,
\begin{equation}
    \bm{M} = \iint_S \bm{r} \times (-p\delta_{mn} + 2\mu_g S_{mn}) d\bm{S},
\end{equation}
where $\bm{r}$ is the displacement vector from the droplet centre-of-mass to a point on its surface, $\delta_{mn}$ is the Kronecker symbol, and $S_{ij}$ is the strain-rate tensor of the ambient airflow \cite{jiang2021inertial}. It is found that the streamwise torque component of $M_x$ remains close to zero at all times, regardless of the turbulence configuration. The evolution patterns of the other two cross-stream torque components $M_y$ and $M_z$ are similar to each other while differing considerably from that of $M_x$, and thus we present the evolution of $\sqrt{M_y^2+M_z^2}$ in fig.~\ref{fig:mz-info}. Initially, $\sqrt{M_y^2+M_z^2}$ remains small and comparable in magnitude to $M_x$ (not shown in the figure) for all simulation cases up to $t/\tau \approx 0.5$, as is the case for the fixed particle study of Wang \emph{et al.} \cite{wang2019rotational} (see e.g. their fig.~3). For $t/\tau \geq 0.5$, the fluctuation in $\sqrt{M_y^2+M_z^2}$ becomes significantly amplified. Since according to fig.~\ref{fig:lam-turb-comp-snapshots}, the droplet is being compressed in the streamwise direction over this period, the results shown in fig.~\ref{fig:mz-urms-sweep} suggest that the droplet experiences much greater hydrodynamic torques as it undergoes flattening. This rapid growth in $\sqrt{M_y^2+M_z^2}$ can potentially lead to droplet rotation and is thus the direct cause of its tilting behaviour, which we will analyse in more detail later in \S\ref{subsec:tilting}, alongside the possible origins of such growth patterns.

\section{Droplet deformation patterns}
\label{sec:drop-deform}
In this section, we provide a quantitative analysis of the deformation patterns of droplets interacting with turbulent ambient airflows up to the point of bag formation. We start with quantifying the global extent and characteristics of droplet deformation using spherical harmonic decomposition in \S\ref{subsec:global}, and move on to examine the droplet tilting behaviour in \S\ref{subsec:tilting}. We finish this section with a discussion on the droplet surface corrugation patterns in \S\ref{subsec:corrug-form}.

\subsection{Global features}
\label{subsec:global}

We first analyse the deformation characteristics of the droplet using spherical harmonic decomposition following Perrard \emph{et al.} \cite{perrard2021bubble}, which allows us to investigate how surface deformation patterns of different length scales evolve as the droplet undergoes strongly nonlinear surface deformation. To assess the overall droplet deformation, we introduce the root-mean-squared deformation $\zeta_\Omega$ measured from an integration over all solid angles $d\Omega \equiv dA/(r-R_0)^2$,
\begin{equation}
    \zeta_\Omega^2 = \frac{1}{4\pi} \iint_S {(r - R_0)}^2 d\Omega,
    \label{for:zeta-omega-def}
\end{equation}

\begin{figure}[htbp]
	\centering
	\subfloat[]{
		\label{fig:zeta-urms-sweep}
		\includegraphics[width=.48\textwidth]{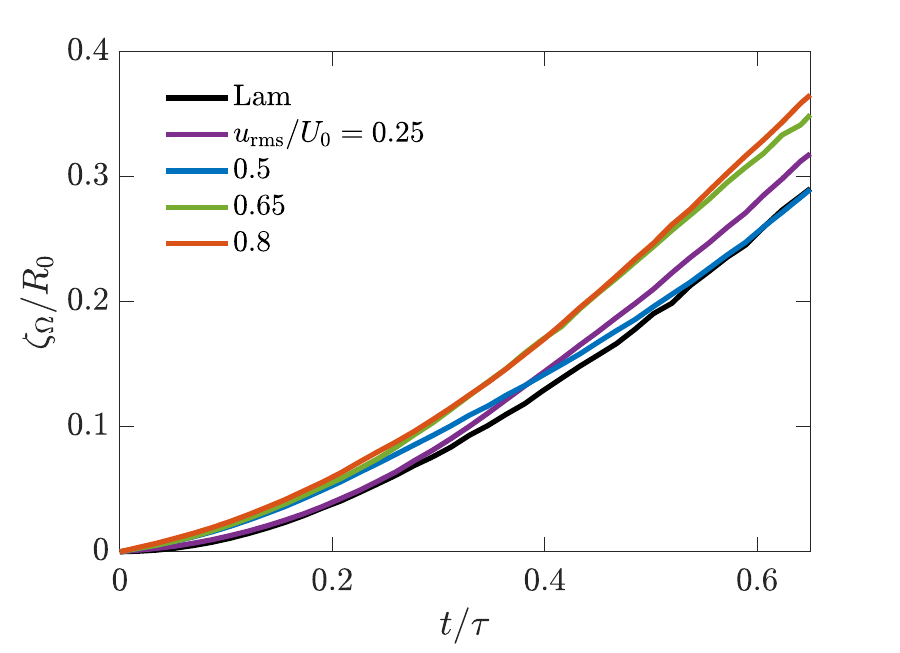}}
	\centering
	\subfloat[]{
		\label{fig:zeta-lla-sweep}
		\includegraphics[width=.48\textwidth]{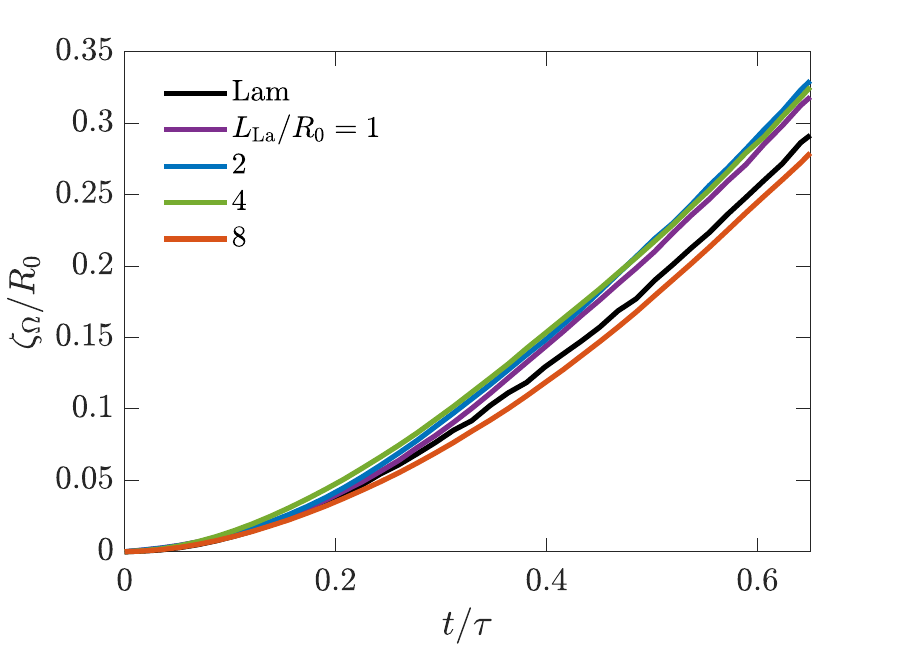}}

        \centering
	\subfloat[]{
		\label{fig:a2-urms-sweep}
		\includegraphics[width=.48\textwidth]{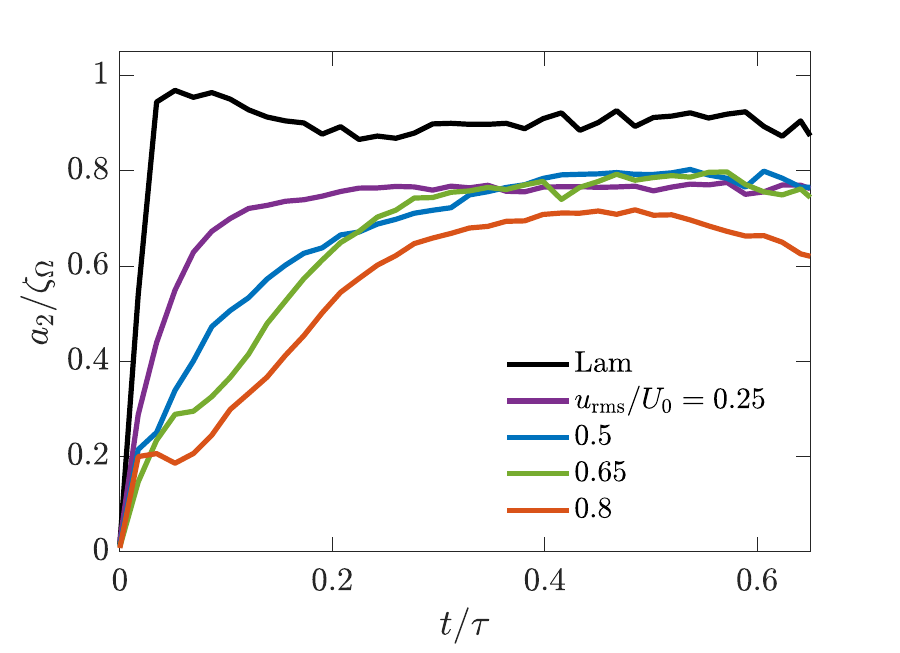}}
	\centering
	\subfloat[]{
		\label{fig:a2-lla-sweep}
		\includegraphics[width=.48\textwidth]{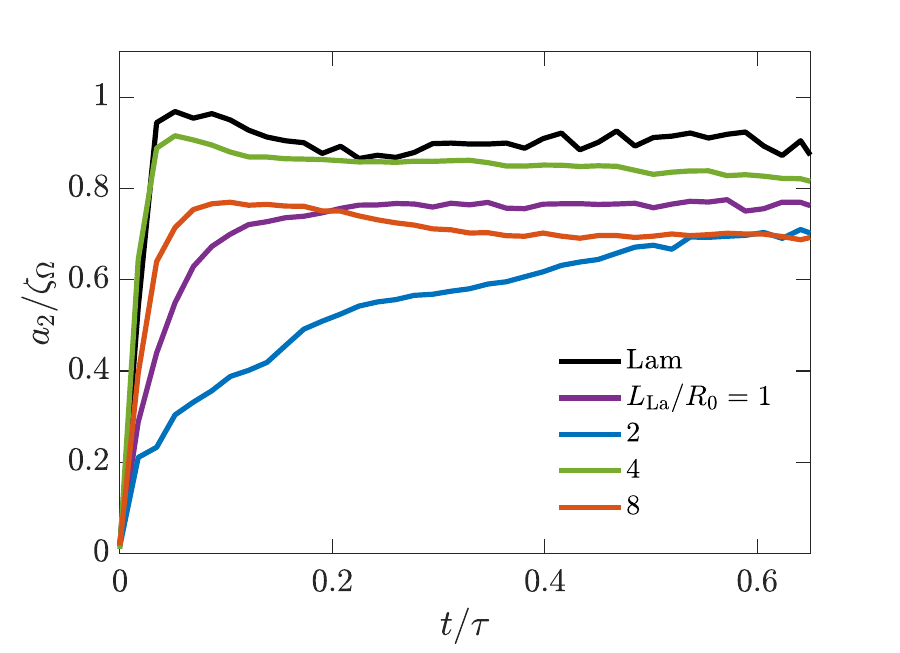}}
	\caption{Evolution of the ensemble-averaged droplet total spherical harmonic deformation $\zeta_\Omega$ (a,b) and mode-2 deformation $a_2$ (c,d) at different values of $u^*$ (a) and $L_0$.}
	\label{fig:sph-decomp}
\end{figure}

Where $R_0$ is the radius of the undeformed droplet, and $r-R_0$ reflects the local deformation extent on the droplet interface, which is assumed to be single-valued. Figs.~\ref{fig:zeta-urms-sweep} and \ref{fig:zeta-lla-sweep} show the ensemble-averaged early-time evolution of $\zeta_\Omega$ for different airphase turbulence configurations. $\zeta_\Omega$ is found to undergo an accelerated increase over time for all cases recorded, without saturating or oscillating trends observed at smaller ${\rm We}$ values in the absence of ambient mean flows \cite{perrard2021bubble, roa2023droplet}. Moreover, turbulence is found to cause a slight increase in the total deformation for most of the turbulent aerobreakup cases in comparison with their laminar aerobreakup counterpart; except $(u^*/U_0, \, L_0/R_0) = (0.25, \, 8)$ where $\zeta_\Omega$ is found to decrease instead, which is probably an effect of the ensemble size. These changes in $\zeta_\Omega$ are not dramatic, which is most likely because the turbulence-induced deformation corresponds to small-amplitude high-mode perturbations, and does not significantly modify the dominant mode-2 perturbations corresponding to droplet flattening, as will be discussed below.

Here we introduce the spherical harmonics $Y_l^m (\theta, \, \phi)$, and the deformation coefficient corresponding to each spherical harmonic mode $a_{l, \, m}$ can be calculated as follows,
\begin{equation}
    a_{l, \, m} = \frac{1}{4\pi} \iint_S (r - R_0) Y_l^m (\theta, \, \phi) d\Omega,
    \label{for:sph-harm-modes}
\end{equation}
which is realised through a Voronoi decomposition of the droplet surface during postprocessing. In the absence of a mean flow and within the limits of linear deformation, the coefficients $a_{l, \, m}$ are solutions to a set of damped oscillator equations \cite{prosperetti1980free, perrard2021bubble}. As harmonic modes with the same value of $l$ feature the same oscillation frequency, we introduce the global coefficient $a_l$ quantifying the total potential energy contained in all spherical harmonic modes $l$,
\begin{equation}
    a_l^2 = \sum_{m=-l}^l a_{l, \, m}^2,
    \label{for:sph-global}
\end{equation}
which is in turn associated with the root-mean-squared global deformation $\zeta_\Omega$ \eqref{for:zeta-omega-def} as follows via the orthogonality of the spherical harmonic bases,
\begin{equation}
    \zeta_\Omega^2 = \sum_{l=1}^{+\infty} a_l^2.
\end{equation}

Among these global coefficients, we select the droplet centre position $\bm{r_c}$ as the origin for the spherical coordinate system such that $a_1 = 0$. The mode-2 deformation coefficient $a_2$ is found to dominate all global coefficients $a_i$ and evolve with an increasing growth rate similar to that of $\zeta_\Omega$. To facilitate comparison between $a_2$ and $\zeta_\Omega$, we plot the evolution of their ratios over time in figs.~\ref{fig:a2-urms-sweep} and \ref{fig:a2-lla-sweep}. For most of the turbulent aerobreakup cases, $a_2/\zeta_\Omega$ reaches a plateau value between 0.7 and 0.8, smaller than its laminar aerobreakup counterpart of 0.9. Since mode-2 spherical harmonic deformation corresponds to prolate-oblate shape changes, this slight decrease in $a_2/\zeta_\Omega$ suggests that while the flattening of the droplet still largely follows a prolate-oblate pattern, the introduction of air-phase turbulence fluctuations excites higher-mode deformation patterns with smaller length scales, as we have already observed in fig.~\ref{fig:lam-turb-comp-snapshots}. Fig.~\ref{fig:a2-urms-sweep} also suggests that with increasing values of $u^*$, the growth of $a_2/\zeta_\Omega$ for $t/\tau \leq 0.3$ becomes slower. This might be associated with stronger surface corrugation patterns as the ambient turbulent fluctuations intensify. The influence of $L_0$ on the evolution of $a_2/\zeta_\Omega$ is less obvious. The evolution patterns of $a_2/\zeta_\Omega$ at $L_0/R_0 \geq 4$ closely resemble their laminar aerobreakup counterpart, which is expected as the droplets experience much milder ambient flow velocity variation at $d_h/d_0 > 1$ (see our discussions of fig.~\ref{fig:back-snapshots}). For $(u^*/U_0, \, L_0/R_0) = (0.25, \, 2)$, $a_2/\zeta_\Omega$ becomes much smaller at very early times as shown in fig.~\ref{fig:a2-lla-sweep}, but have exceeded 0.5 by $t/\tau = 0.3$. These smaller values of $a_2/\zeta_\Omega$ at early times correspond to a phase reversal in some of the individual realisations, i.e., for these cases the droplets first elongate upon insertion, and higher-mode deformation arises as the droplet relaxes from this initial elongation and begins to flatten.

\begin{figure}[htbp]
	\centering
	\subfloat[]{
		\label{fig:asp-ratio-urms-sweep}
		\includegraphics[width=.48\textwidth]{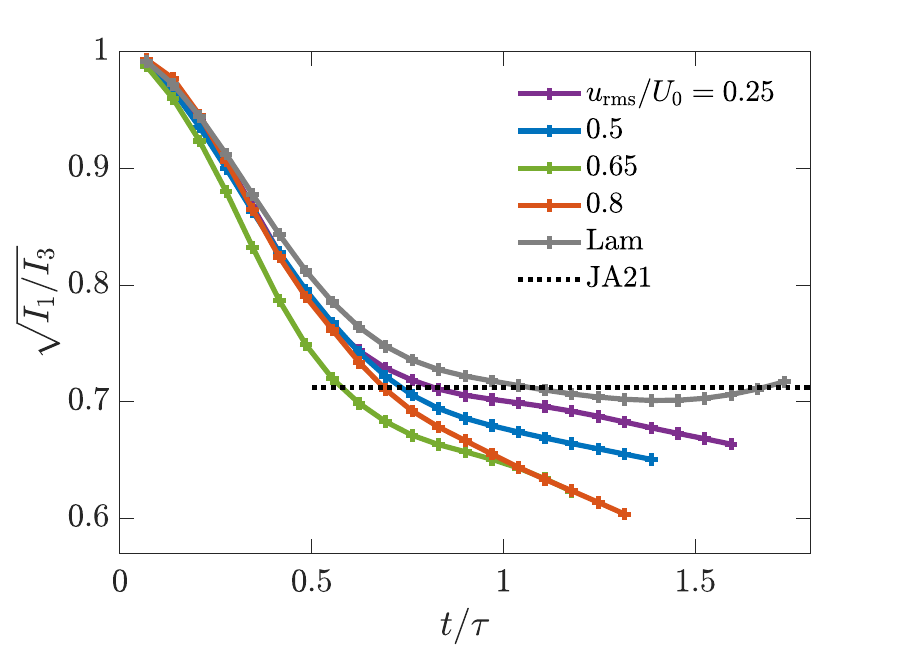}}
	\centering
	\subfloat[]{
		\label{fig:asp-ratio-lla-sweep}
		\includegraphics[width=.48\textwidth]{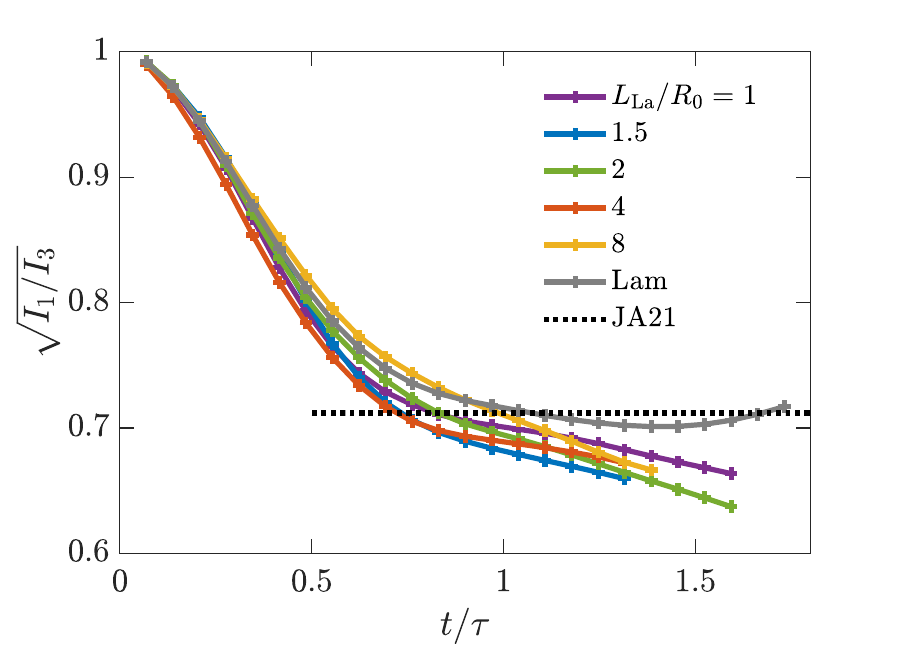}}

	\caption{Evolution of the ensemble-averaged droplet aspect ratio $\sqrt{I_1/I_3}$ at different values of $u^*$ (a) and $L_0$.}
	\label{fig:asp-evol}
\end{figure}

Given that the early-time droplet deformation can still be largely described as a prolate-oblate shape change despite corrugation formation, we now seek to quantify the global influence of turbulence corresponding to the mode-2 spherical harmonics, while leaving local higher-order corrugation features for the curvature distribution analysis in \S\ref{subsec:corrug-form}. It should be noted that, since the droplet can become distorted and/or tilted during its deformation process under strong air-phase turbulence, the definitions of bag length and width \cite{Jackiw2021, tang2022bag, ling2023detailed} as the streamwise and spanwise extents of the bag are not strictly applicable here. Instead, following Cannon \emph{et al.} \cite{cannon2024morphology}, we define the droplet aspect ratio $\sqrt{I_1/I_3}$, which is the ratio between the smallest and largest eigenvalues of the moment-of-inertia tensor for the droplet,
\begin{equation}
    \bm{T_I} = \int_V \rho ({\bm{r}}^2 \bm{I} -\bm{r} \otimes \bm{r}) dV,
    \label{for:moi-tensor-def}
\end{equation}
and show their evolution in fig.~\ref{fig:asp-evol}. We note here that $I_1$ and $I_3$ are proportional to the lengths of the minor and major axes of the spheroid, respectively.

It is first observed that $\sqrt{I_1/I_3}$ initially decreases over time and stabilises around $t/\tau = 0.75$, for laminar and turbulent airflows satisfying $u^*/U_0 \leq 0.5$. This corresponds to the flattening stage of the droplet where the bag width increases and the bag length decreases. At later stages, the laminar flow case shows a slight increase in the aspect ratio, corresponding to the late-time bag blowout phenomenon where the bag length significantly increases. The introduction of air-phase turbulence does not significantly affect the evolution of $\sqrt{I_1/I_3}$ at early times except for $u^* = 0.65$; while at mid-to-late times it is found to cause a decrease in $\sqrt{I_1/I_3}$ compared with the laminar airflow case. Taking into account the observation of Zhao \emph{et al.} \cite{zhao2019effect} that liquid bags are stretched longer and wider in turbulent ambient flows, a decrease in $\sqrt{I_1/I_3}$ then suggests that the increase in bag width exceeds that of its length. Moreover, for turbulent ambient flows featuring $u^*/U_0 \geq 0.65$, fig.~\ref{fig:asp-ratio-urms-sweep} shows a continued decrease of $\sqrt{I_1/I_3}$, which is possibly because the air-phase turbulence is now strong enough to distort the flattened droplet so significantly that both its orientation and bag blowout dynamics are modified, as observed in figs.~\ref{fig:back-0.65-1} and \ref{fig:back-0.8-1}.

We also present in fig.~\ref{fig:asp-evol} the aspect ratio at the onset of bag formation predicted by Jackiw and Ashgriz \cite{Jackiw2021} for ${\rm We} = 15$, which is determined as follows. According to that study, the major axis lengths for bag morphologies are approximately $2R_i/d_0 \approx 1.6$; whereas the minor axis length is given as follows,
\begin{equation}
    \frac{d_i}{d_0} = \frac{4}{{\rm We}_{\rm rim} + 10.4},
    \label{for:ja-rim-diameter}
\end{equation}
where the rim Weber number is defined as ${\rm We}_{\rm rim} \equiv \rho_l \Dot{R}^2 d_0/\sigma$. Based on an analysis of pressure balance at the drop surface, Jackiw and Ashgriz \cite{Jackiw2021} proposed the following prediction for the bag spanwise growth rate $\Dot{R}$,
\begin{equation}
    \frac{\Dot{R}}{d_0/2} = \frac{1}{\tau} \frac{a^2}{4} \left( 1 - \frac{128}{a^2 We} \right) T_{bal},
    \label{for:ja-span-growth}
\end{equation}
where $a = 6$ is the stretching rate at the frontal stagnation point of the droplet, and the initial flow balancing time $T_{bal}$ is set as $\tau/8$. As such, we are able to determine the theoretical aspect ratio $\sqrt{I_1/I_3}$ assuming an ellipsoidal shape with axis lengths $a_i = b_i = R_i$ and $c_i = d_i/2$,
\begin{equation}
    \sqrt{\frac{I_1}{I_3}} = \sqrt{\frac{b_i^2 + c_i^2}{a_i^2 + b_i^2}} = 0.712,
    \label{for:ja-aspect-ratio}
\end{equation}
which we plotted in fig.~\ref{fig:asp-evol} for reference. It is found that the prediction of \eqref{for:ja-aspect-ratio} is very close to the minimum in our laminar airflow case, thus validating our aspect ratio calculations. Overall, we consider the deviation of turbulent aerobreakup aspect ratios from \eqref{for:ja-aspect-ratio} to result from the increased droplet fore-aft pressure difference, which modifies the spanwise growth rate $\Dot{R}$ \eqref{for:ja-span-growth}, while the generation of higher-mode interfacial deformation might also play a role.

\subsection{Droplet tilting behaviour}
\label{subsec:tilting}

\begin{figure}[htbp]
	\centering
	\subfloat[]{
		\label{fig:tilting-2}
		\includegraphics[width=.32\textwidth]{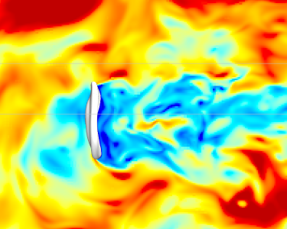}}
	\centering
	\subfloat[]{
		\label{fig:tilting-3}
		\includegraphics[width=.32\textwidth]{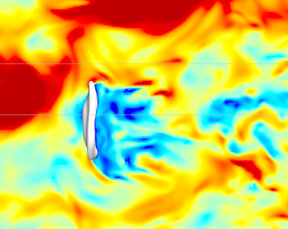}}
    \centering
	\subfloat[]{
		\label{fig:tilting-4}
		\includegraphics[width=.32\textwidth]{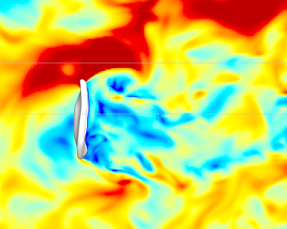}}

    \centering
	\subfloat[]{
		\label{fig:tilting-5}
		\includegraphics[width=.32\textwidth]{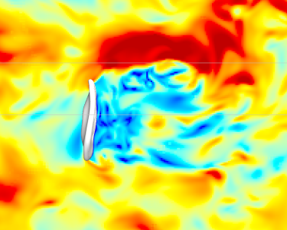}}
	\centering
	\subfloat[]{
		\label{fig:tilting-6}
		\includegraphics[width=.32\textwidth]{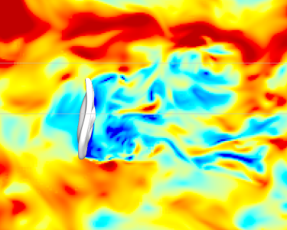}}
    \centering
	\subfloat[]{
		\label{fig:tilting-7}
		\includegraphics[width=.32\textwidth]{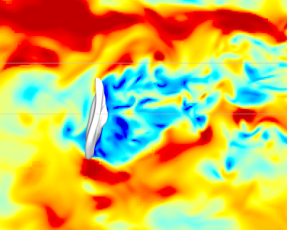}}
	\caption{Snapshots showing the tilting of a droplet with air-phase turbulence characteristics $u^*/U_0 = 0.25$, $L_0/R_0 = 2$, where the background flow field is coloured based on the magnitude of the local airflow velocity. Warmer colour represents higher magnitudes of airflow velocity in these snapshots. From (a) to (f): $t/\tau = 1.03$, 1.09, 1.14, 1.20, 1.26, and 1.32.}
	\label{fig:tilting-snapshots}
\end{figure}

As the droplet flattens in the ambient air-phase turbulence, it may exhibit the tilting behaviour formerly reported by Jiao \emph{et al.} \cite{jiao2019direct}; namely, the principal axes of the deforming droplet no longer align with the streamwise direction. Similar phenomena have been observed in previous experimental studies of droplet aerobreakup \cite{Jackiw2021, jackiw2022prediction} and also investigations of sea spume generation \cite{troitskaya2023statistical}; which are analogous to the tumbling behaviour of anisotropic solid spheroids where the particle rotates around a non-symmetry axis \cite{cui2023effect}. In our current study, this behaviour is usually first observed at intermediate times as shown in fig.~\ref{fig:tilting-snapshots}, when the droplet has fully flattened and not yet evolved into bag-rim structures at later times; although for one particular realisation with $(u^*/U_0, \, L_0/R_0) = (0.25, \, 2)$ this occurred at very early times when the droplet is still largely spherical. Here, we have coloured the background flow field using the magnitude of airflow velocity $u^*/U_0$. These snapshots reveal that the droplet tilting phenomenon is closely associated with turbulence intermittency: while the principal axis of the droplet initially aligns largely with the streamwise direction, as shown in fig.~\ref{fig:tilting-2}, a cluster of high-speed air parcels approaches (fig.~\ref{fig:tilting-3}) and bypasses (figs.~\ref{fig:tilting-4}-\ref{fig:tilting-5}) the droplet asymmetrically from above. This high-speed air parcel modifies the pressure distribution on the droplet surface to form a net torque (see our discussions in Sec.~\ref{sec:drop-turb-dynamics}), causing it to tilt over in fig.~\ref{fig:tilting-6}. It is also noted that this high-speed air parcel does not directly impact the droplet, which might correspond to the earlier observations of Vela-Mart{\'\i}n and Avila \cite{vela2021deformation} that non-local actions by outer eddies away from the drop surface contribute significantly to the surface energy increment. For most simulation cases, the droplet will retain this oblique orientation afterwards as the interaction with these high-speed air parcels is a rare event.

To quantify the tilting phenomenon described above, we measure the droplet principal axis $\bm{e_1} \equiv (\theta_{11}, \, \theta_{12}, \, \theta_{13})$, which can be obtained by computing the eigenvectors of the moment of inertia tensor $\bm{T_I}$ \eqref{for:moi-tensor-def}. Here, $\theta_{ij}$ is the angle between the droplet principal axis $\bm{e_i}$ and the lab-frame coordinate $\bm{j}$; and we are particularly interested in the orientation angle $\theta_{11}$ between the droplet minor axes and the streamwise direction, since its change most clearly reflects the droplet tilting dynamics.

We then explore the possible connections between the tilting behaviour and the turbulent airflow characteristics at the scale of the droplet diameter $d_0$. Following Masuk \emph{et al.} \cite{masuk2021simultaneous, masuk2021orientational}, we introduce the slip velocity $\bm{u_{slip}}$,
\begin{equation}
    \bm{u_{slip}} = \bm{\Bar{u}_i(x_0)} - \bm{u_c}.
    \label{for:u-theta-slip-def}
\end{equation}
Here, $\bm{\Bar{u}_i(x_0)}$ is the coarse-grained air-phase velocity at $x_0$, computed by averaging all velocity records within a shell enclosing the droplet given by $1.5R_0 \leq r \leq 2R_0$; and we have checked that changing the upper and lower boundaries of this sampling shell does not have any discernible effect on the value of $\Bar{u_i(x_0)}$. As $\bm{u_c}$ is the droplet centre-of-mass velocity, the slip velocity $\bm{u_{slip}}$ represents the coarse-grained speed of the airflow felt by the droplet.

\begin{figure}[htbp]
	\centering
	\subfloat[]{
		\label{fig:u-slip-evol-urms-sweep}
		\includegraphics[width=.48\textwidth]{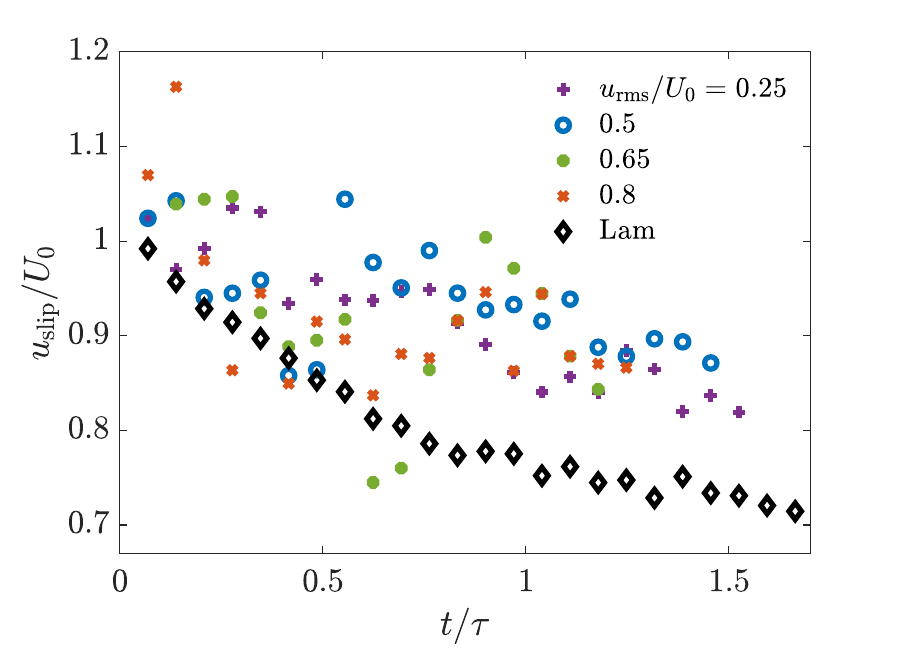}}
	\centering
	\subfloat[]{
		\label{fig:u-slip-evol-lla-sweep}
		\includegraphics[width=.48\textwidth]{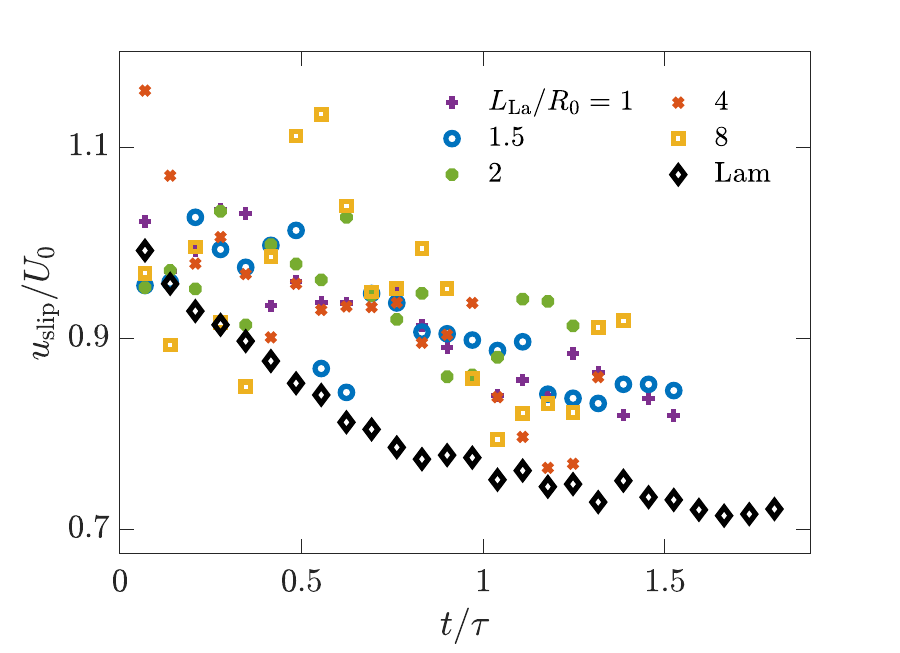}}

	\caption{Evolution of the slip velocity $u_{slip}$ over time at different air-phase turbulence configurations.}
	\label{fig:u-slip-evol}
\end{figure}

Fig.~\ref{fig:u-slip-evol} shows the evolution of the magnitude of slip velocity $u_{slip}$, where a decreasing trend is observed for all cases reported. This decrease is due to the acceleration of the droplet by the ambient airflow, and is most prominent for the early-time flattening stage where $t/\tau \leq 0.6$. Interestingly, we observe that at the same non-dimensional time $t/\tau$, the slip velocities for the turbulent airflow cases mostly exceed their laminar airflow counterpart, and there does not appear to be any strong dependence of $u_{slip}$ on the turbulence configurations. This increase in $u_{slip}$ is most likely due to the high-speed turbulent fluid parcels bypassing the droplet, as we observed in fig.~\ref{fig:tilting-snapshots}; which is also associated with the increased centre-of-mass acceleration and deformation rates we discussed earlier in \S\ref{sec:drop-turb-dynamics} and \S\ref{subsec:global}.

\begin{figure}[htbp]
	\centering
	\subfloat[]{
		\label{fig:theta11-urms-sweep}
		\includegraphics[width=.48\textwidth]{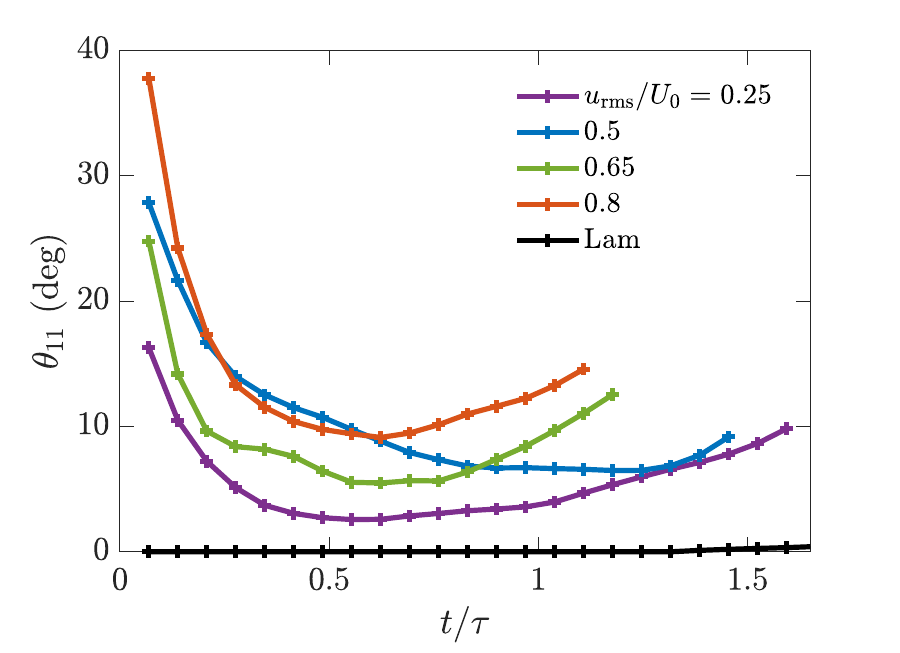}}
        \centering
	\subfloat[]{
		\label{fig:theta11-lla-sweep}
		\includegraphics[width=.48\textwidth]{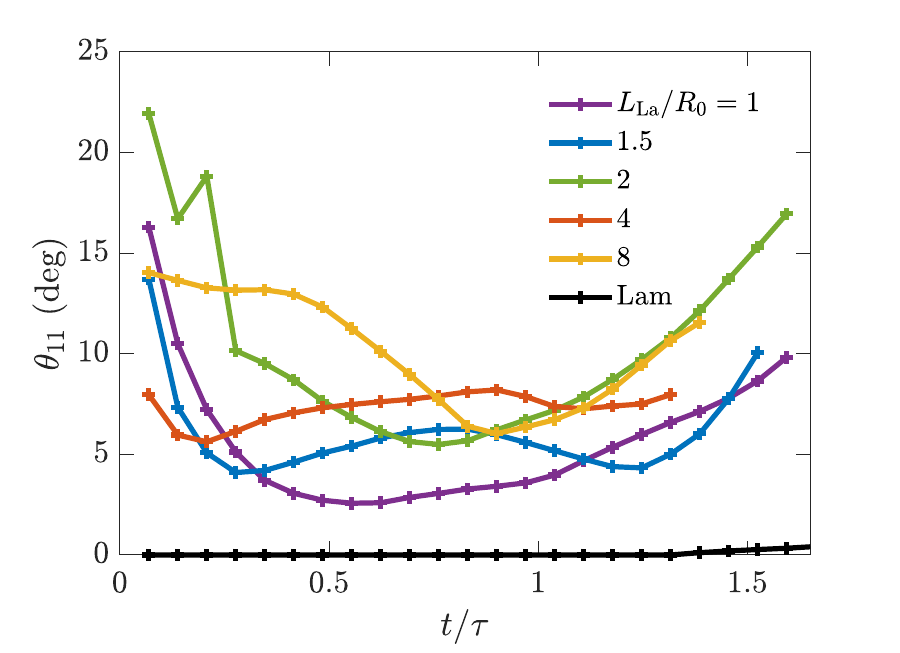}}
  
    \centering
	\subfloat[]{
		\label{fig:theta-slip-urms-sweep}
		\includegraphics[width=.48\textwidth]{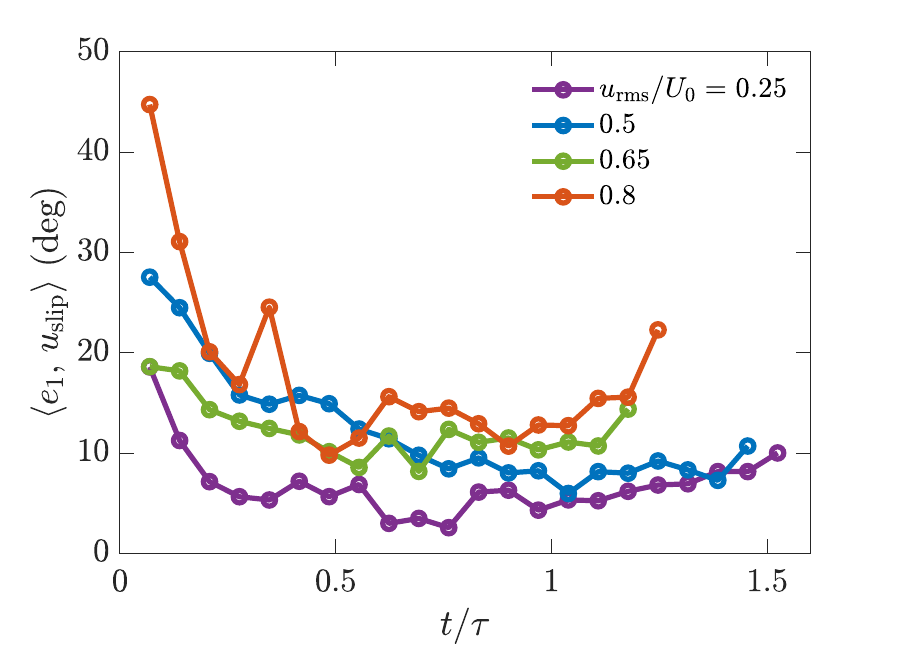}}
	\centering
	\subfloat[]{
		\label{fig:theta-slip-lla-sweep}
		\includegraphics[width=.48\textwidth]{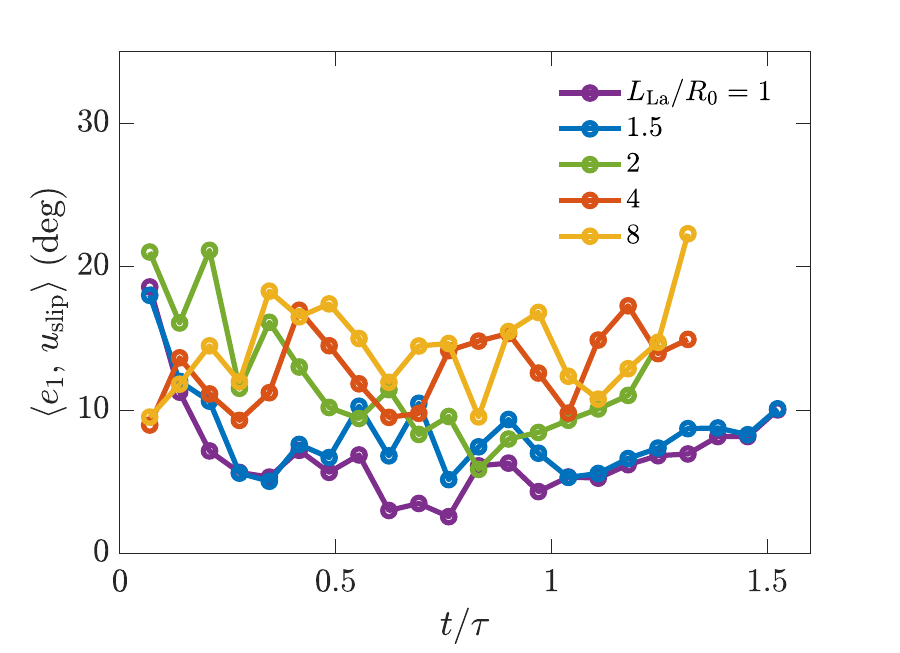}}

	\caption{Evolution of ensemble-averaged droplet orientation angle $\theta_{11}$ (a,b), and the alignment angle $\left< \bm{e_1}, \, \bm{u_{slip}} \right>$ between the droplet principal axis $\bm{e_1}$ and the slip velocity $\bm{u_{slip}}$ (c,d).}
	\label{fig:theta11-theta-slip-urms-lla-sweep}
\end{figure}

We now present the evolution patterns of the droplet orientation angle $\theta_{11}$ at different turbulence configurations in figs.~\ref{fig:theta11-urms-sweep} and ~\ref{fig:theta11-lla-sweep}. While $\theta_{11}$ remains zero for the laminar airflow case, when turbulence is introduced, we first observe large initial values of $\theta_{11}$, which are most likely caused by small-scale droplet surface perturbations imprinted by the turbulent airflow just after the insertion of the drop. We have already noted the effects of these small perturbations when discussing the decrease in $a_2/\zeta_\Omega$ observed in figs.~\ref{fig:a2-urms-sweep} and \ref{fig:a2-lla-sweep}, which decay rapidly as the droplet flattens. For most of the cases shown in figs.~\ref{fig:theta11-urms-sweep} and \ref{fig:theta11-lla-sweep}, we observe an increase in $\theta_{11}$ for $t/\tau \geq 1$; since the droplet has already finished flattening by this stage, we consider this increase in $\theta_{11}$ to represent the late-time tilting behaviour observed in fig.~\ref{fig:tilting-snapshots}. It is also within this late stage that $\theta_{11}$ appears to mildly increase with increasing $u^*$ and $L_0$, although exceptions exist when $(u^*/U_0, \, L_0/R_0) = (0.5, \, 1)$ and $(0.25, \, 2)$.

The previous study by Masuk \emph{et al.} \cite{masuk2021orientational} reveals a strong preferential alignment between bubbles deforming in turbulence and the slip velocity. This is measured by the alignment angle $\left< \bm{e_1}, \, \bm{u_{slip}} \right>$ between the minor axis of the flattening drop and the slip velocity, which is defined as,
\begin{equation}
    \left< \bm{e_1}, \, \bm{u_{slip}} \right> \equiv \arccos \left( \frac{\bm{e_1} \cdot \bm{u_{slip}}}{\norm{\bm{e_1}} \norm{\bm{u_{slip}}}} \right),
\end{equation}
whose evolution we present in figs.~\ref{fig:theta-slip-urms-sweep} and \ref{fig:theta-slip-lla-sweep}. While the values of $\left< \bm{e_1}, \, \bm{u_{slip}} \right>$ calculated are generally close to those of $\theta_{11}$, larger fluctuations are observed in its evolution, which are expected due to the presence of variations in the ambient turbulent airflow and vortex shedding in the wake of the drop. Interestingly, despite the presence of scatter and the apparent increase of the alignment angle with increasing $L_0$ shown in fig.~\ref{fig:theta-slip-lla-sweep}, we still observe a largely decreasing trend in $\left< \bm{e_1}, \, \bm{u_{slip}} \right>$ at early time, similar to that of $\theta_{11}$. This suggests a close connection between the two quantities, as the coarse-grained airflow velocity felt by the droplet becomes increasingly aligned with its minor primary axis $\bm{e_1}$ as it flattens. This agrees with Cui \emph{et al.} \cite{cui2023effect}, as they observed that under the influences of the fluid inertial torque, the symmetry axes of oblate spheroids tend to align with the slip velocity.

Referring back to fig.~\ref{fig:mz-urms-sweep}, one observes that the amplitude of the hydrodynamic torque acting on the droplet begins to rapidly increase at around the same time when $\theta_{11}$ and $\left< \bm{e_1}, \, \bm{u_{slip}} \right>$ minimise. We believe this is due to the increased susceptibility of the flattened droplet to ambient flow field perturbations originating from the bypassing of high-speed ambient flow parcels as shown in fig.~\ref{fig:tilting-snapshots}, even though the magnitude of the slip velocity felt by the droplet continues to decrease, as shown in fig.~\ref{fig:u-slip-evol}. To better demonstrate this argument, we note that previous studies prescribe the following functional form for hydrodynamic torques experienced by immersed oblate particles \cite{jiang2021inertial, cui2023effect},
\begin{equation}
    |\bm{M}_H| = \frac{1}{2} \rho_g |\bm{u_{slip}}|^2 a^3 F(\lambda) \sin 2\left< \bm{e_1}, \, \bm{u_{slip}} \right>,
    \label{for:torque-prescp}
\end{equation}
where $a$ is the length of the droplet major axis, and $F(\lambda)$ is a shape function dependent on the droplet aspect ratio $\lambda$. This prescription is strictly valid only for Stokes flow scenarios where the particle Reynolds number $Re_p \equiv \rho_g |\bm{u_{slip}}| a / \mu_g \ll 1$, but available results support its usage up to $Re_p \approx 10^2$ \cite{ouchene2020numerical, jiang2021inertial}. We do not expect our simulations, which have $Re_p$ around $10^3$, to match this result; but we note a similarity in how it connects the turbulent properties of the surrounding flow to the torque experienced by the droplet. Since it is found in fig.~\ref{fig:theta11-theta-slip-urms-lla-sweep} that when the droplet fully flattens, the alignment angle $\left< \bm{e_1}, \, \bm{u_{slip}} \right>$ remains small and does not change significantly across different turbulence configurations, whereas the slip velocity amplitude $|\bm{u_{slip}}|$ keeps decreasing, the rapid growth of torque exerted on the droplet may be attributable to a significant growth in $a^3 F(\lambda)$, which is determined by the droplet geometry. Indeed, the shape function $F(\lambda)$ increases rapidly with decreasing aspect ratio within the range of $0.1 \leq \lambda \leq 1$, as shown in fig.~3 of Ref.~\cite{cui2023effect}, which supports our reasoning that the rapid torque growth and the droplet tilting behaviour are associated with decreasing droplet aspect ratio due to flattening. At the current stage, a more rigorous examination of the droplet tilting dynamics remains difficult compared with the tumbling behaviour of solid particles \cite{voth2017anisotropic, cui2023effect}, since the droplet is deforming and tilting at the same time. However, we consider that the discussion of $\theta_{11}$, and the comparison with both $\left< \bm{e_1}, \, \bm{u_{slip}} \right>$ and the hydrodynamic torque shed light on the influences of nearby extreme turbulence events on the droplet dynamics.

\subsection{Formation of surface corrugations}
\label{subsec:corrug-form}

In this section, we direct our attention towards more local morphological features present on the drop surface; more specifically the distributions of the surface velocity increment $\delta u_s$ \cite{perrard2021bubble} and the local droplet interface curvature $\kappa$ \cite{mangani2022influence, qi2024breaking}. Here, the surface velocity increment $\delta u_s$ is defined on a reference envelope $\Gamma$ encompassing the droplet, as the difference between the radial components of the local air-phase velocity $\bm{u}$ and the mean air-phase velocity of all points located on this envelope; namely
\begin{gather}
    \delta u_s(R, \, \theta, \, \phi) = \bm{\Tilde{u}} (\bm{R}) \cdot \Hat{r}, \\
    \bm{\Tilde{u}} (\bm{R}) = \bm{u} - \frac{1}{4\pi} \iint_{\Gamma} \bm{u}(\bm{R}, \, \theta, \, \phi) \, d\Omega.
    \label{for:delta-u-def}
\end{gather}
The distance between any point on $\Gamma$ and the droplet surface is fixed as $\Delta l = 0.05R_0$. It should be noted that, in contrast to Perrard \emph{et al.} \cite{perrard2021bubble} where $\Gamma$ is a fixed spherical surface, in our case $\Gamma$ follows the droplet and deforms along with it due to the presence of the mean flow. Positive $\delta_s$ indicates the ambient airflow moving at velocities higher than the average value, while negative $\delta_s$ values suggest lower-than-average ambient velocities. For the early stages of droplet aerobreakup characterised by spanwise flattening, it is expected that the peripheral region of the drop mainly features positive $\delta_s$, while the frontal and back faces of the drop feature negative $\delta_s$ instead. 

\begin{figure}[htbp]
	\centering
	\subfloat[]{
		\label{fig:del-us-dist-pre-ins}
		\includegraphics[width=.48\textwidth]{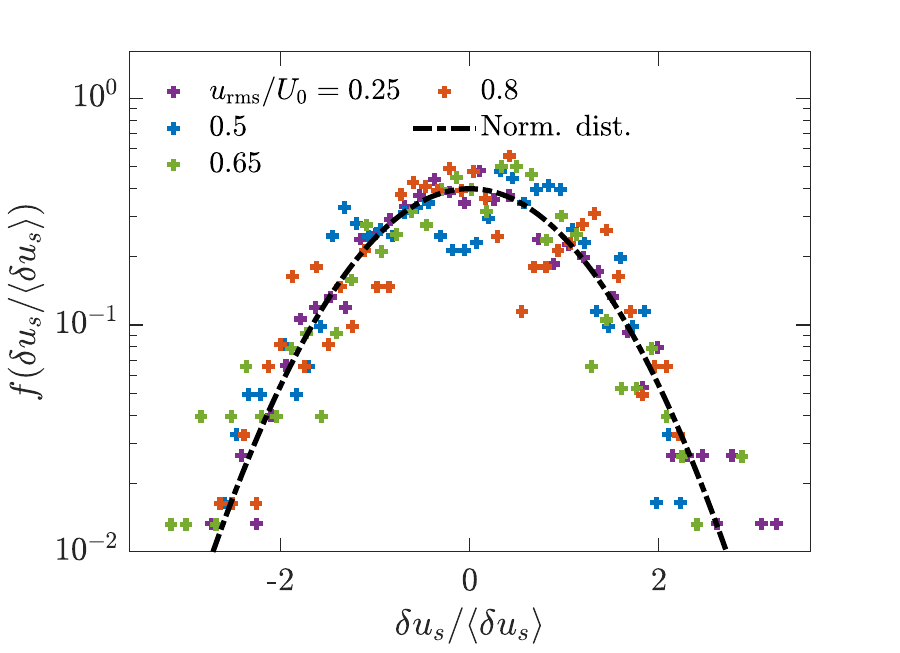}}
    \centering
	\subfloat[]{
		\label{fig:del-us-kappa-corr}
		\includegraphics[width=.48\textwidth]{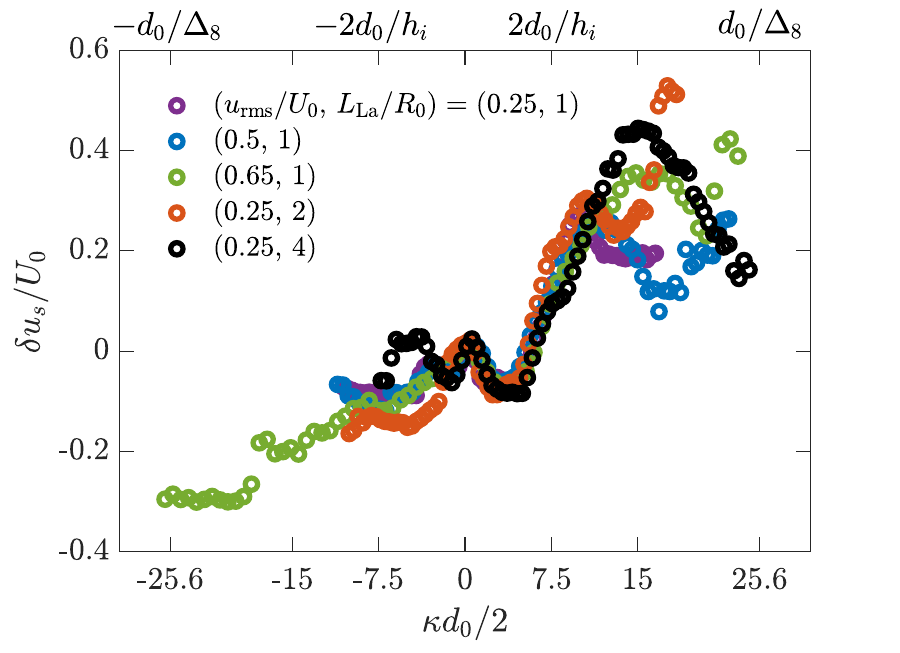}}

    \centering
	\subfloat[]{
		\label{fig:kappa-visual-0.65-1-1}
		\includegraphics[width=.3\textwidth]{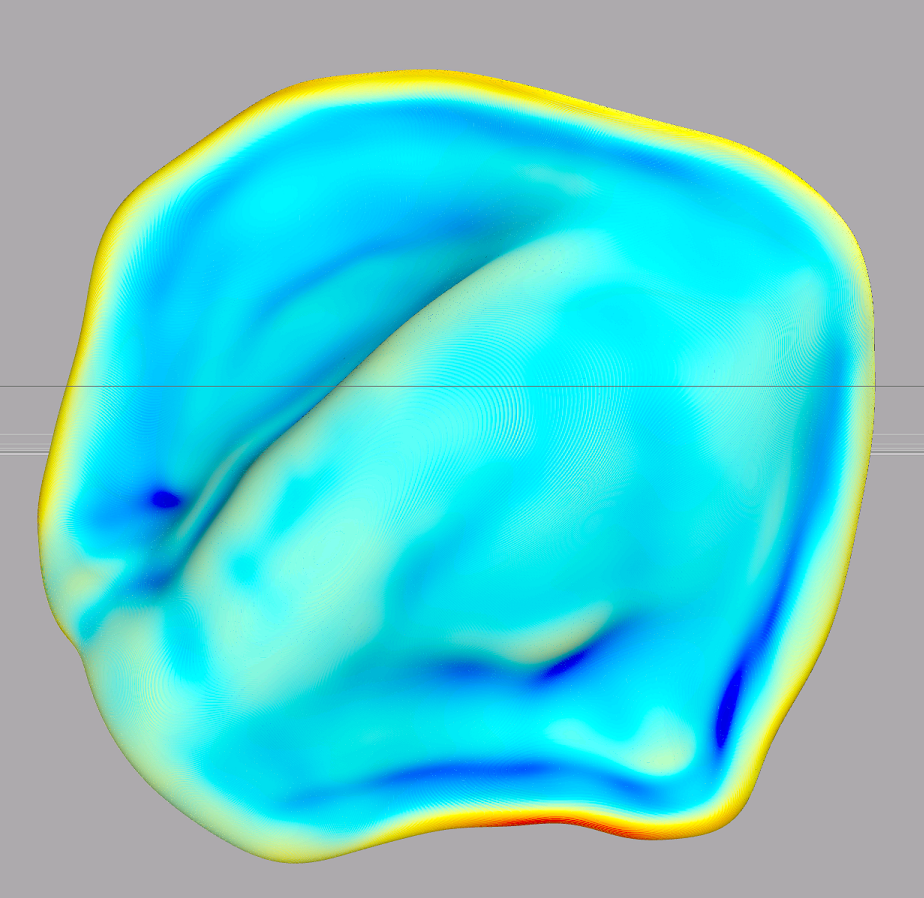}}
    \centering
	\subfloat[]{
		\label{fig:kappa-visual-0.65-1-2}
		\includegraphics[width=.3\textwidth]{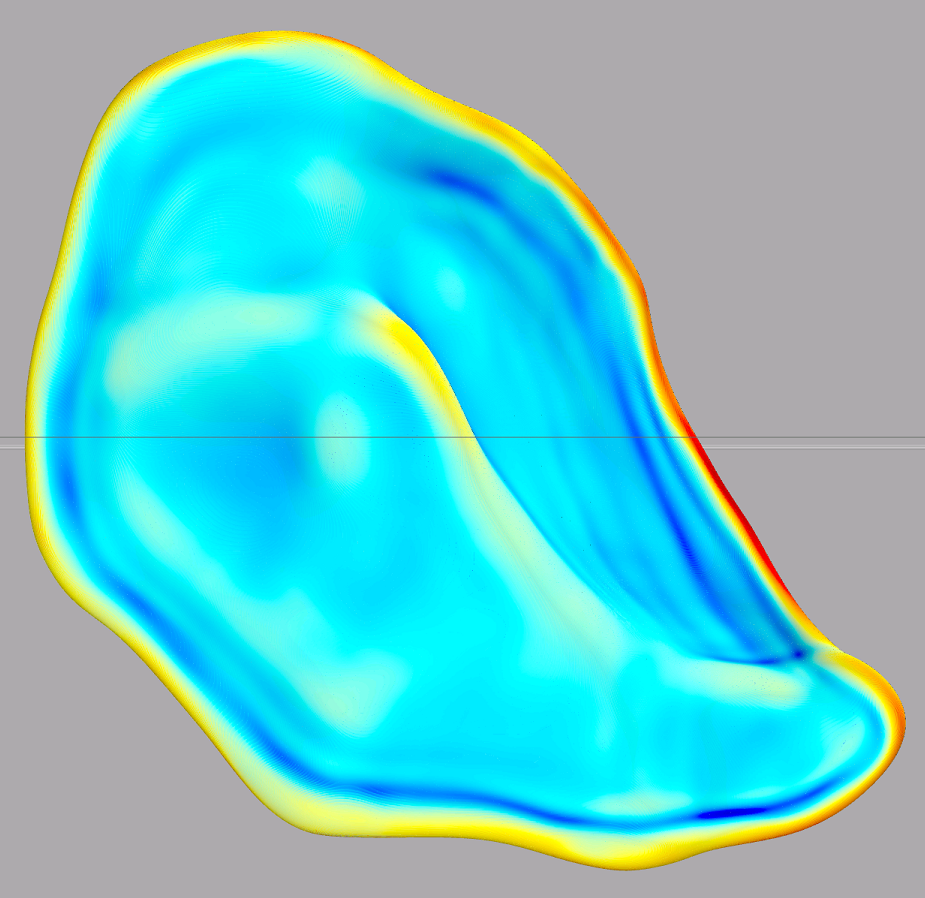}}
    \centering
	\subfloat[]{
		\label{fig:kappa-visual-0.65-1-3}
		\includegraphics[width=.3\textwidth]{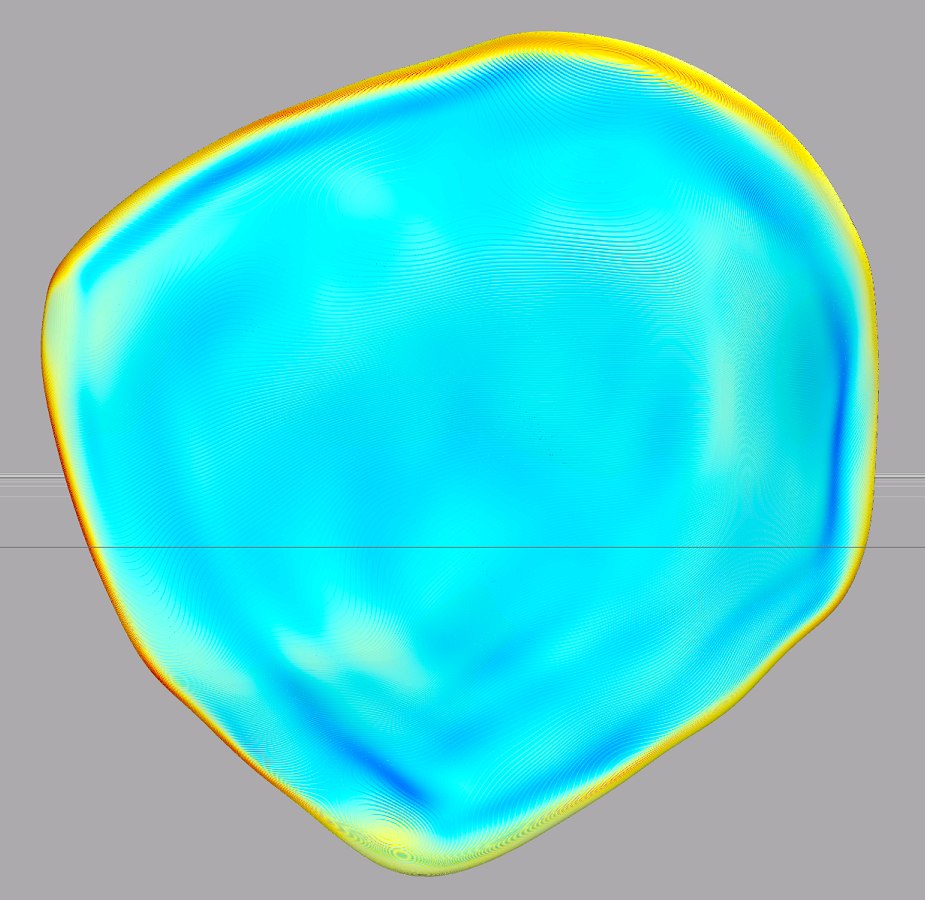}}
 
	\caption{(a): Probability distribution functions of the surface velocity increment $\delta u_s$ immediately before droplet insertion, measured at different $u^*$ values. (b): Correlation between $\delta u_s$ and local interface curvature $\kappa$, sampled at $t/\tau = 1.109$ and at different $u^*$ values. The secondary axis on the top of fig.~\ref{fig:del-us-kappa-corr} shows reference curvature values corresponding to the turbulence grid size $d_0/\Delta_8$ and droplet peripheral rim $2d_0/h_i$. (c)-(f): Colour-mapped droplet surfaces showing the spatial distribution of local interface curvature $\kappa$ for $(u^*/U_0, \, L_0/R_0) = (0.65, \, 1)$ and $t/\tau = 1.11$, where warmer colour corresponds to larger $\kappa$.}
	\label{fig:del-us-info}
\end{figure}

In fig.~\ref{fig:del-us-dist-pre-ins} we present the probability distribution functions of $\delta u_s$ normalised by its standard deviation $\left < \delta u_s \right >$ at different air-phase turbulence configurations before the insertion of the droplet, where the turbulent airflow has reached a statistical steady-state. The velocity increments are computed on the spherical surface $r = 1.05R_0$ and are found to abide by a normal distribution. In fig.~\ref{fig:del-us-kappa-corr} we inspect the correlation between $\delta u_s$ and the droplet local curvature $\kappa$ at a fixed time $t/\tau = 1.109$ after droplet insertion, when the droplet has fully flattened. It is found that $\delta u_s$ mostly increases with $\kappa$, except for $(u^*/U_0, \, L_0/R_0) = (0.25, \, 4)$ and $\kappa d_0/2 \geq 15$. Moreover, this positive correlation does not appear to depend on the specific turbulence configurations within the range of $-2d_0/h_i \leq \kappa d_0/2 \leq 2d_0/h_i$, where $2d_0/h_i = 2d_0/d_i$ is the curvature of the peripheral rim according to \eqref{for:ja-rim-diameter}. Figs.~\ref{fig:kappa-visual-0.65-1-1}-\ref{fig:kappa-visual-0.65-1-3} present droplet surfaces at the point of bag initiation, which have been coloured to reveal the spatial distribution of the local curvature $\kappa$. Large values of $\kappa$ are observed primarily at the outer edge of the peripheral rim, where large $\delta u_s$ is also expected. Conversely, negative $\kappa$ is found either on the `neck' connecting the bag film and peripheral rim or the dimples on the bag film, and elsewhere on the bag film $\kappa$ remains close to zero. This suggests that on the droplet surface, the development of highly-curved structures is closely associated with extreme events in ambient turbulence; whereas structures with medium to low local curvature are more likely associated with mean flow effects.

\begin{figure}[htbp]
	\centering
	\subfloat[]{
		\label{fig:del-us-dist-evol}
		\includegraphics[width=.48\textwidth]{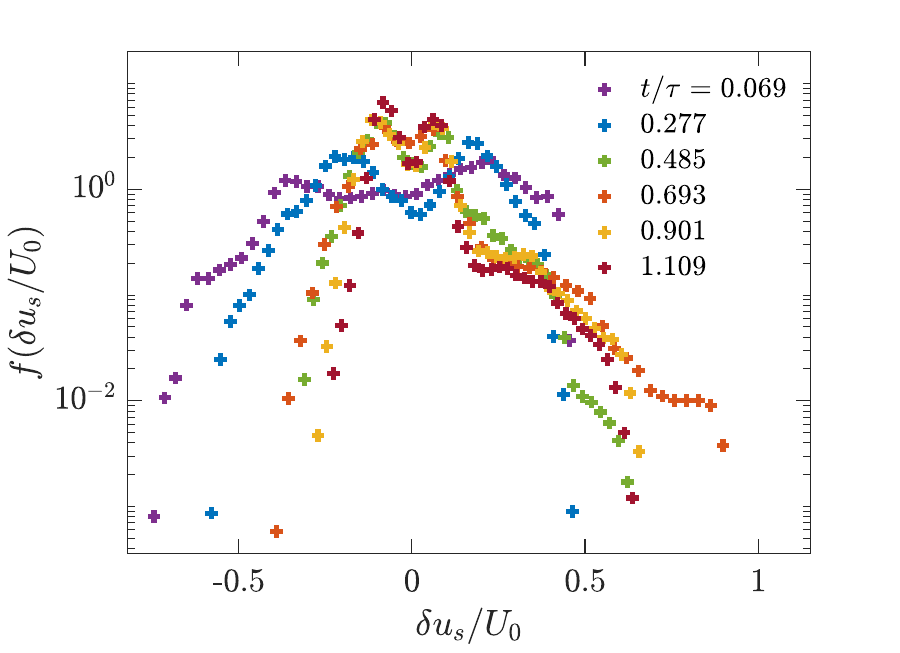}}
	\centering
	\subfloat[]{
		\label{fig:kappa-dist-evol}
		\includegraphics[width=.48\textwidth]{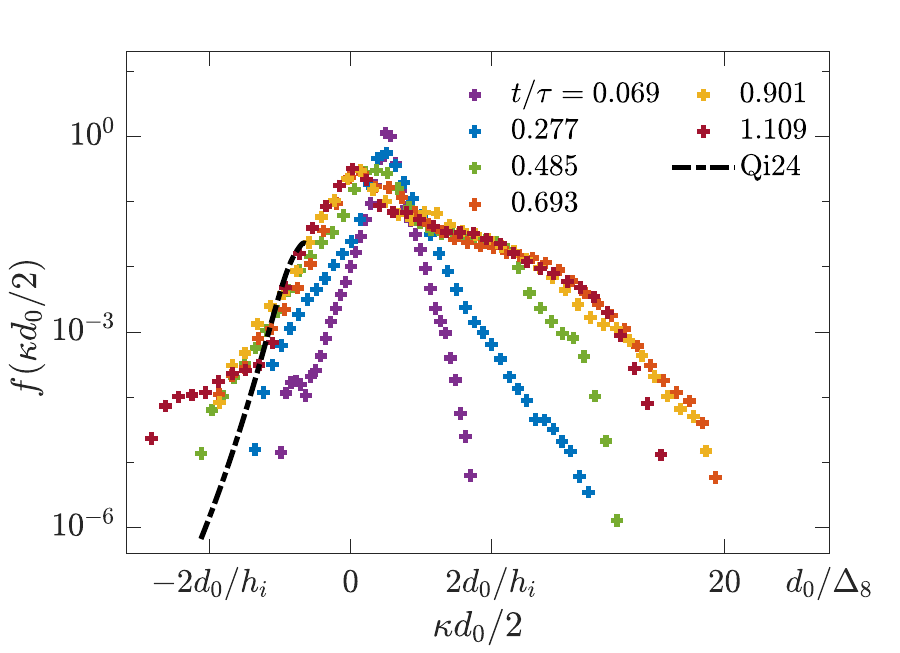}}

	\caption{Evolution of the ensemble-averaged probability distribution functions of surface velocity increment $\delta u_s$ (a) and local interface curvature $\kappa$ (b) over time for $u^*/U_0 = 0.25$, $L_0/R_0 = 1$. We also include in fig.~\ref{fig:kappa-dist-evol} the theoretical prediction of Qi \emph{et al.} \cite{qi2024breaking} for comparison, where the fitting parameter $C_1$ in \eqref{for:eddy-surface} is set as 0.4.}
	\label{fig:del-kappa-evol}
\end{figure}

We further inspect the evolution of $\delta u_s$ over time in fig.~\ref{fig:del-us-dist-evol}. Initially, the distribution of $\delta u_s$ spans over a quite wide range of $-0.75 \leq \delta u_s/U_0 \leq 0.5U_0$ with a flat plateau in the middle bordered by steep tails on either side, reflecting a large range of velocities present in the separated airflow around a largely spherical droplet. As time elapses, the distribution of $\delta u_s$ narrows down with an increasingly steep left tail. Interestingly, while there are already two distinctive peaks present in the distribution of $\delta u_s$ at a relatively early time $t/\tau = 0.069$, which gradually approach one another as time elapses, the monotonically decreasing right tail has developed a shoulder around $\delta u_s/U_0 = 0.25$ by $t/\tau = 0.901$, which persists up to $t/\tau = 1.109$. Taking into account that the two early-time peaks are almost symmetric about $\delta u_s = 0$, we consider that they most likely correspond to the stagnation and separation points at the drop surface, since it is at these locations that the ambient airflow velocity minimises and maximises, respectively. The formation of the shoulder around $\delta u_s/U_0 = 0.25$ suggests a breakdown of symmetry in the probability distribution function so that large positive $\delta u_s$ values are more likely to be observed. Since positive $\delta u_s$ is associated with the droplet periphery, the formation of this shoulder can be understood as an indicator of the torque growth and droplet tilting patterns we discussed earlier in \S\ref{sec:drop-turb-dynamics} and \S\ref{subsec:tilting}.

In contrast to the narrowing trend observed in the distributions of $\delta u_s$, fig.~\ref{fig:kappa-dist-evol} suggests that the distributions of the local interface curvature $\kappa$ broadens over time. Initially, the distribution of $\kappa$ is symmetric and peaks around $\kappa d_0/2 = 2$, which suggests the presence of small-scale perturbations on the largely spherical droplet surface due to interactions with the turbulent airflow. As the droplet deforms over time, the distribution of $\kappa$ broadens and develops a skew towards large positive curvature values, representing bumps or ripple-like irregularities \cite{mangani2022influence}. The peak of $\kappa$ distributions has shifted to $\kappa = 0$ by $t/\tau = 0.901$, which most likely corresponds to the flat surface on either side of the disk-shaped droplet before the bag blows out.

\begin{figure}[htbp]
	\centering
	\subfloat[]{
		\label{fig:kappa-urms-sweep}
		\includegraphics[width=.48\textwidth]{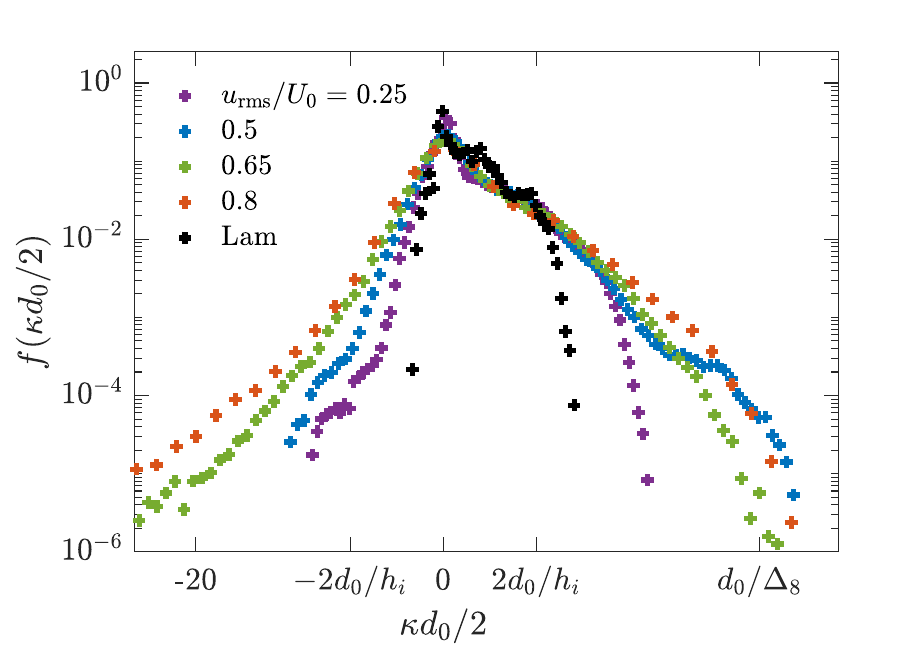}}
	\centering
	\subfloat[]{
		\label{fig:kappa-lla-sweep}
		\includegraphics[width=.48\textwidth]{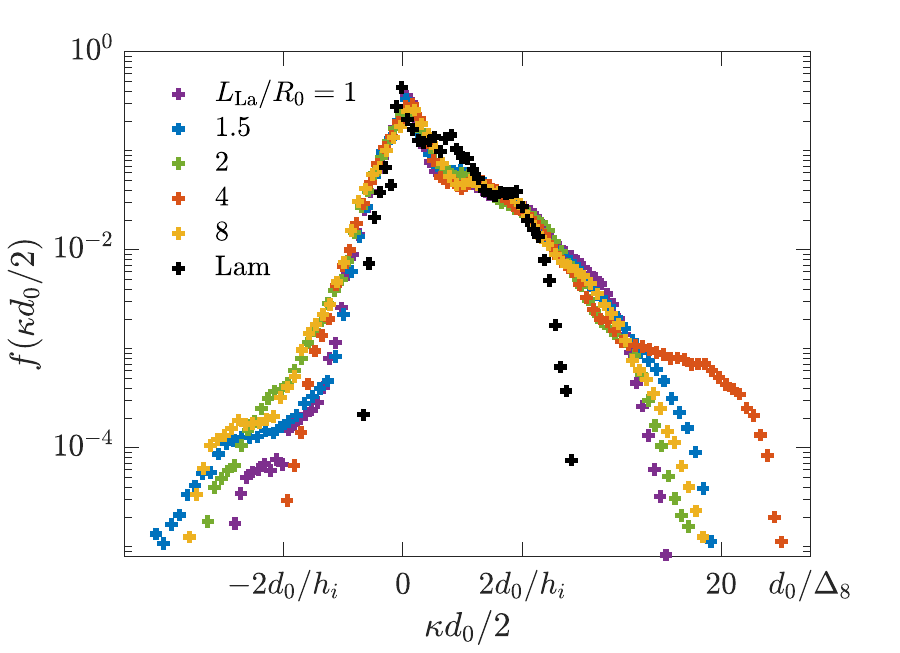}}
  
        \centering
	\subfloat[]{
		\label{fig:del-u-urms-sweep}
		\includegraphics[width=.48\textwidth]{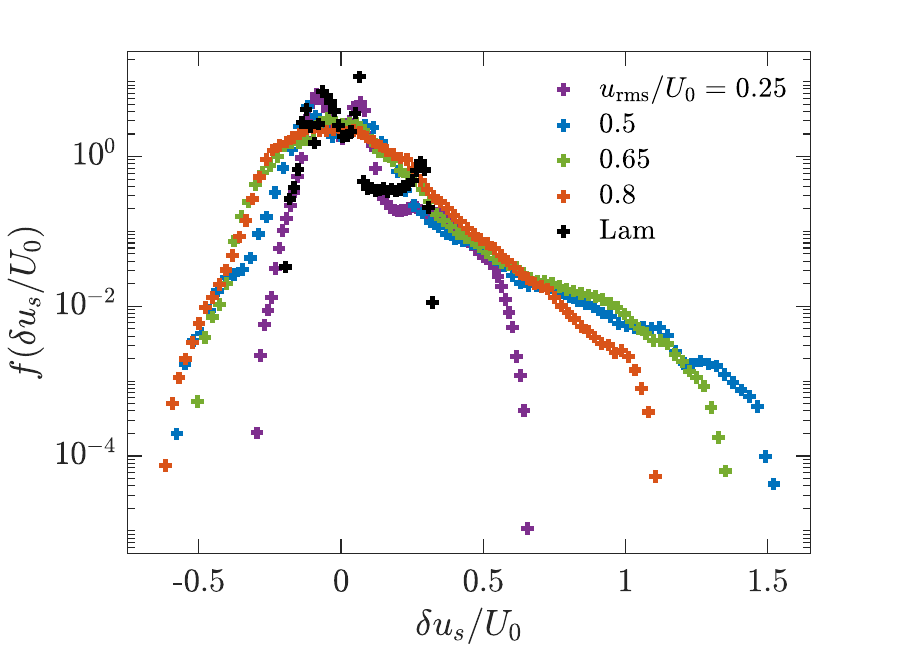}
  }
	\centering
	\subfloat[]{
		\label{fig:del-u-lla-sweep}
		\includegraphics[width=.48\textwidth]{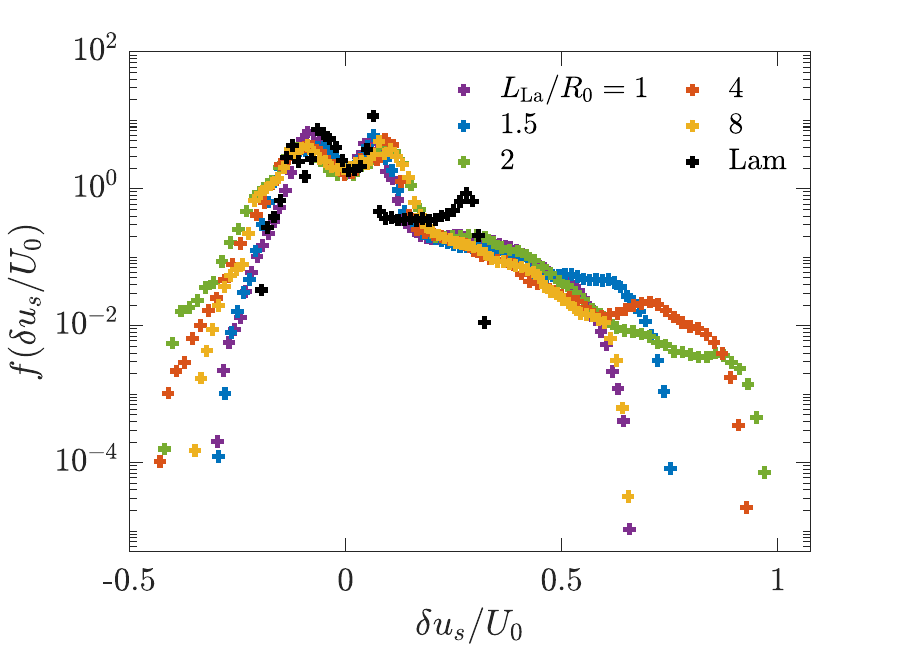}}
	\caption{Ensemble-averaged probability distribution functions of local interface curvature $\kappa$ (a,b) and surface velocity increment $\delta u_s$ (c,d) sampled at $t/\tau = 1.11$ for air-phase turbulence with different fluctuating velocities $u^*$ (a) and injection length scales $L_0$ (b).}
	\label{fig:del-kappa-urms-lla-sweep}
\end{figure}

We also include in fig.~\ref{fig:kappa-lla-sweep} the theoretical prediction of Qi \emph{et al.} \cite{qi2024breaking} for turbulence-bubble interactions, where it is found to partially describe the intermediate-time curvature distribution of negative $\kappa$ values. This theory assumes that the concave regions on the droplets result from their interactions with energetic, sub-droplet-scale eddies with size $D_e < d_0$, whose inertia $\rho_g u_e^2$ is sufficient to overcome the increase in surface energy $\sigma/D_e$ caused by the local deformation; namely
\begin{equation}
    u_{e, \, d} = \sqrt{C_1 \sigma / \rho_g D_e},
    \label{for:eddy-surface}
\end{equation}
where $C_1$ is a fitting parameter. The distribution of instantaneous eddy velocity $u_e$ is given as follows,
\begin{equation}
    P(u_e | D_e) = \frac{3\sqrt{2}}{2} \varepsilon_e^{2/3} D_e^{-1/3} P(\varepsilon_e), 
\end{equation}
where the distribution of the local energy dissipation rate $\varepsilon_e$ is in turn approximated as a log-normal function according to a multi-fractal model. Consequently, the distribution of the sizes of the eddies strong enough to deform the drop surface is expressed as
\begin{equation}
    P(D_e) \propto D_e^2 \omega_c \int_{u_{e,d}}^\infty P(u_e | D_e) du_e,
\end{equation}
where the droplet surface area deformed by the eddy is assumed to scale with $D_e^2$, and the frequency of collision between the droplet and the eddy is approximated as $\omega_c \propto {\left< \varepsilon \right>}^{1/3} D^2 D_e^{-11/3}$ by analogy with the gas kinetic theory. As the local deformation curvature can be estimated by $\kappa \approx -2/D_e$, this allows one to write the distribution of the local negative curvature $\kappa$ as
\begin{equation}
    P(\kappa) \propto \kappa^{-2} P(D_e).
    \label{for:kappa-dist-theory}
\end{equation}

Similar to Qi \emph{et al.} \cite{qi2024breaking}, we find that when setting $C_1 = 0.4$, \eqref{for:kappa-dist-theory} matches partially with the left tail of the simulation results up to $\kappa d_0/2 = -2.5$, where the former shows a kink. While the agreement here suggests that collision with sub-droplet-scale eddies is an important physical mechanism responsible for the dimple formation in our study, we note that the contribution of other mechanisms, such as the balance between the deforming effects of the mean flow and surface tension, cannot be accounted for by this model and still awaits further analysis. Moreover, we note that while fig.~3 in Ref. \cite{qi2024breaking} also shows a skew towards positive curvature values, the right tail of their curvature distribution functions are concave-shaped and extends to very large positive values ($\kappa d_0/2 \approx 40$); whereas our results at $t/\tau = 1.109$ feature convex-shaped right tails that tend to fall off abruptly at a finite value around $\kappa d_0/2 \approx 20$. This is likely because droplets take much longer to deform and elongate, and feature much fewer `sharp' regions at the time of bag formation compared with bubbles with negligible inertia.

In fig.~\ref{fig:del-kappa-urms-lla-sweep} we further show the influence of air-phase turbulence on the distribution patterns of $\kappa$ and $\delta u_s$. Figs.~\ref{fig:kappa-urms-sweep} and \ref{fig:kappa-lla-sweep} suggest that the right tails of the $\kappa$ distributions are most strongly influenced by air-phase turbulence; as compared with their laminar airflow counterpart, all turbulent aerobreakup cases feature an increased number of large positive curvature records. This indicates that local small-scale structures forming on the bag, e.g. the protruding nodes observed in figs.~\ref{fig:back-0.65-1} and \ref{fig:back-0.8-1}, become increasingly prominent as $u^*$ increases. According to Jackiw and Ashgriz \cite{jackiw2022prediction}, the formation of these nodes and the subsequent formation of node drops can be attributed to either RT or RP instability developing on the periphery rim. The influence of air-phase turbulence on the left tails of the curvature distributions is not as prominent as the right tails, as the left tails only slightly broaden when $u^*$ increases. This is most likely because the formation of convex regions on the drop surface with large local curvatures (rims and nodes) benefits jointly from the suction of low ambient gas pressure and capillary pinching effects, whereas creating the concave regions (dimples) requires strong small-scale eddies to act against capillary forces, as we have discussed for fig.~\ref{fig:kappa-dist-evol}. Interestingly, we observe that the curvature distribution within the range of $0 \leq \kappa d_0/2 \leq 8$ does not appear to depend on the turbulence characteristics.

Figs.~\ref{fig:del-u-urms-sweep} and \ref{fig:del-u-lla-sweep} show the velocity increment distributions at different air-phase turbulence configurations. Similar to the curvature distributions, it is found that the introduction of air-phase turbulence leads to a broadening effect, which becomes the strongest at the right tail for $\delta u_s / U_0 \geq 0.5$. Also, similar to the curvature distribution for $0 \leq \kappa d_0/2 \leq 8$, the velocity increment distribution for $0 \leq \delta u_s/U_0 \leq 0.5$ does not appear to depend strongly on the turbulence characteristics. We consider that both these turbulence-independent regimes correspond to the same droplet morphological feature present in all simulations carried out in this study, which is the flattened frontal and back faces of the droplet; since it is here that the most probable values of both $\kappa$ and $\delta u_s$ are found.

\section{Influence of the liquid-gas viscosity ratio $\mu_r$}
\label{sec:visc-ratio}
\begin{figure}[htbp]
    \centering
	\subfloat[]{
		\label{fig:kappa-comp-urms-0.25}
		\includegraphics[width=.48\textwidth]{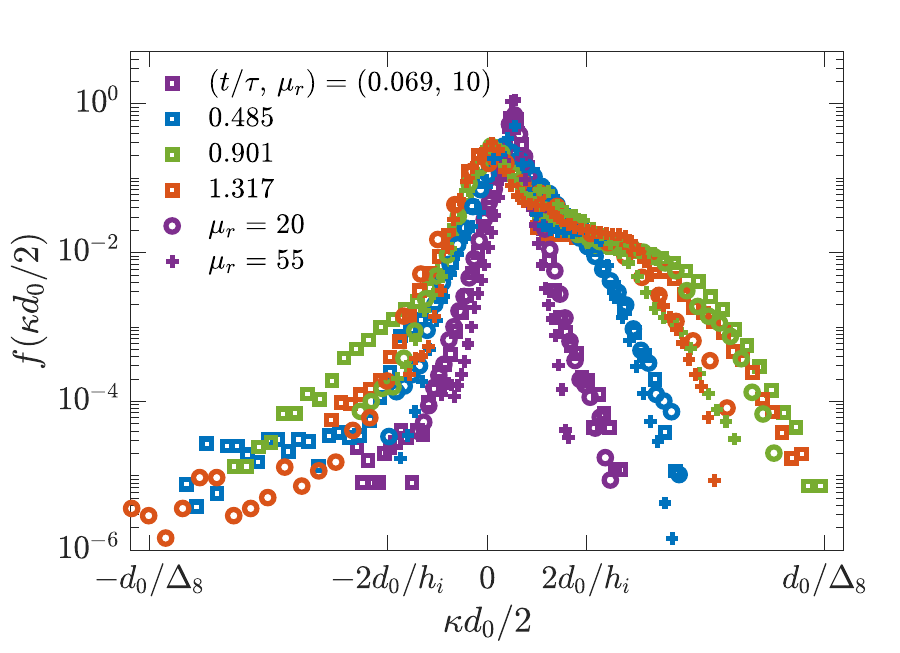}}
	\centering
	\subfloat[]{
		\label{fig:delu-comp-urms-0.25}
		\includegraphics[width=.48\textwidth]{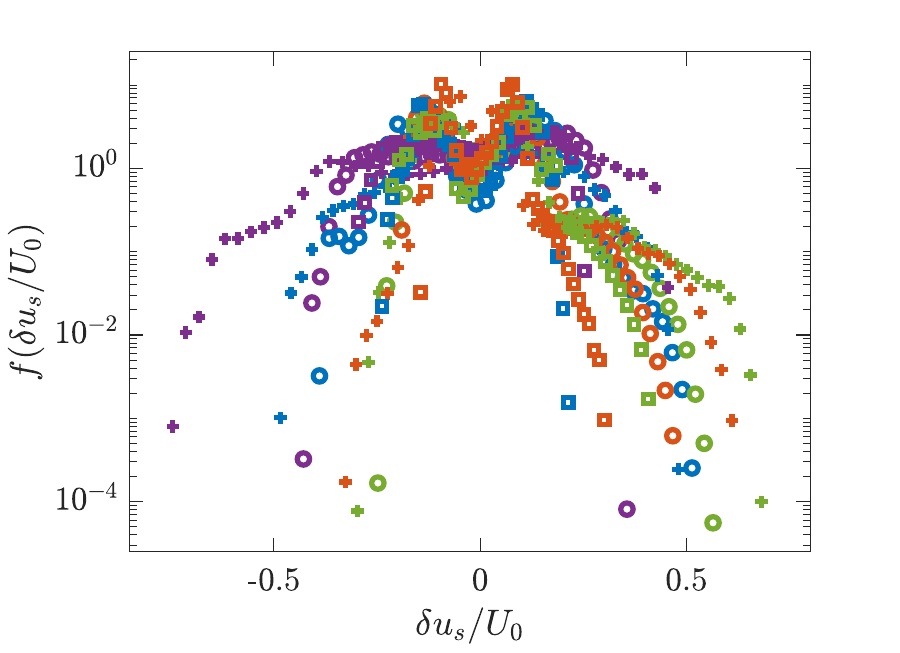}}

    \centering
	\subfloat[]{
		\label{fig:kappa-comp-urms-0.5}
		\includegraphics[width=.48\textwidth]{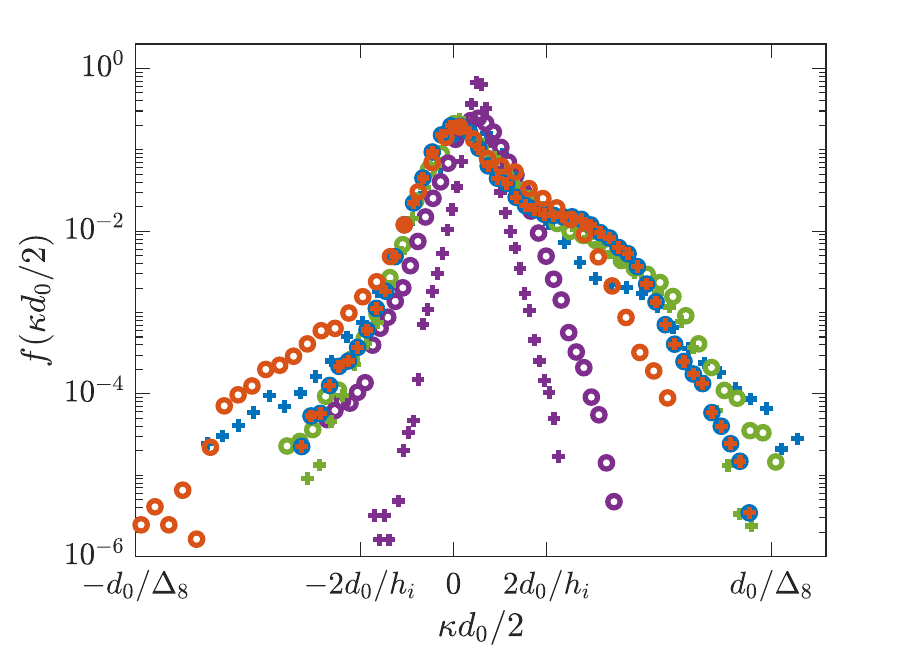}}
	\centering
	\subfloat[]{
		\label{fig:delu-comp-urms-0.5}
		\includegraphics[width=.48\textwidth]{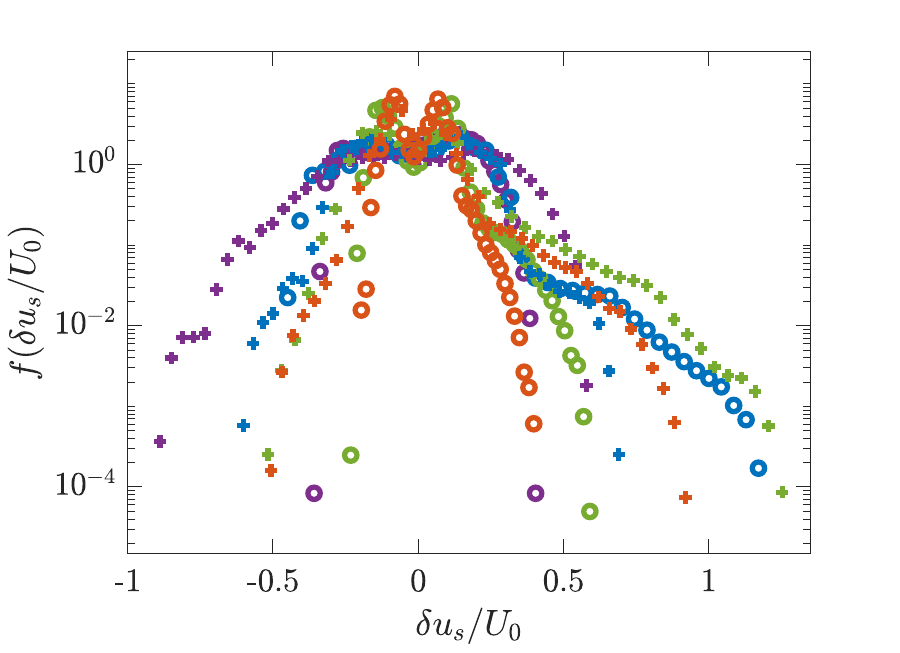}}

    \centering
	\subfloat[]{
		\label{fig:kappa-comp-urms-0.65}
		\includegraphics[width=.48\textwidth]{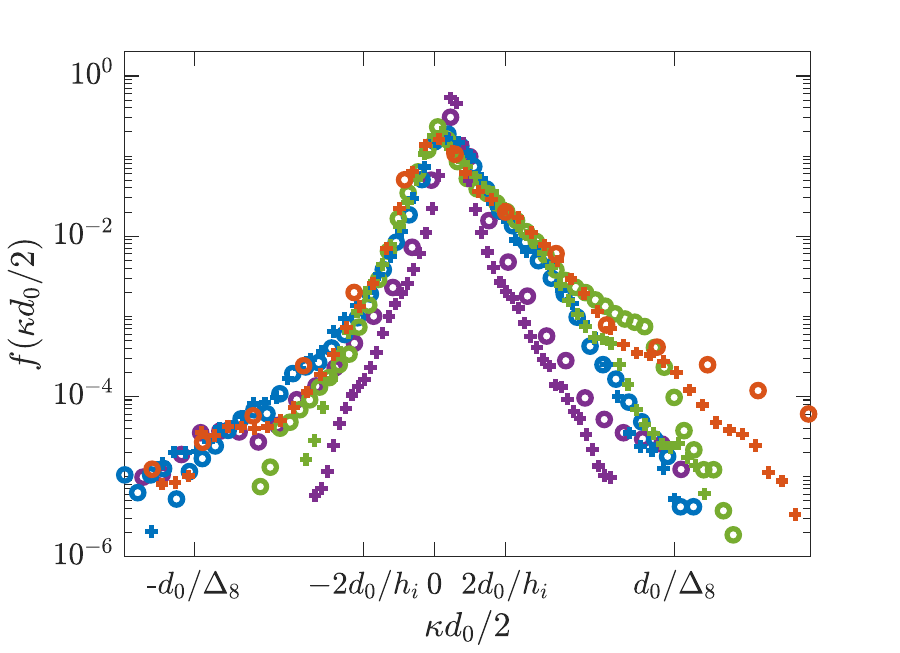}}
	\centering
	\subfloat[]{
		\label{fig:delu-comp-urms-0.65}
		\includegraphics[width=.48\textwidth]{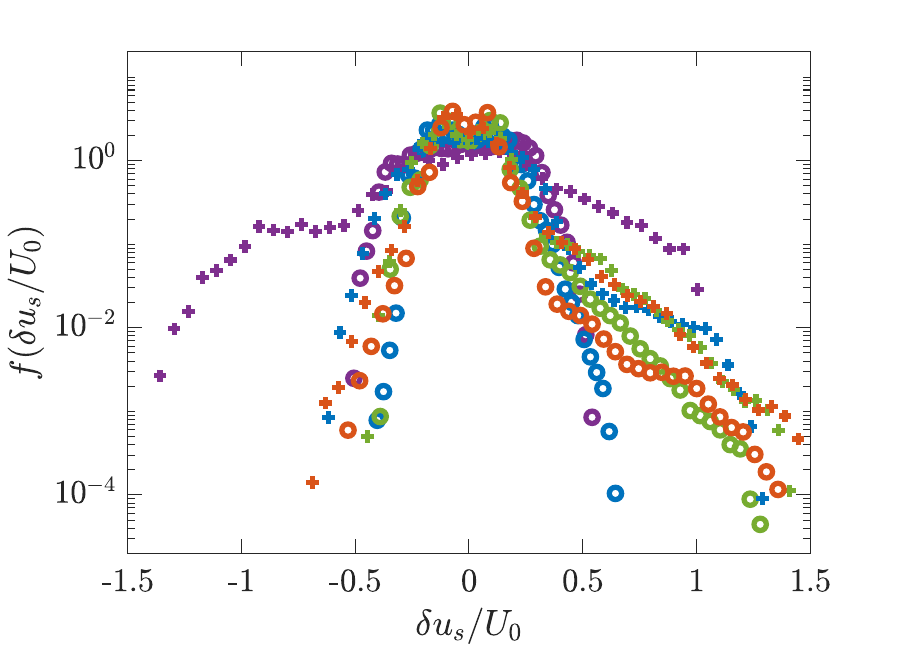}}

	\caption{Comparison of ensemble-averaged probability distribution functions of local curvature and velocity increment at $t/\tau = 1.11$ and $\mu_r = 10$, 20 and 55. (a,b): $u^*/U_0 = 0.25$; (c,d): $u^*/U_0 = 0.5$; (e,f): $u^*/U_0 = 0.65$; while for all simulation cases $L_0/R_0 = 1$.}
	\label{fig:delu-kappa-mur-comp}
\end{figure}

As can be seen from \eqref{for:lambda-re-eta-defs}, the Kolmogorov microscale $\eta$ is directly linked with the air-phase kinematic viscosity $\nu_g$; therefore a decrease of the viscosity ratio $\mu_r$ can cause an increase in $\eta$ and allow the simulation to better resolve air-phase turbulence dissipation. The effects of $\mu_r$ on the air-phase turbulence characteristics and the droplet centre-of-mass properties have already been covered in \S\ref{subsubsec:turb-gen} and \S\ref{sec:drop-turb-dynamics}, respectively. Here, we present distributions of droplet interface curvature $\kappa$ and surface velocity increment $\delta u_s$ obtained at different times with reduced $\mu_r$ values of 20 and 10 (only for $u^*/U_0 = 0.25$), and compare with those obtained for $\mu_r = 55$ in fig.~\ref{fig:delu-kappa-mur-comp}. This allows us to provide some insight into the effects of the liquid-gas viscosity contrast on the deformation characteristics of the droplet and its interaction with the ambient turbulence.

It is observed that for all three turbulence configurations shown in fig.~\ref{fig:delu-kappa-mur-comp}, the local interface curvature distribution does not show any consistent change with decreasing $\mu_r$, with the only noticeable difference occurring in the left and right tails from time to time. Alongside results presented in \S\ref{subsubsec:turb-gen} and \S\ref{sec:drop-turb-dynamics}, these data suggest that the liquid-gas viscosity ratio and small eddies at the Kolmogorov microscale do not significantly affect the droplet dynamics and global deformation patterns. Furthermore, the insensitivity of the droplet behaviour to $\mu_r$ justifies our use of $L_{turb} = 8$ for resolving only large eddies in the turbulent bulk flow. On the other hand, the velocity increment distributions show a stronger dependence on $\mu_r$; especially at a very early time $t/\tau = 0.069$ immediately after droplet insertion, as we observe the distribution of $\delta u_s$ broadens as $\mu_r$ increases. This is expected since larger values of $\mu_r$ correspond to larger gas-phase Reynolds numbers $Re$, which in turn lead to stronger turbulence and more extreme velocity values recorded around the droplet.

\section{Concluding remarks}
\label{sec:discussions}
We have presented novel direct numerical simulations of droplet deformation in a turbulent mean flow, systematically investigating various aspects of the dynamic and morphological droplet evolution patterns. Utilising a robust synthetic turbulence generation method \cite{xie2008efficient}, we obtained ambient airflows with validated turbulence statistics, and reproduced various aspects of droplet behaviour observed in pioneering works by Zhao \emph{et al}. \cite{zhao2019effect} and Jiao \emph{et al.} \cite{jiao2019direct}, including the increase in bag size, tilting and formation of high-curvature local interfacial geometries. Our results shed light on the detailed physical mechanisms governing the droplet acceleration and deformation processes, thus laying the foundation for future works investigating the detailed turbulent bag breakup dynamics and associated fragment statistics. Here, we further discuss the possible implications of introducing air-phase turbulence on the fragment statistics resulting from bag breakup, which will be the focus of more detailed follow-up work. 

Firstly, our figs.~\ref{fig:ux-rdot-urms-lla-sweep} and \ref{fig:acc-dist-evol} suggest that introducing turbulence causes the droplet to further accelerate in the streamwise direction, with some degree of randomness in the radial direction resulting from turbulence lift effects. We therefore expect promoted droplet surface instability development and draining of bag films at later times, leading to an earlier perforation time of the latter; which is already reported in previous experimental studies \cite{zhao2019effect}.

Secondly, as is observed in the snapshots in fig.~\ref{fig:back-snapshots}, the flattened droplet exhibits diverse morphological features not observed in laminar aerobreakup simulations \cite{tang2022bag, ling2023detailed}. Among these are the distortion of the bordering rim and formation of protruding nodes, which we expect to eventually give rise to `node drops' similar to the experimental observations of Jackiw and Ashgriz \cite{jackiw2022prediction}, and result in a reduction of the final volume fraction of rim drops resulting from Plateau-Rayleigh breakup \cite{Jackiw2021}. Additionally, the formation of small-scale, highly-curved features on the droplet at late times, as observed in figs.~\ref{fig:del-kappa-evol} and \ref{fig:del-kappa-urms-lla-sweep}, suggests a more corrugated bag film which will disrupt the hole expansion process governed by the Taylor-Culick mechanism \cite{tang2022bag}, modifying the development of hole rim instabilities and the subsequent drop shedding processes \cite{vledouts2016explosive, jackiw2022prediction, neel2020fines, tang2024fragmentation}. We therefore expect the final fragment size distribution arising from turbulent aerobreakup to become broader compared with laminar aerobreakup scenarios, since the parent features of such fragments become more varied. Nonetheless, it remains a challenge to properly resolve the bag film perforation process and obtain high-fidelity fragment statistics, even with the aid of the MD algorithm for establishing grid-converged size distributions for large fragments \cite{chirco2021manifold, tang2022bag, kulkarni2024atomizing}.

Lastly, we note that the tilting behaviour of the droplet, as observed and analysed in figs.~\ref{fig:tilting-snapshots} and \ref{fig:theta11-theta-slip-urms-lla-sweep}, implies more widespread trajectories and spatial distributions of the fragments compared with the laminar bag breakup case, as they are now ejected from a highly inclined and corrugated bag surface. A more detailed understanding of the tilting dynamics would therefore deepen our understanding of the breakup of bag-shaped sea surface perturbations which also feature oblique orientations, as discussed by Troitskaya \emph{et al.} \cite{Troitskaya2017, Troitskaya2018, troitskaya2023statistical}.

\section{Acknowledgment}
Insights from Professor R. J. Hearst have helped in improving
the analysis of air-phase turbulence characteristics in this work. The authors would like to thank EPSRC for the computational time made available on the UK supercomputing facility ARCHER2 via the UK Turbulence Consortium (EP/R029326/1). Use of the University of Oxford Advanced Research Computing (ARC) facility is also acknowledged. K. Tang is supported by a Research Studentship from the University of Oxford.

\bibliography{references}

\end{document}